\RequirePackage{ifpdf}
\documentclass[12pt]{JHEP3}
\pdfoutput=1
\usepackage{amsmath}
\usepackage{graphicx}
\usepackage{color}
\let\normalcolor\relax

\def\eqref#1{(\ref{#1})}

\def\be{\begin{equation}}
\def\ee{\end{equation}}
\def\bea{\begin{eqnarray}}

\def\eea{\end{eqnarray}}
\def\nn{\nonumber}

\def\l{\lambda}

\def\lab{\label}
\def\f{\phi}

\def\le{\left}
\def\ri{\right}

\def\half{\frac12}

\def\6{\partial}

\def\a{\alpha}

\def\lab{\label}

\def\re{\mathrm{Re}}
\def\im{\mathrm{Im}}
\def\morder#1{\mathcal{O}\left(#1\right)}
\def\Sq{S}
\def\Sqt{\widetilde S}
\def\flf{\Xi}

\title{Fluctuations in non-conformal holographic plasma at criticality}

\author{Panagiotis Betzios$^{a}$, Umut G\"ursoy$^{b}$, Matti J\"arvinen$^{b}$, Giuseppe~Policastro$^{c}$
~\\ \\
$^a$ Crete Center for Theoretical Physics, Institute for Theoretical and Computational Physics, 
Department of Physics, University of Crete 71003, \\
Heraklion, Greece.
~\\ \\
$^b$ Institute for Theoretical Physics, Utrecht University \\
Princetonplein 5, 3584 CC Utrecht, The Netherlands
~\\ \\
$^c$ Laboratoire de Physique Th\'eorique de l'\'Ecole Normale Sup\'erieure, CNRS, \\
Universit\'e PSL, Sorbonne Universit\'es, Universit\'e Pierre et Marie Curie, \\
24 rue Lhomond, 75005 Paris, France \\
}

\abstract{We continue the study initiated in [arXiv:1708.02252] of the fluctuations of a strongly-coupled non-conformal plasma described holographically by Einstein gravity coupled to a dilaton with an exponential potential. The plasma approaches a critical point of a continuous phase transition in a specific limit, where the metric becomes a linear-dilaton background. This results to an analytic description of the quasi-normal mode spectrum, that can be extended perturbatively in the deviation away from the critical point. 
In the previous paper we showed that at criticality the quasinormal frequencies coalesce into a branch cut on the real axis. In this paper we give a more extended and complete discussion of these results. We compare in detail the numerical and analytical approximations in order to confirm their validity; we study (numerically and in a WKB approximation) the momentum dependence of the modes, in order to determine the cross-over scale that limits the validity of the hydrodynamic approximation, and which becomes arbitrarily low at the critical point; and we discuss in detail the procedure we use to complete the theory in the UV by gluing a slice of AdS geometry, and the extent to which it should provide a good approximation to a smooth UV-complete situation. }

\preprint{CCTP-2018-7\\
ITCP-IPP 2018/5\\
LPTENS/18/11}
\keywords{\vspace{-1cm} AdS/CFT, Quark-Gluon Plasma, thermalization}

\begin{document}



\section{Introduction}

The prime example of a holographic theory is the maximally supersymmetric ${\cal N}=4$ SYM theory, for which we have a host of results derived from the gravitational description on its properties in the hydrodynamic regime. Despite usefulness of this theory as a benchmark, it has some drawbacks  that limit its applicability to real-world systems, for example systems closely related to the quark-gluon plasma (for a recent review see  \cite{CasalderreySolana:2011us}). Among these drawbacks is the fact that it is exactly conformally invariant, in contrast to 
most interesting and realistic systems. It is, therefore, important to understand the effect of breaking of conformality on the transport properties of the system, in order for instance to improve the understanding of the behavior of the quark-gluon plasma close to the deconfinement phase transition \cite{Ryu:2015vwa}. 

In this paper we continue the study of the behaviour of the plasma phase of a non-conformal field theory, holographically dual to Einstein gravity coupled to a 
scalar field with a potential of the form $e^{-8X \phi/3}$ where $X$ is a constant that measures deviation from the conformal limit $X=0$. We refer to this non-conformal plasma as the {\em Chamblin-Reall plasma}~\cite{ChamblinReall}. The choice of this class  of models results from a compromise between the desire for a realistic setup for the description of a real-world QCD quark-gluon plasma, and simplicity that is more amenable to 
analytic analysis. 
In the models considered in the Improved Holographic QCD program \cite{ihqcd1, ihqcd2} the  choice of the potential is dictated by several requirements 
(asymptotic freedom, linear confinement in the vacuum, spectra of glueballs, matching the equation of state) and as a result one obtains  a more complicated 
potential that leads to gravitational backgrounds which can only be solved numerically.  By contrast, the models we consider admit analytic black hole solutions, which considerably simplifies  their study and allows obtain more precise results. On the other hand these models capture one essential feature of the realistic models, namely the lack of conformal invariance, that can be tuned by changing the  parameter $X$ in the potential.  

In 
\cite{Gursoy:2015nza} three of the present authors found a time-dependent black hole solution, corresponding to a boost-invariant flow of the Chamblin-Reall plasma, which allowed us characterise the rate of approach to equilibrium. We found that the rate of cooling could be parametrically slower than in the conformal case, with temperature decaying in time according to a power law $T \sim \tau^{-s/4}$, where the exponent $s$ relates to the parameter in the potential as $s = 4/3(1-4X^2)$, with $s=4/3$ corresponding to the conformal case. 
It was also shown in \cite{Gursoy:2015nza} that these gravitational findings are consistent with predictions from the hydrodynamics Ansatz, assuming that the temperature would follow adiabatically the evolution of the energy density and the pressure  determined by the equation of state.  

Contrary to the conformal plasma with a fixed critical exponent  $s=4/3$, in the more general case of the Chamblin-Reall plasma $s$ can be made arbitrarily small by letting $X$ approach a critical value $X_c = -1/2$. This particular critical value corresponds to a theory that exhibits a continuous confinement/deconfinement transition at a finite temperature \cite{Gursoy:2010kw,Gursoy:2010jh}. In these papers, criticality in the limit $X\to X_c = -1/2$ was established and it was noticed that Hawking-Page transition between asymptotically AdS black-brane and the thermal gas solutions in Einstein-dilaton gravity becomes continuous (second or higher order). Moreover, the string frame metric in the vicinity of the transition becomes a {\em linear dilaton background} of bosonic string theory, a fact that becomes instrumental in studying two-point functions of Polyakov loops in the vicinity of the transition \cite{Gursoy:2010kw}. 

These previous results and observations prompted us to consider more closely the near-critical regime. A useful probe of the system  beyond the thermodynamical and hydrodynamical regimes is given by the spectrum of quasi-normal modes,  that reflect the non-hydrodynamic fast-relaxing 
processes of the system.  We have reported on the main features of the spectrum of fluctuations  in \cite{Betzios:2017dol};  in that paper we considered the sector of spin-two modes at zero momentum, derived an analytic expression for the correlator valid near the critical point, and showed that 
 in the $X\to -1/2$ limit the quasi-normal poles condense into a branch cut on the real axis; we also discussed a UV completion of the model, obtained by 
gluing a slice of AdS near the boundary, and showed that the QNM form two distinct sets, that can be identified respectively with modes associated to the CR geometry in the IR, and modes associated with the UV part.

 In the present paper we give a more extensive and complete picture; in particular we compare in detail the analytical and numerical approximations, we study the momentum and frequency dependence of the QNMs, and study in great detail the approach to criticality at $X_c = -1/2$ and the behavior of fluctuations of the system in this limit. 

We find it convenient to parametrize the conformal breaking by a parameter $\xi = 4(1-X^2)/(1-4X^2) $, related to $X$ such that  $\xi \to \infty$ as $X \to X_c$. In this paper, we solve the fluctuation equations analytically in a perturbative expansion in $\xi^{-1}$, that is in the vicinity of criticality, and compare with numerical results that can be obtained for any $\xi$. 
First, we observe that the fluctuations in the critical limit are controlled by the linear dilaton geometry as one would expect from \cite{Gursoy:2010kw,Gursoy:2010jh}. We prove that this is the case by comparing and precisely matching the well-known reflection amplitude on the linear dilaton blackhole \cite{Witten:1991yr, Dijkgraaf:1991ba, Alexandrov:2003ut, Nakayama:2004vk} from which we obtain the QNM spectrum analytically.  We also observe a very interesting connection between the Chamblin-Reall blackhole in the critical limit $\xi\to\infty$ and the large D expansion of a D-dimensional AdS black hole \cite{Emparan:2013moa,Emparan:2015rva}. We further dwell on this connection in section \ref{sec::conclusion} at the end of the paper. 

Contrary to the results of previous works \cite{Janik:2015waa,Buchel:2015saa,Ishii:2015gia}  that found a mild dependence of the QNM on the breaking of conformal invariance in different models,  we find that the position of the QNM depends strongly on $\xi$. In fact the imaginary part goes to zero as $1/\xi$, all the modes approach the real axis, and presumably merge in this limit to form a branch cut on the real axis, similarly to what happens for the BTZ black holes at extremality \cite{Birmingham:2001pj}.  What is remarkable is that here the branch cut appears in a limit in which the temperature remains non-zero, indicating a sort of dissipationless fluid. Moreover, we find that the hydrodynamic modes do not scale with $\xi$, and thus decouple from the low-energy description in the critical limit.

We have also analyzed the dependence of QNM spectra on momentum. 
Generically the hydrodynamic mode is expected to dominate the long-time dynamics, but there is a scale of momentum for which the hydrodynamic and quasi-normal modes cross, so that above this momentum the hydrodynamic mode ceases to be the longest-lived excitation. In ${\cal N}=4$ SYM, for instance,  the crossover scale is at $k\approx 1.3 \, (2 \pi T)$ \cite{Amado:2008ji,Landsteiner:2012gn}. In our case, the crossover scale is of the order  $1/\sqrt{\xi}$ and so it is pushed at arbitrarily low momentum as the critical limit is approached, which signals a dramatic breakdown of the hydrodynamic description.

It seems clear that the vicinity to a critical point should be responsible for this behavior, but we should remark that generically close to a second order phase transition there is a divergent correlation length and correspondingly some new gapless modes appear that have to be included in the hydrodynamics, however they are just a small discrete set of new poles that can approach the origin, so the situation in our case is quite different. It would be interesting to understand better the critical limit and the extent to which one can have an effective description for this pressureless, dissipationless fluid. 

In particular the connection we find  with the studies of the linear dilaton BH at criticality, indicates that the infinite number of gapless modes might be due to a hidden infinite symmetry present in the model, that would indicate the presence of integrability at criticality (while in most second order phase transition points one generically just expects the presence of conformal symmetry). In studies of non-critical string theory on the linear dilaton background this symmetry is related to $W_\infty$, and the hydrodynamic description is in terms of an incompressible fermi-fluid possessing an infinite number of conserved quantities~\cite{Ginsparg:1993is}.
 
We could also determine analytically the large-$q$ behavior of the quasinormal modes, and found that $\text{Im}\,  \omega \sim q^{1-\alpha}, \alpha = {2 \xi \over 2+ \xi} = {4 \over 3} {1-X^2 \over 1 - 2 X^2}  $.  As emphasized in  \cite{Fuini:2016qsc},  this information characterizes the shape of weakly-damped, narrow spikes or shock waves that propagate through the plasma and contribute to the long-time dynamics. 
 The study of the momentum dependence of the modes also reveals an interesting level-crossing phenomenon: the hydro mode, which starts as the lowest mode at $q=0$, can be found to have swapped place with one or more of  the non-hydrodynamic modes at large $q$, and for large enough $\xi$ it appears to stay below all the other QNM.

We find that, even though one can safely take the critical limit in the IR of the Chamblin-Reall geometry, it is necessary to include effects of finite $\xi$ in order to have a well-posed boundary problem. Indeed the strict $\xi = \infty$ limit is not meaningful for the purpose of defining correlators in the dual field theory. We can make sense of the holographic correlators in the theory by connecting the IR geometry of the Chamblin-Reall solution to an asymptotically AdS geometry in the UV \cite{Gursoy:2015nza}  and  study the modification of the QNM spectrum due to the introduction of  this UV regulator. This can be done by  modifying the potential in the UV or by other means such as a hard wall. In this article we concentrate on an approach where we cutoff the CR geometry and attach to it a slice of AdS geometry, and glue accordingly also the fluctuations around the background, in which case the QNMs can still be solved analytically. 
We found that this gluing procedure leads to two branches of QNMs: a first set at low frequency that corresponds to the first few modes of the CR geometry, and at higher frequencies a new set of modes with almost constant imaginary part. As the temperature is lowered from $T_c$ to zero, the first set becomes longer and approach the real axis, so to reproduce in the zero-temperature limit the modes of the CR black hole, whereas the second set is pushed to higher and higher frequencies\footnote{A similar taming of the holographic description for 2D de-Sitter solutions was considered in \cite{Anninos:2017hhn} where the authors also glued an IR $dS_2$ black-hole geometry to a UV $AdS_2$.}. After the gluing, the limit $\xi \to \infty$ is regular and we can show explicitly and analytically, how the QNMs corresponding to the CR geometry accumulate to form a branch cut on the real axis of the complex frequency plane.

The paper is organized as follows. In the next section we present the Chamblin-Reall blackhole solution for an arbitrary value of $\xi$ and discuss its thermodynamic properties. We also present the critical geometry that arise in the limit $\xi \to \infty$ and show that it coincides with the linear dilaton blackhole. Sections \ref{sec:flucts}, \ref{sec:Xonehalf} and \ref{sec:UVcompletion} contain our main findings. In section \ref{sec:flucts}, we derive and solve the fluctuation equations numerically and obtain the QNM spectra of Chamblin-Reall blackhole for arbitrary $\xi$. In particular we investigate the dependence of the spectra on $\xi$ and momentum.  In section \ref{sec:Xonehalf} we focus in detail the critical limit $\xi\to\infty$ and obtain the QNM spectra analytically. In section \ref{sec:UVcompletion}, we glue the IR geometry of the Chamblin-Reall black hole to a UV regulator that is an asymptotically AdS geometry and study both the QNM spectra modified by such gluing procedure and obtain the holographic correlation functions that now are well-defined after this gluing procedure. Section \ref{sec::conclusion}  contains a discussion of our results and an outlook. 

\section{CR backgrounds} \label{sec:bg}

The model we consider is defined by the five-dimensional Einstein-dilaton gravity
\begin{equation}\label{action}
 {\cal A} = \frac{1}{16\pi G_5} \int d^{5}x \sqrt{-g}\left( R - \frac43 (\6\phi)^2 
+ V(\phi)
  \ri)+\,\,\, G.H.
\end{equation}
with the potential
\be \label{potdef}
 V(\phi) = \frac{12(1-X^2)}{\left(4X^2\right)^2\ell^2} e^{-\frac{8 X}{3} \phi}
\ee
where\footnote{The parameter $X$ matches the phase variable of~\cite{ihqcd1,ihqcd2}, defined in the domain wall coordinates as $3 X = \frac{d\phi}{du}/\frac{dA}{du}$, which is a constant for the CR backgrounds.} $-1<X<0$, and $\ell$ is a positive parameter of length dimension. 
\\
\\
This system admits an exact black-brane solution~\cite{ChamblinReall}, which we will refer to as the CR geometry. The metric can be written in the domain wall coordinates as~\cite{Gursoy:2015nza}  
\be\lab{dw}
ds^2 = e^{2A(u)}\le(-f(u) dt^2 + dx_idx^i)\ri) + \frac{du^2}{f(u)}\, 
\ee
with 
\bea \label{BHmetX}
A &=& A_0 +\frac{1}{4X^2} \log \le(\frac{u_0-u}{\ell}\ri) \nonumber \\
f & = &  1-\le(\frac{u_0-u}{u_0-u_h}\ri)^{-\frac{1-X^2}{X^2}} \\ 
 \l &\equiv& e^{\f} = \le(\frac{u_0-u}{\ell}\ri)^{\frac{3}{4X}} \ ,
\eea
where $A_0$, $u_0$ and $u_h$ are integration constants. The boundary is located at $u=-\infty$, horizon at $u=u_h$ and there is a curvature singularity at $u=u_0$. 
\\
\\
In order to study fluctuations, it is convenient to switch from $u$ to the conformal coordinate $r$, defined by requiring that the warp factors of $dt^2$ and $dr^2$ are the same. This leads to
\be \label{relldef}
 r = \frac{4X^2e^{-A_0}\ell}{1-4X^2} \le(\frac{u_0-u}{\ell}\ri)^{-\frac{1-4X^2}{4X^2}} \equiv \ell' \le(\frac{u_0-u}{\ell}\ri)^{-\frac{1-4X^2}{4X^2}} \ ,
\ee
for $X\neq -1/2$ and 
\be \label{relldef12}
 \frac{r}{\ell} = -e^{-A_0}\log\le(\frac{u_0-u}{\ell}\ri) \ ,
\ee
for $X=-1/2$. We fixed the integration constant in (\ref{relldef}) such that the boundary is at $r=0$ for $-1/2<X<0$ and defined a new length scale $\ell'$ for later convenience. The threshold value $X=-1/2$ plays a central role in our work and will be the focus of the next subsection.
\\
\\
We also find convenient to work with the ingoing Eddington-Finkelstein coordinates by switching to the new time coordinate $dv = dt - dr/f(r)$, 
where the blackening factor is
\be \label{fxidef}
 f(r) = 1 - \left(\frac{r}{r_h}\right)^\xi \ , \qquad r_h = \ell'  \le(\frac{u_0-u_h}{\ell}\ri)^{-\frac{1-4X^2}{4X^2}}\ , \qquad \xi \equiv \frac{4(1-X^2)}{1-4X^2} \ .
\ee
Moreover we use the dimensionless radial coordinate $\hat r = r/\ell'$. Putting everything together, the solution is
\bea \label{grcoord}
 ds^2 &=& e^{2A_0}\hat r^{-\frac{2}{1-4X^2}}\left[-2 \ell' d\hat r dv -f(r) dv^2 + \delta_{ij} dx^{i} dx^{j}\right] \\
 \l &=& \hat r^{-\frac{3X}{1-4X^2}} = \hat r^{-X(\xi-1)}
\eea
The coordinate $\hat r$ runs from $0$ at the boundary to $\hat r_h =r_h/\ell'$ at the horizon. 

The temperature of the brane solution is given by
\be \label{Tdef}
T = \frac{\xi}{4\pi \hat r_h \ell'}  = \frac{1-X^2}{4\pi X^2\ell} e^{A_0} \le(\frac{u_0-u_h}{\ell}\ri)^{\frac{1-4X^2}{4X^2}} \, .
\ee
The entropy density is, 
\be \label{entdef}
S =  \frac{1}{4G_5}\ e^{3A_0}\ \hat{r}_h^{-\frac{1}{1-4X^2}}=\frac{1}{4G_5}\ e^{3A_0}\le(\frac{u_0-u_h}{\ell}\ri)^{\frac{1}{4X^2}}\, .
\ee
These geometries belong to the neutral hyperscaling violating geometries explored in \cite{Gouteraux:2011ce}. 

\subsection{CR solution for $X=-1/2$}\label{Xhalf}

We noticed the special value of $X=-1/2$ above where equation (\ref{relldef}) has a coordinate singularity. In fact this value was singled out and studied in detail in \cite{Gursoy:2010jh} where an emergent IR conformal behavior was observed. In the same paper it was also pointed out that the corresponding vacuum solution for this choice of $X$ is the product of the linear dilaton background of 2D non-critical string theory with $R^3$. Thus it should be governed by an exact CFT (Liouville theory / WZW model). Here we study in more detail the CR solution exactly at $X=-1/2$ in the regular domain-wall coordinates that are valid for $-1 < X < 0$.
The potential is
\begin{equation}\label{exppot}
    V \,  =  \frac{9}{\ell^2} e^{\frac43 \, \phi}\, .
\end{equation}
The metric of the black-hole solution is,
\bea
    ds^2 &=& e^{2A_0}\le(\frac{u_0-u}{\ell}\ri)^{2}
    \le\{dx_idx^i -    
\le(1-\le(\frac{u_0-u}{u_0-u_h}\ri)^{-3}
\ri)dt^2\ri\}\nonumber\\
   {}&& + 
\le(1-\le(\frac{u_0-u}{u_0-u_h}\ri)^{-3}
\ri)^{-1}du^2\label{BHmet}.
\eea
There is an event horizon located at $u_h$. 
The dilaton reads
\begin{equation}\label{dilsoll}
    \l \equiv e^{\f}  = \le(\frac{u_0-u}{\ell}\ri)^{-\frac{3}{2}}\, .
\end{equation}
The corresponding vacuum solution is obtained by sending $u_h\to u_0$. One can easily check that these solutions are indeed related to the 2d black hole and linear dilaton background once passed to the string frame \cite{Gursoy:2010jh}, as follows. First we pass to the conformal coordinates using  (\ref{relldef12})\footnote{We set $A_0=0$ for simplicity in the rest of this section.}:
\be
    ds^2 = e^{-\frac{2r}{\ell}} \le( dr^2 \le(1- e^{\frac{3(r-r_h)}{\ell}}\ri)^{-1} - dt^2 \le(1- e^{\frac{3(r-r_h)}{\ell}}\ri) + dx_i dx^i \ri); \,\,\, \f = \frac{3r}{2\ell} \label{dilBH}\, .
\ee
The vacuum solution is now obtained by sending $r_h\to\infty$ which replaces the blackening factors above with unity. The string frame metric is related to the above by $ds_{st}^2 = \exp(4\f/3) ds^2$ so that the conformal prefactor becomes unity in the string frame. This solution then precisely corresponds to the product of the linear dilaton BH in 2-d~\cite{Witten:1991yr, Alexandrov:2003ut, Nakayama:2004vk} in the leading in $\a'$ approximation, times $R^3$. The temperature (\ref{Tdef}) is fixed by the integration constant $A_0$: 
\begin{equation}\label{tempsp}
 T = \frac{3 e^{A_0}}{4\pi \ell}\,.
\end{equation}
We observe that the ratio $\ell/\ell_s e^{-A_0}$ controls the size of the spacetime in string units, the temperature and the central charge of the worldsheet CFT, if one identifies the Einstein-Dilaton action as the low energy effective action of non-critical string theory (Liouville theory). Therefore this combination drops out of all dimensionless quantities thus we can set $A_0=0$ with no loss of generality. 
\\
\\
In the paper~\cite{Giddings:1991mi} it is shown how to embed the 2d {\em cigar} part of the geometry (\ref{dilBH}) in  10d superstring theory using a WZW model product coset construction. The associated metric in string frame is in that case
\be
ds_{st}^2 = k \left(dr^2 \le(1- e^{\frac{3(r-r_h)}{\ell}}\ri)^{-1} - dt^2 \le(1- e^{\frac{3(r-r_h)}{\ell}} \ri) \right) + k d \Omega_3^2 + dx^i dx_i \, ,
\ee
together with the linear dilaton and an $H_3$ flux piercing the $S^3$.
Comparing such a solution to our background we have truncated some extra coordinates such as the $S^3$ that corresponds to the extra $SU(2)_k$ of the $SL(2,R)_k/U(1) \times SU(2)_k$ WZW model. Some further discussion of similar solutions from the point of view of Little String Theory can be found in~\cite{Aharony:2004xn}. 
To fully read the spectrum of the worldsheet sigma model, one would need to resort to CFT techniques, here we will be content with studying field fluctuations on this background (so-called minisuperspace approximation). 
The fluctuation equations of the mini-superspace modes can be found in~\cite{Nakayama:2007sb} and have as a solution the hypergeometric functions $_2F_1$. In the next section we will study the fluctuations on our background and match them in a specific limit with these mini-superspace eigenfunctions.

\section{Fluctuations around the CR solution} \label{sec:flucts}

The fluctuation equations around the CR geometry with a generic $X$ within the range $-1/2 < X < 0$ can be obtained by making the following  Ansatz for fluctuations of the metric and the dilaton as
\bea
 \delta g &=& e^{2 A_0}\hat r^{-\frac{2}{1-4X^2}}\left[- H_{vv} dv^2 + 2 H_{vi} dv dx^i + H_{ij}  dx^{i} dx^{j}   \right] \\
 \delta \l &=& \hat r^{-\frac{3X}{1-4X^2}} \psi \ .
\eea
Here we have already used gauge transformations to set the fluctuations $H_{rr}$, $H_{rv}$, and $H_{ri}$ to zero. We look for modes with fixed frequency, momentum in $x^1$ direction, and arbitrary dependence on $r$:
\be
 H_{\mu\nu} (\hat r,v,x^1) = \tilde H_{\mu\nu}(\hat r) e^{-iv\omega+ikx^1}\ ,\qquad \psi (\hat r,v,x^1) = \tilde \psi(\hat r) e^{-iv\omega+ikx^1} \ .
\ee
There are six propagating degrees of freedom, arising from a spin two mode and a scalar mode. Because the momentum in the $x^1$ direction partially breaks rotational symmetry, the modes are classified as follows (see also~\cite{Janik:2016btb}).
The simplest modes are the spin two modes transverse to the momentum, given by $\tilde H_{23}$ and $\hat H_\mathrm{as} \equiv (\tilde H_{22}-\tilde H_{33})/2$. There are also two shear modes, given by
\be
 \hat H_{2} = k \tilde H_{v2} + \omega \tilde H_{12} \ , \qquad   \hat H_{3} = k \tilde H_{v3} + \omega \tilde H_{13} \ .
\ee
The remaining two degrees of freedom are more complicated combinations of the dilaton and the metric fluctuations:
\bea
\zeta_1 &=&  \frac{1}{2}\left(\tilde H_{22}+\tilde H_{33}\right) - \frac{2}{3X} \tilde \psi\\\nn
\zeta_2 &=& -k^2 \tilde H_{vv}+2 k \omega \tilde H_{v1}+\omega ^2\tilde H_{11}-\frac{k^2 \left(\left(2 X^2-1\right) \left(\frac{\hat r}{\hat r_h}\right)^\xi-1\right)+\omega ^2}{2}\left(\tilde H_{22}+\tilde H_{33}\right) \ .
\eea
These expressions are determined by gauge covariance (having already killed some of the $\tilde H$ as mentioned above).

As it turns out, the transverse spin two modes and $\zeta_1$ satisfy a relatively simple equation
\be \label{nonhydroeq}
- \left(\ell'^2 k^2 \hat r + (\xi-1) \ell' i \omega \right) \flf(\hat r)+\left(2 i \ell' \hat r  \omega +f(\hat r)-\xi\right)\flf'(\hat r) +\hat r f(\hat r) \flf''(\hat r) = 0 
\ee
with $\flf = \tilde H_{23},\ \hat H_\mathrm{as},\ \zeta_1$. Notice that this is also the equation of motion for a massless scalar field in the CR background. The equation for the shear channel is
\bea \label{sheareq}
0&=& \left(-\ell'^2 k^2 \hat r - (\xi-1) \ell' i \omega  +\frac{i\ell' \omega k^2\xi\, \hat r^\xi \hat r_h^{-\xi}}{f(\hat r) k^2 -\omega ^2}\right) \hat H_i(\hat r)\\
&&+\left(  2 i \ell'  \omega  \hat r+f(\hat r)-\xi+\frac{k^2 f(\hat r) \xi\, \hat r^\xi \hat r_h^{-\xi}}{f(\hat r) k^2 -\omega ^2} \right) \hat H_i'(\hat r)+\hat r f(\hat r)\hat H_i''(\hat r) 
\eea
with $i=1,2$. Finally the remaining equation, which is identified as the sound channel, couples $\zeta_2$ to $\zeta_1$:
\bea \label{soundeq}
0&=&\frac{k^2 (\xi -4) \xi ^2 \hat r^{2 \xi -1}\hat r_h^{-2\xi} \left(k^2 \xi -2 (\xi -1) \omega ^2\right)}{(\xi-1)\left[k^2 ((2-\xi ) f(\hat r)-\xi )+2 (\xi -1) \omega ^2\right]} \zeta_1(\hat r) \nn\\
&&+  \left[-\ell'^2 k^2 \hat r - (\xi-1) \ell' i \omega  -\frac{k^2 (\xi-2) \xi\, \hat r^\xi \hat r_h^{-\xi}  \left(\xi  \hat r^{\xi -1}\hat r_h^{-\xi}+2 i \ell' \omega \right)}{k^2 ((2-\xi ) f(\hat r)-\xi )+2 (\xi -1) \omega ^2}\right] \zeta_2(\hat r)\\\nn
&&+\left[  2 i \ell'  \omega  \hat r+f(\hat r)-\xi-\frac{2 k^2 (\xi-2 ) \xi  f(\hat r)\hat r^\xi \hat r_h^{-\xi}}{k^2 ((2-\xi ) f(\hat r)-\xi )+2 (\xi -1) \omega ^2}\right] \zeta_2'(\hat r)+\hat r f(\hat r)\zeta_2''(\hat r) \ .
\eea
The coupling between $\zeta_2$ and $\zeta_1$ is trivial (in the sense that $\zeta_1$ appears in the dynamic equation for $\zeta_2$, but not vice versa) only for the exponential dilaton potential. For more generic potentials, $\zeta_{1,2}$ satisfy a nontrivially coupled system of two differential equations (see e.g.~\cite{Janik:2016btb}).
Notice that this coupling vanishes as $k \to 0$ and actually all fluctuation equations become identical in this limit. 

All coefficients of the three fluctuation equations \eqref{nonhydroeq}--\eqref{soundeq} become real for purely imaginary $\omega$ (and real $k$). Consequently, correlators extracted from these will be real on the vertical axis of the complex $\omega$-plane, and transform by complex conjugation under $\omega \mapsto -\mathrm{Re}\omega + i\, \mathrm{Im} \omega$, which is the expected behavior on general grounds.  

Another general feature is that the location of the horizon only affects the modes trivially: $\hat r_h$ can be factored out of the fluctuation equations by rescaling $\hat r \mapsto \hat r_h \hat r$, $\omega \mapsto \omega/\hat r_h$, and $q \mapsto q/\hat r_h$. 
Equivalently, the location of the quasi normal modes depends only on the rescaled frequency and momentum
\be \label{rescaledpom}
 q = \frac{k}{2 \pi T} = \frac{2 k\ell' \hat r_h}{\xi}\ , \qquad \varpi = \frac{\omega}{2 \pi T} = \frac{2 \omega\ell' \hat r_h}{\xi} \ .
\ee

We will find useful to have the fluctuation equation also in Schr\"odinger form; we restrict our study to the equation~\eqref{nonhydroeq} for the transverse spin-two (and one scalar field) fluctuations, which does not involve any of the hydrodynamic modes. Defining a new radial coordinate 
\be \label{wdef}
 w =\left(\frac{\hat r}{\hat r_h}\right)^\xi \ ,
\ee
which runs from $0$ at the boundary to $1$ at the horizon, and redefining the fluctuation as 
$\flf(w) = e^{g(w)} h(w)$, where $g(w)$ satisfies
\be
 g'(w) = \frac{w- i \varpi w^{1/\xi}}{2 w(1-w)} \,, 
\ee
the equation~\eqref{nonhydroeq} becomes 
\bea\label{eqSchr}
 -h''(w) &+ &V(w) h(w) =  0 \\
 V(w) & =&  - \frac{\left(\varpi^2-q^2\right) w^{\frac 2\xi }+q^2 w^{\frac{\xi +2}{\xi }}+w^2}{4 (1-w)^2 w^2} \ . \nonumber
\eea
Notice that the potential is real for real (or purely imaginary) $\varpi$.  Near the horizon we find that
\be
 V(w) = -\frac{1+\varpi^2}{4(1-w)^2} + \mathcal{O}\left({\frac{1}{1-w}}\right)
\ee
which (combined with the factor $e^g$) gives the expected behavior $\flf \sim \mathrm{const.}$ for the ingoing mode and $\flf \sim (1-w)^{i\varpi}$ for the outgoing mode.

For later use we also consider the limit of large $\xi$. In particular, we notice that the limits $w \to 0$ and $\xi \to \infty$ do not commute. When $w^{1/\xi}\ll 1$, i.e., $\hat r \ll \hat r_h$, the potential
\be
 V(w) = -\frac{\le(\varpi^2-q^2\ri)w^{2/\xi}}{4 w^2}\le[1+\mathcal{O}\left(w\right)\ri]
\ee
can be treated as a subleading correction to the fluctuation equation, leading to the usual normalizable and nonnormalizable solutions $\flf \sim C_1 +C_2 w$. When $e^{-\xi}\ll w \ll 1$ (so that $\hat r_h - \hat r \ll 1$), same terms in the potential behave as
\be \label{Vinterm}
 V(w) \simeq -\frac{\varpi^2-q^2}{4 w^2}
\ee
with corrections suppressed by $w$ and $|\log(w)/\xi|$. This latter form leads to oscillating solutions for $\varpi^2-q^2>1$, so that if one takes the limit $\xi \to \infty$ first, it is not possible to find normalizable modes.

\begin{figure}[!tb]
\begin{center}
\includegraphics[width=0.49\textwidth]{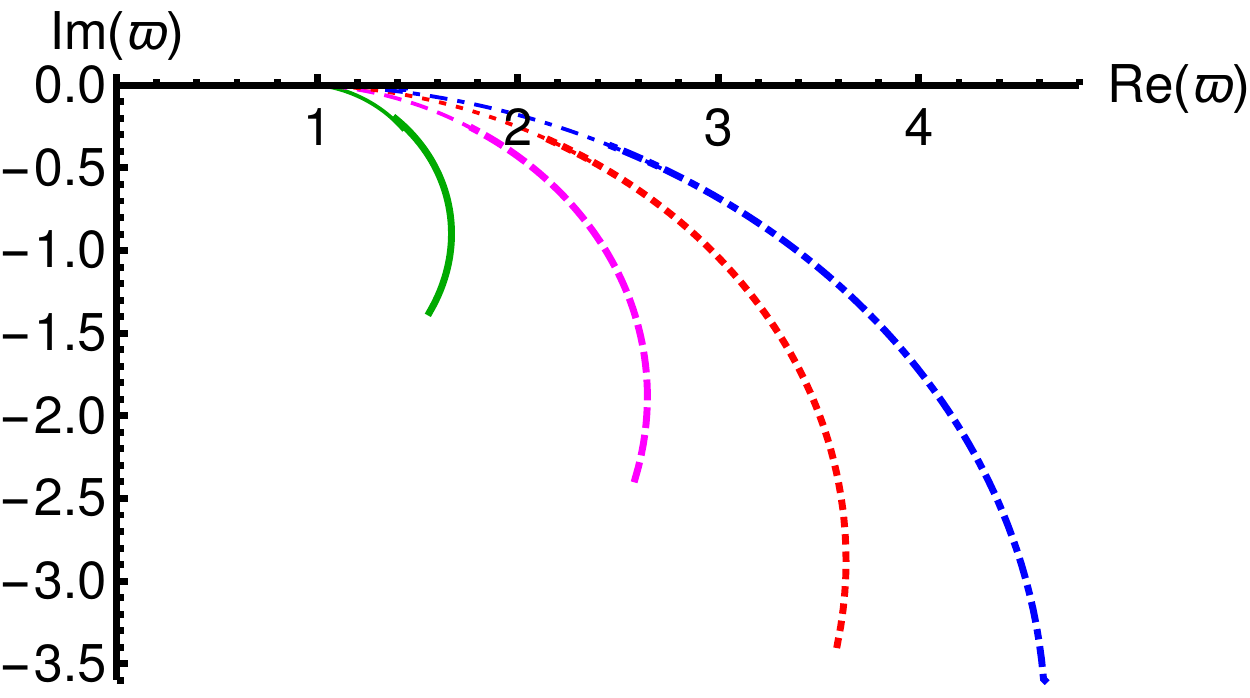}%
\hspace{2mm}\includegraphics[width=0.49\textwidth]{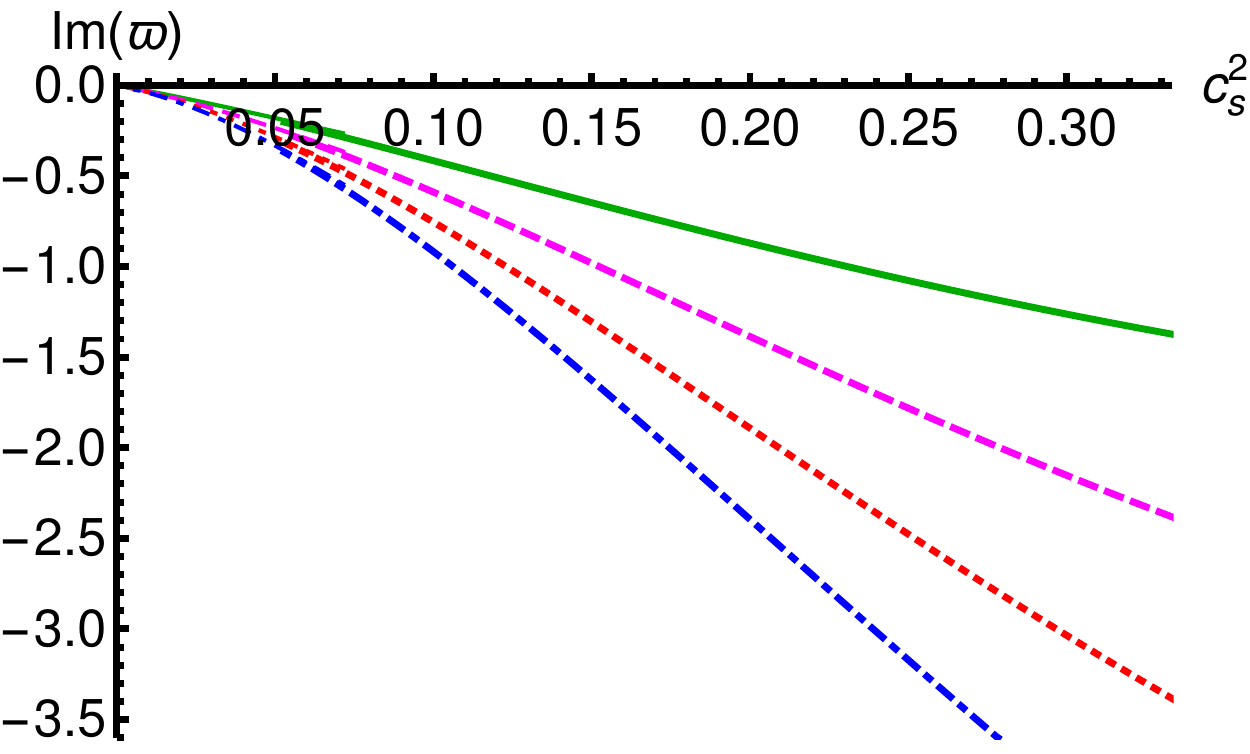}
\end{center}
\caption{Dependence of the four lowest nonhydrodynamic quasi normal modes on $X$ at $q=0$. Thick lines were obtained by directly solving the fluctuation equations (for the transverse spin two modes) numerically for $-0.46<X<0$, and the thin lines are based on the analytic approximation of Sec.~\protect\ref{sec:Xonehalf}. Left: the trajectories of the modes on the complex $\varpi$-plane. Right: The imaginary parts of the modes as a function of $c_s^2 = (1-4X^2)/3$.}
\label{fig:modexdep}
\end{figure}

\begin{figure}[!tb]
\begin{center}
\includegraphics[width=0.49\textwidth]{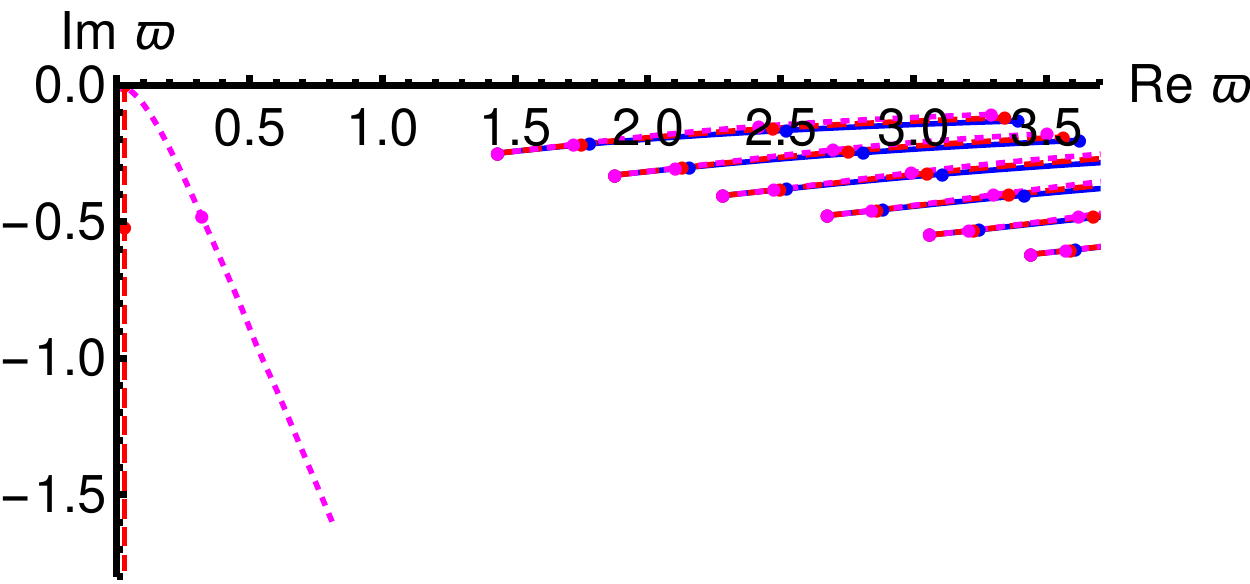}%
\hspace{2mm}\includegraphics[width=0.49\textwidth]{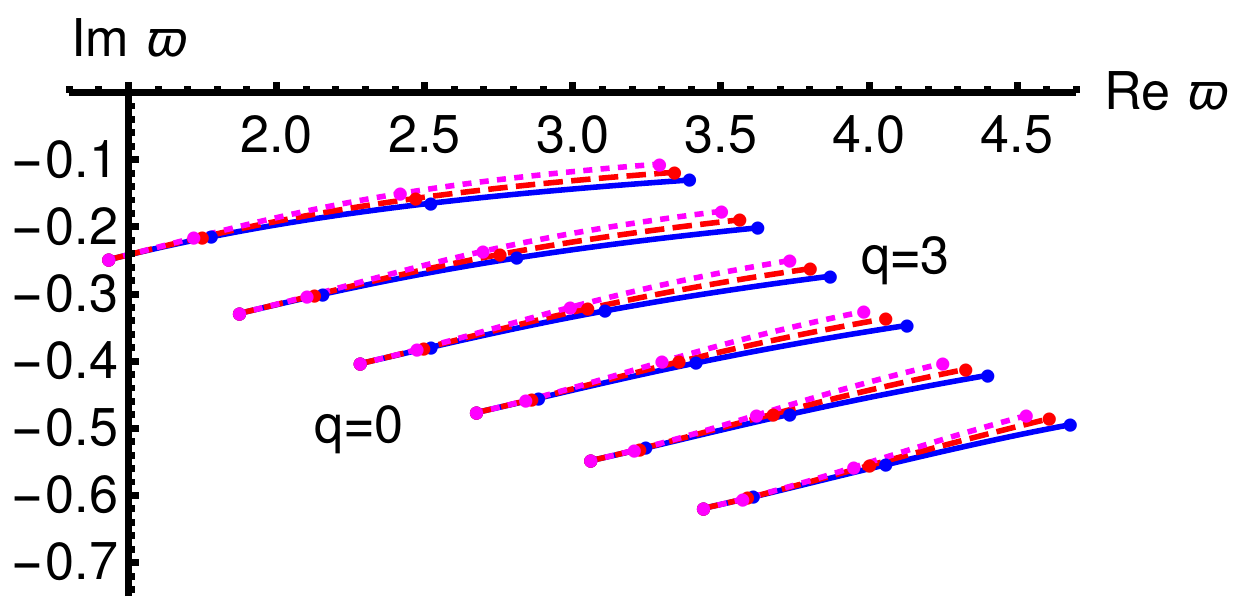}
\end{center}
\caption{Dependence of the quasi normal modes on $q$ on the complex $\varpi$ plane for $X=-0.45$. The quasi normal modes from Eqs.~\protect\eqref{nonhydroeq}, \protect\eqref{sheareq}, and \protect\eqref{soundeq} are shown as solid blue, dashed red, and dotted magenta curves, respectively. The momentum $q$ varies from $q=0$ to $q=3$ along the curves. The hydro modes lie at the origin for $q=0$. The dots are at $q=0$, 1, 2, and $3$. Left: overall plot showing both the hydrodynamic and nonhydrodynamic modes. Right: a zoom in the region with the lowest nonhydrodynamic modes. }
\label{fig:qdepx045complexplane}
\end{figure}

\begin{figure}[!tb]
\begin{center}
\includegraphics[width=0.49\textwidth]{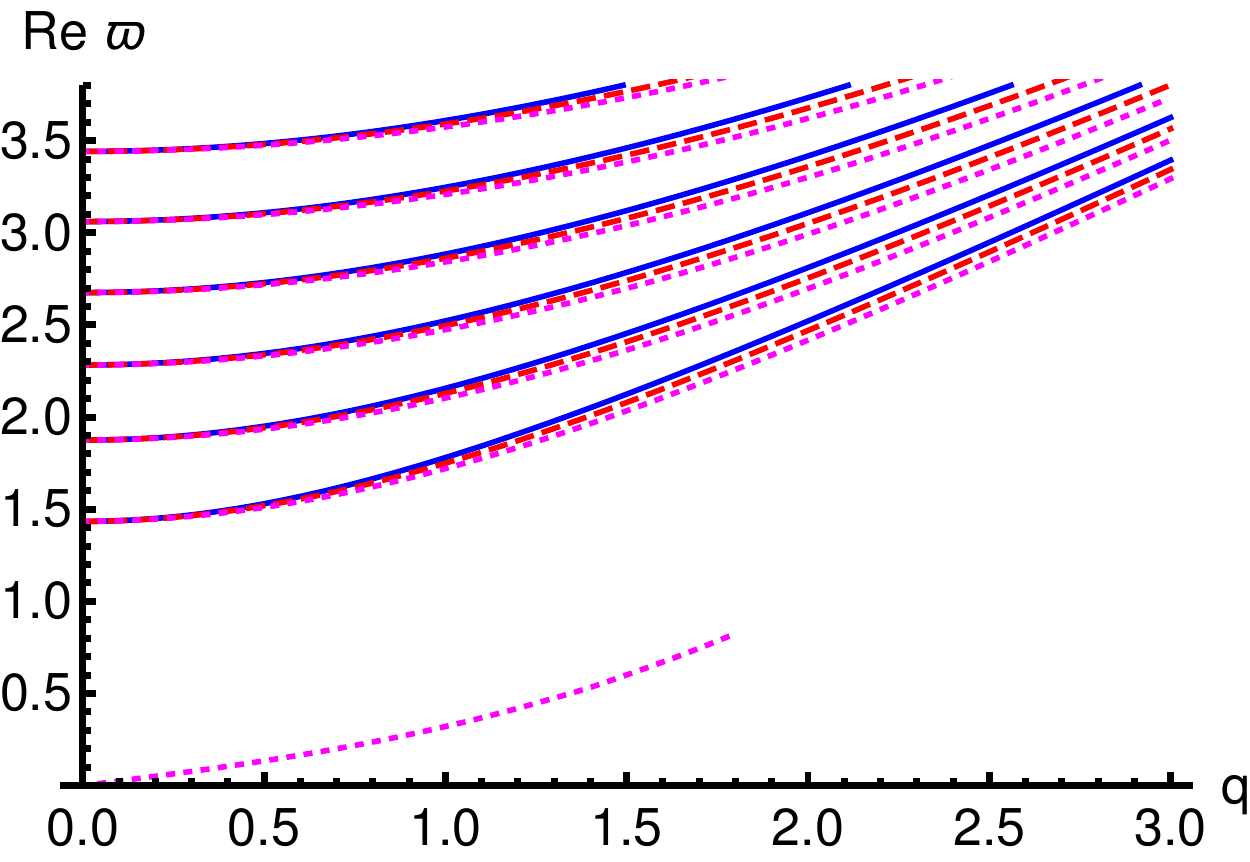}%
\hspace{2mm}\includegraphics[width=0.49\textwidth]{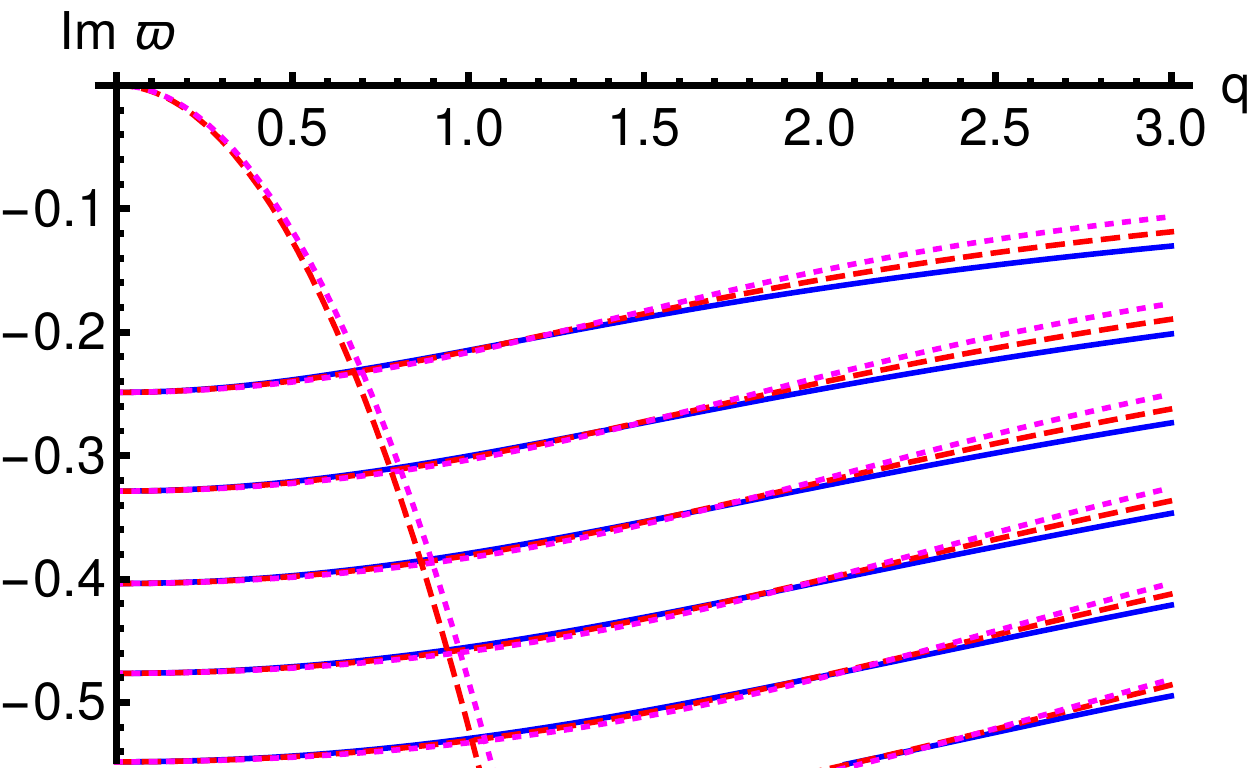}
\end{center}
\caption{Dependence of the quasi normal modes on $q$ for $X=-0.45$. Left: dependence of $\mathrm{Re}\,\varpi$ on $q$. Right:  dependence of $\mathrm{Im}\,\varpi$ on $q$. Notation as in Fig.~\protect\ref{fig:qdepx045complexplane}.}
\label{fig:qdepx045reim}
\end{figure}

\subsection{Numerical analysis at generic $X$} \label{sec:numerics}

We have solved the fluctuation equations numerically for various values of $X$ within the range $-1/2<X<0$. We substituted $\flf(\hat r) = e^{- i \omega \hat r} K(\hat r)$ in order to reduce the exponential dependence on $\hat r$ for $\mathrm{Im}\,\omega < 0$, and solved the resulting equations by estimating the $\hat r$-derivatives by a pseudospectral approximation with $50$ grid points chosen from a Gauss-Labotto grid. The correlators  of the energy-momentum tensor in the various channels were then extracted from the coefficient of the terms $\propto \hat r^\xi$ at the boundary. More precisely, we used the definitions of Eqs.~\eqref{besselfs} and\eqref{sourceredef} in Appendix~\ref{sec:bdry} for the source and the vev terms. Notice that these expressions hold near the boundary up to highly suppressed corrections for fluctuations in each sector which makes it much easier to extract the correlators for an arbitrary potential.

In Fig.~\ref{fig:modexdep} (the left-hand side appeared also in \cite{Betzios:2017dol}) we show the modes at zero momentum as functions of $X$; we see that as $X$ approaches the critical value, each pole appear to move to the point $\varpi = 1$; taken together, the poles form a line that approach the real axis and in the critical limit 
should form a branch cut (see also~\cite{Grozdanov:2016vgg}). This is difficult to check as the numerics become more difficult, hence the need for a more analytic treatment to which we will turn in the next sections. On the right hand side of the plot, we show that the poles have an approximately linear dependence on the deviation of the speed of sound from its conformal value, as observed also by \cite{Janik:2015waa}.

In Figs.~\ref{fig:qdepx045complexplane} and~\ref{fig:qdepx045reim} we show the dependence of the quasi-normal modes on the momentum.  
We used as a reference value $X= -0.45$, a value relatively close to the critical value $X=-1/2$. 
The imaginary parts of the hydrodynamic modes obey\footnote{For the sound mode, the precise coefficient predicted by hydrodynamics is $1/3 + 2X^2/3$ which tends to $1/2$ as $X \to -1/2$.} $\mathrm{Im}\,\varpi \simeq -0.5 q^2$ at small $q$, whereas the imaginary parts of the nonhydro modes behave as $\mathrm{Im}\,\varpi \sim 1/\xi$, as we will prove below. Therefore the imaginary parts cross for $q \sim 1/\sqrt{\xi}$, and consequently hydrodynamics breaks down for smaller and smaller $q$ as $\xi$ increases. The real part of the sound mode satisfies $\mathrm{Re}\,\varpi \simeq c_s q$ where $c_s = \sqrt{(1-4X^2)/3}$.

\begin{figure}[!tb]
\begin{center}
\includegraphics[width=0.49\textwidth]{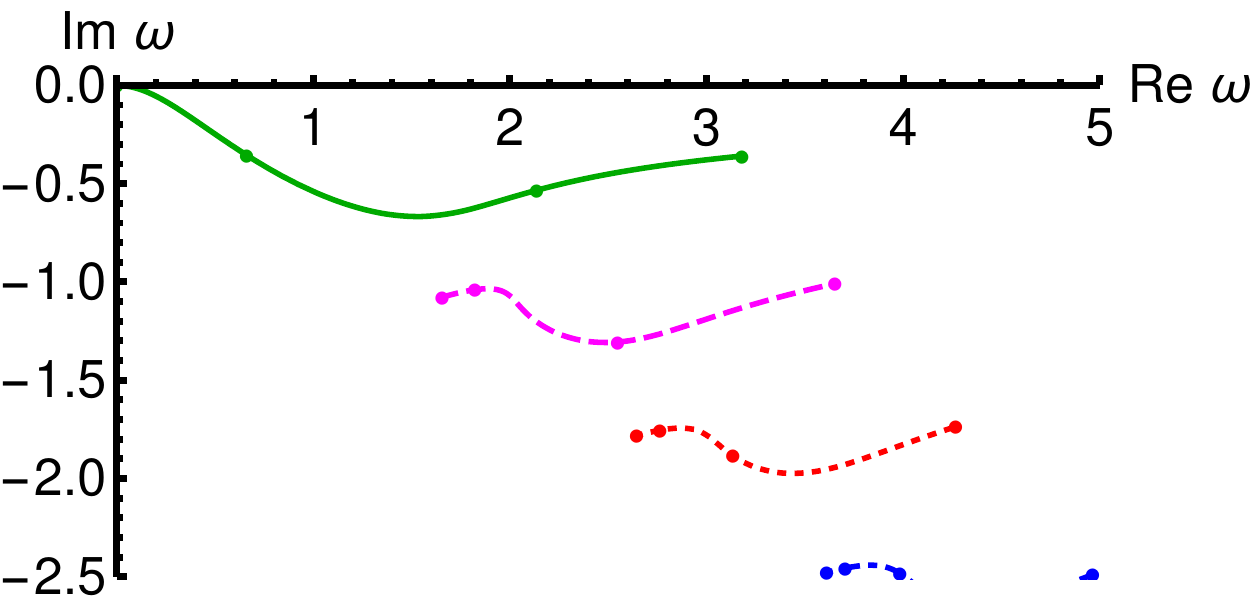}%
\hspace{2mm}\includegraphics[width=0.49\textwidth]{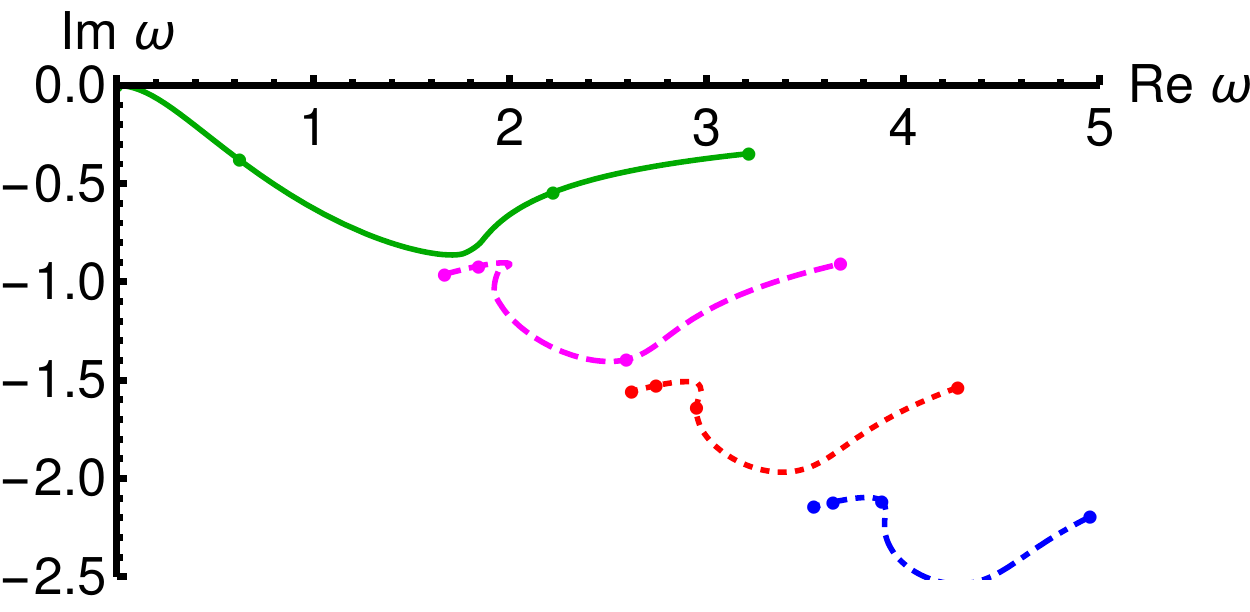}
\includegraphics[width=0.49\textwidth]{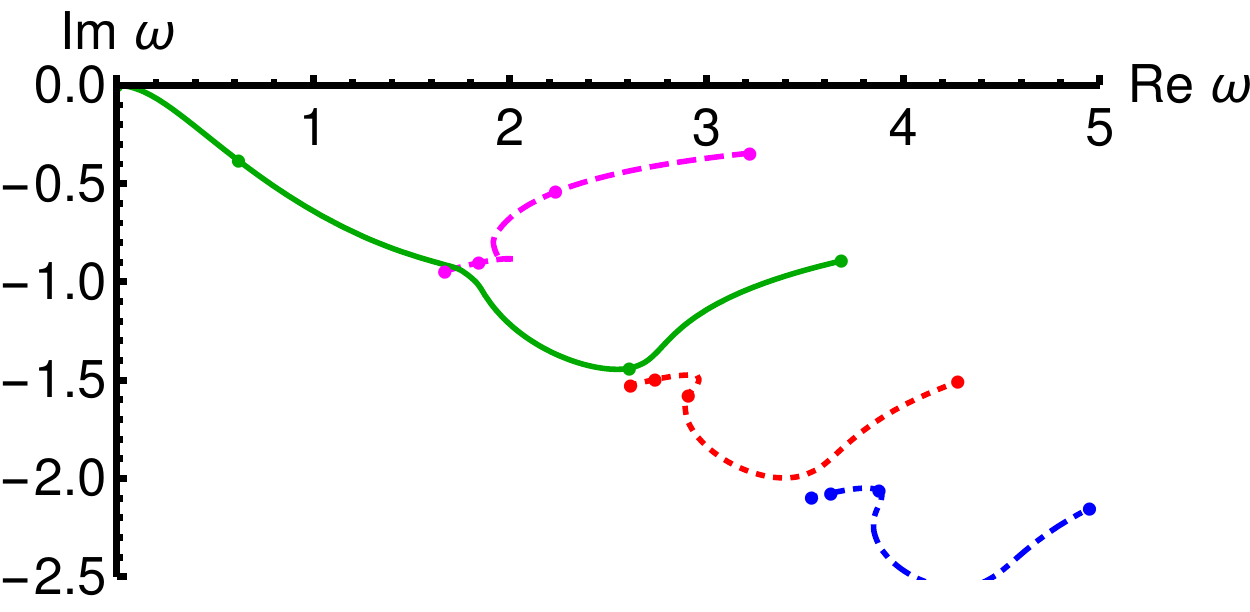}%
\hspace{2mm}\includegraphics[width=0.49\textwidth]{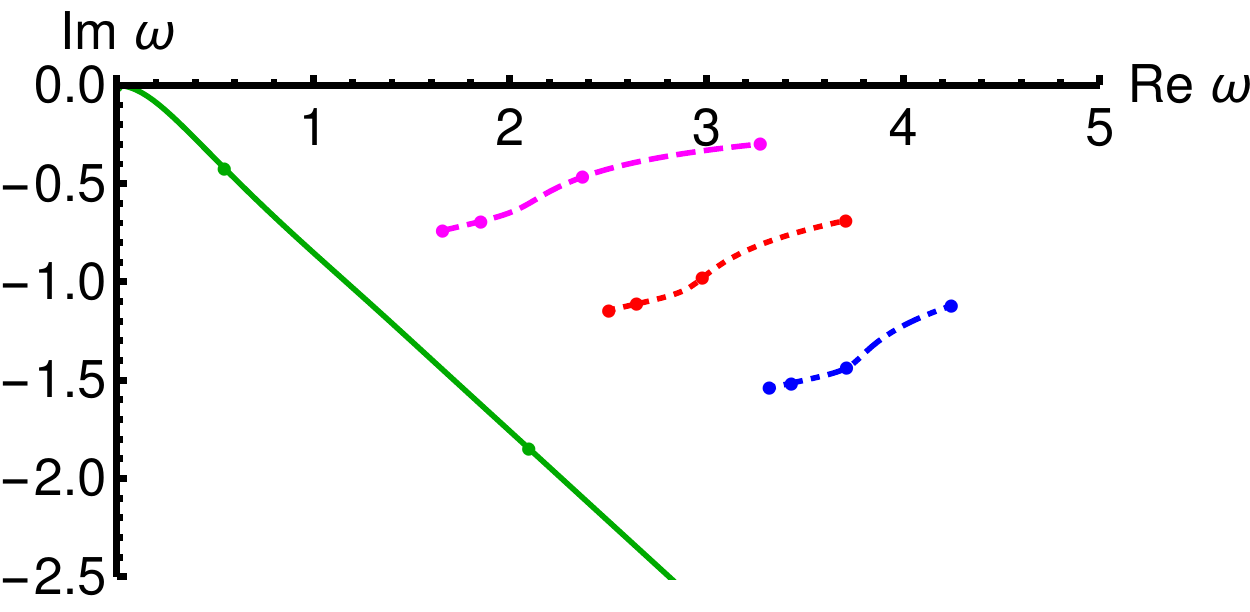}
\end{center}
\caption{Dependence of the sound channel quasi normal modes on $q$ for $X=-0.25$ (top left), $X=-0.29$ (top right), $X=-0.295$ (bottom left) and $X=-0.35$ (bottom right). }
\label{fig:soundchannelcrossing}
\end{figure}

There is an interesting level crossing structure in the sound channel around $X = -0.3$ (see Fig.~\ref{fig:soundchannelcrossing}). For $-0.29 \lesssim X < 0$, the imaginary part of $\varpi$ for the sound mode (that is the mode for which $\varpi \to 0$ a $q \to 0$) is smaller than $\mathrm{Im}\varpi$ for the other modes, whereas for $-1/2<X \lesssim -0.3$, $\mathrm{Im}\varpi$ of the sound mode crosses all other modes and becomes subdominant. Consequently, near $X=-0.3$, there is a sequence of level crossings between the sound mode and all the nonhydro modes. The first crossing takes place between $X=-0.29$ and $X=-0.295$, see the top right and bottom left plots in Fig.~\ref{fig:soundchannelcrossing}. This level crossing is somewhat similar to the behavior observed for the quasi-nomal modes of the scalar field, as a function of the temperature, in a flow between two different conformal points \cite{Attems:2016ugt}, although the details are different. 

\begin{figure}[!tb]
\begin{center}
\includegraphics[width=0.49\textwidth]{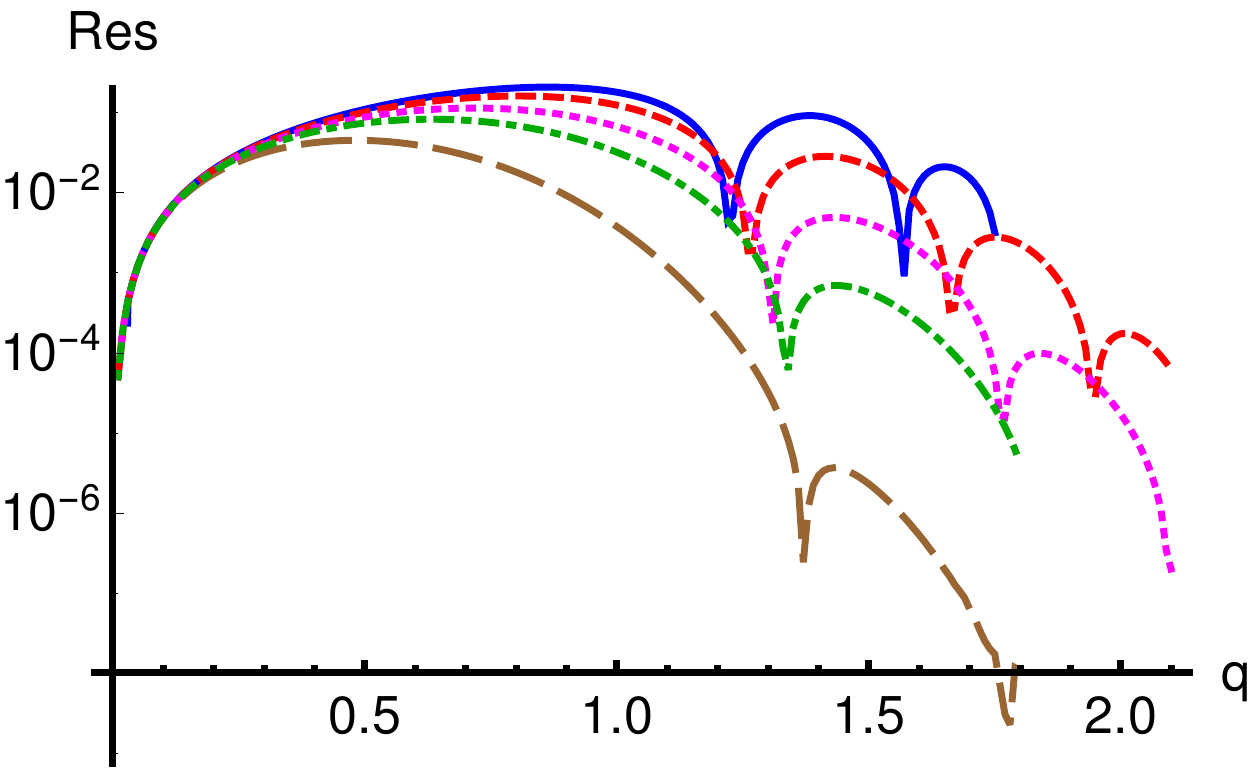}%
\hspace{2mm}\includegraphics[width=0.49\textwidth]{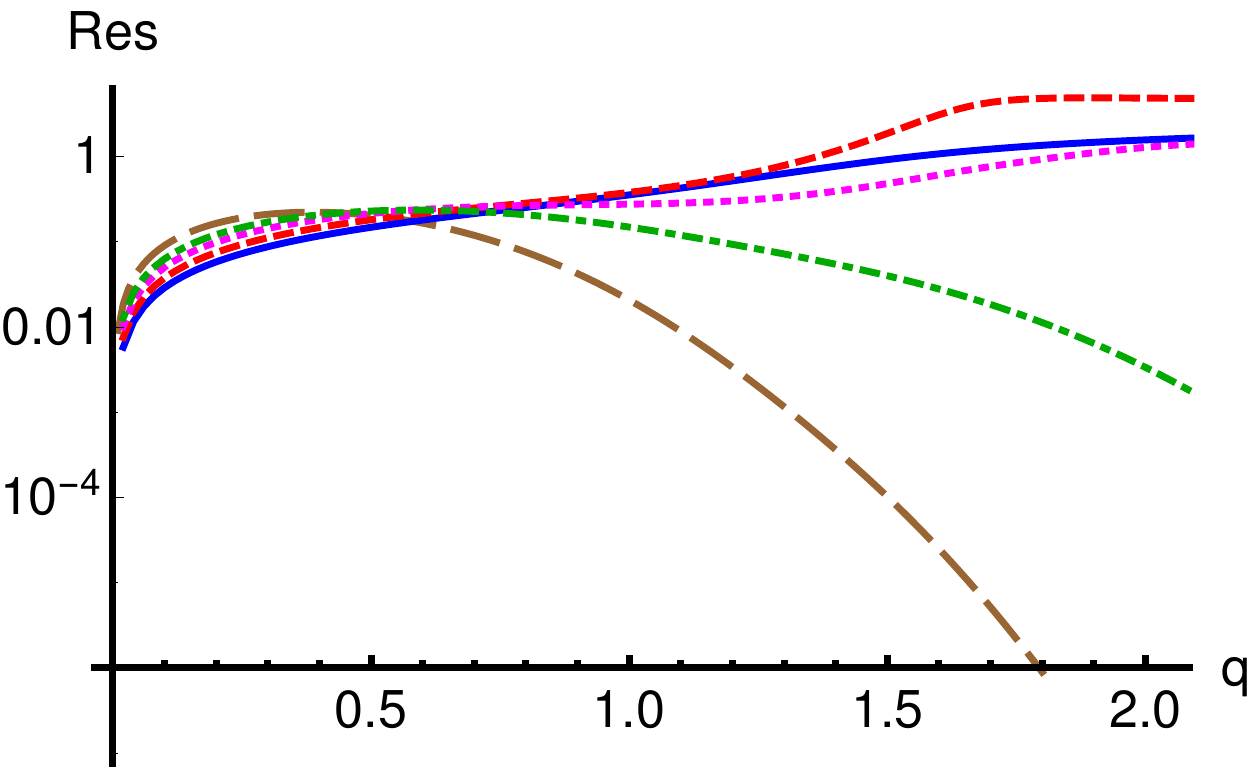}
\end{center}
\caption{The (absolute values of the) residues of the hydrodynamic modes as a function of $q$. Left: shear mode. Right: sound mode. The values of $X$ for the solid blue, dashed red, dotted magenta, dash-dotted green, and long-dashed brown curves are $X=0$, $X=-0.25$, $X=-0.35$, $X=-0.4$, and $X=-0.45$, respectively. }
\label{fig:residues}
\end{figure}

We have also determined residues of the hydrodynamic modes numerically. Results are shown in Fig.~\ref{fig:residues} for various values of $X$ and as a function of $q$. The value of $\hat r_h$ only affects the overall normalization of the residues, and here we set $\hat r_h = 1$.
We see that the residues vanish as $q \to 0$, indicating the expected decoupling of the hydrodynamic modes in this limit. The residues of the shear modes (left plot) oscillate as a function of $q$, which has also been observed in the case of the AdS geometry ($X=0$)~\cite{Amado:2008ji,Landsteiner:2012gn}. We notice however that for $X \ne 0$, the oscillations do not seem to be linked to the crossing of the imaginary parts of the quasi normal modes (to the contrary to what was found for $X=0$). This is clear because as we have pointed out,  
the values of $q$ at the crossings behave as $\sim 1/\sqrt{\xi}$ so that they decrease with increasing $|X|$, but the values of $q$ at the nodes in Fig.~\ref{fig:residues} increase with increasing $|X|$ instead.

In general we notice that the values of the residues decrease more rapidly with $q$ for $q \gtrsim 1$ as $|X|$ grows. In particular, for small $|X|$, the residues of the sound mode increase with $q$ while for larger $|X|$ they decrease with $q$. Notice that this reflects the different behavior of the mode at high $q$ due to the crossing depicted in Fig.~\ref{fig:soundchannelcrossing}.

\begin{figure}[!tb]
\begin{center}
\includegraphics[width=0.49\textwidth]{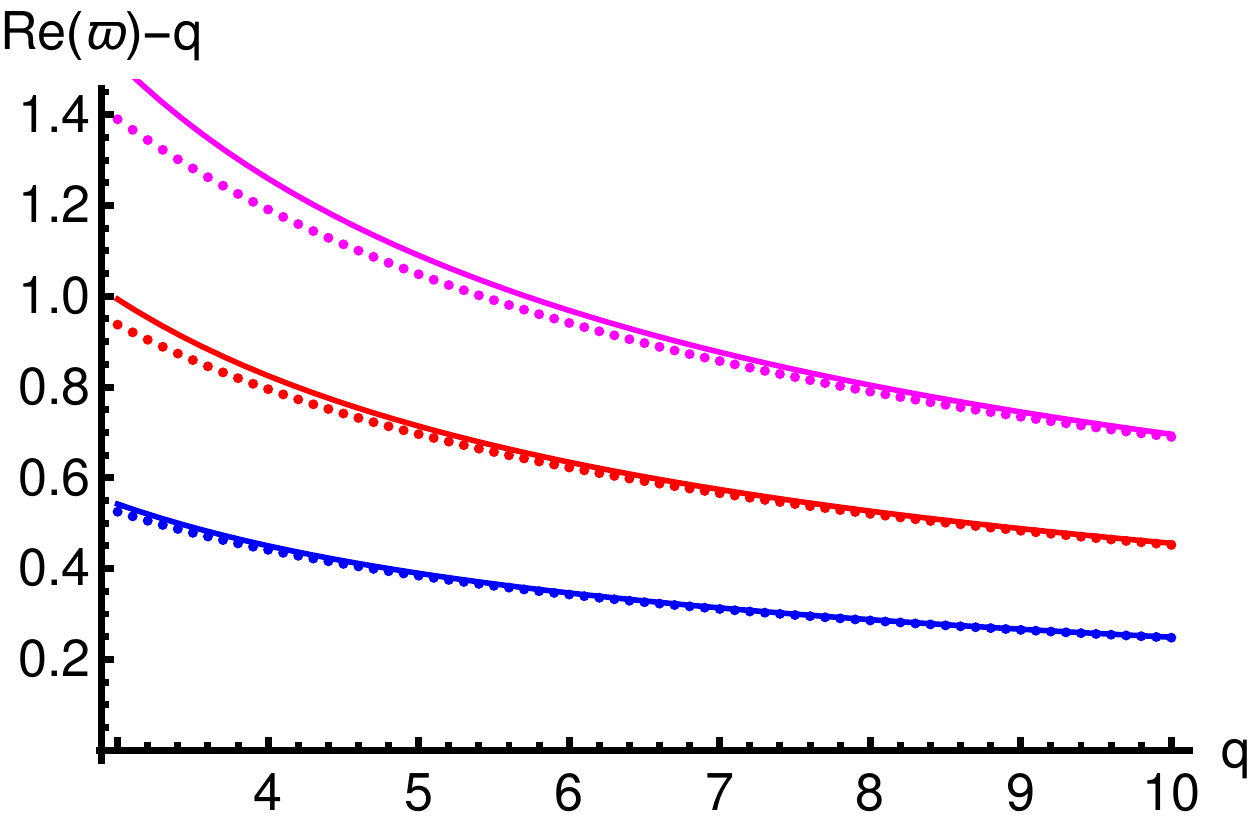}%
\hspace{2mm}\includegraphics[width=0.49\textwidth]{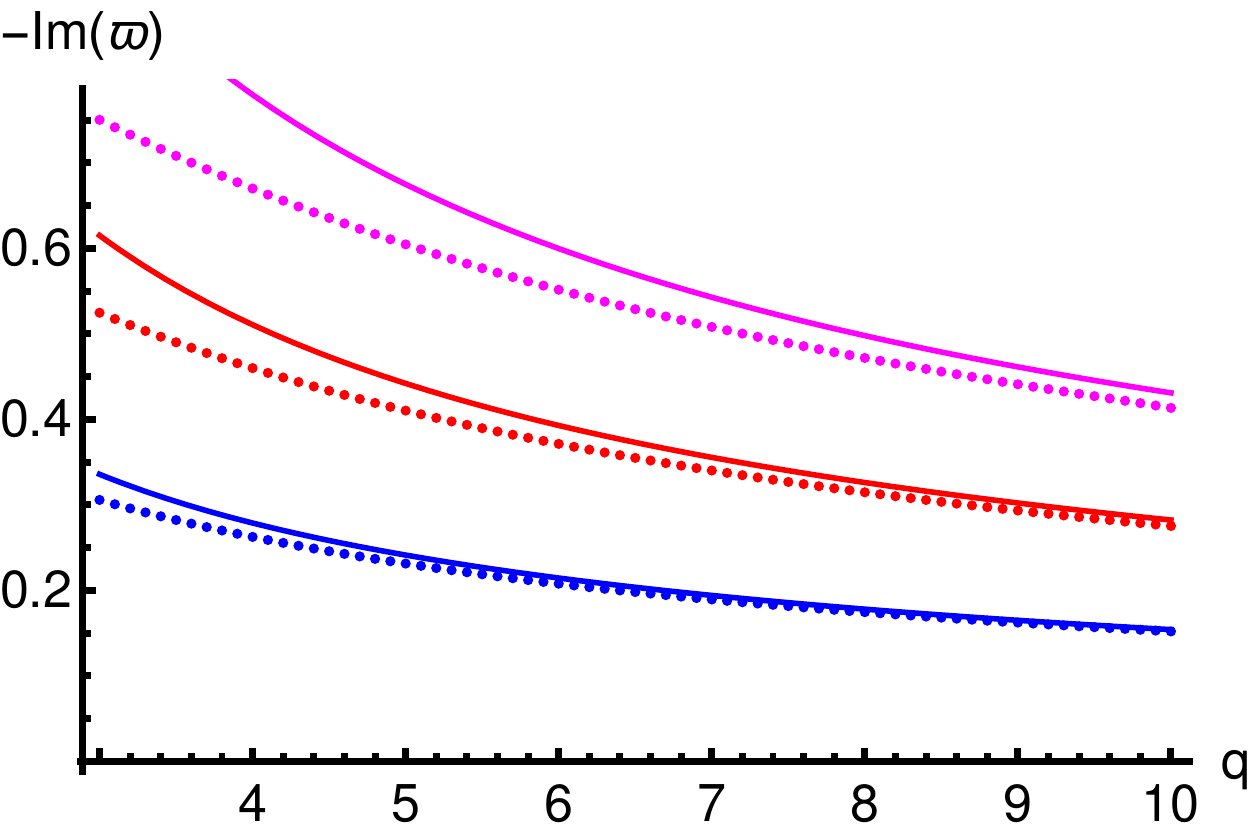}
\end{center}
\caption{The dependence of the three lowest transverse spin-two quasi normal modes on $q$ for $X=-0.4$  (dots) at large values of $q$ and the comparison to the WKB approximation of Appendix~\protect\ref{WKB} (solid curves). Left: real parts (with the linear term $\varpi \approx q$ subtracted). Right:  imaginary parts.}
\label{fig:largeq}
\end{figure}

One can determine the large momentum behavior of the modes analytically using a WKB analysis~\cite{Fuini:2016qsc}. This is done in appendix \ref{WKB} and we find: 
\be
 \frac{\varpi}{q}-1 \propto q^{-\frac{2\xi}{\xi+2}}e^{-\frac{2\pi i}{\xi+2}}\ , \qquad (q \to \infty)\ ,
\ee
where the proportionality constant is positive.
The agreement between the WKB approximation and the numerical results for the locations of the QNMs is demonstrated in Fig.~\ref{fig:largeq}.

\section{Analytic results in the limit $X \to -1/2$}\label{sec:Xonehalf}

As it turns out, the fluctuation equations in the transverse spin-two channel~\eqref{nonhydroeq} can be solved analytically in the limit $X \to -1/2$, equivalently $\xi \to \infty$. In order to make sense of this limit, we need to also decide which quantities to hold fixed in the limit. First we notice how the energy scales behave in this limit. We have fixed the units of the radial coordinate (by rescaling it with $\ell'$) such that the dilaton $\l \sim 1$ at $\hat r \sim 1$.  The temperature from~\eqref{Tdef} behaves as $T \sim \xi/\ell' \hat r_h $. The temperature is the scale that determines the location of the QNMs. Notice that there is the factor $1-4X^2$ in the definition of $\ell'$ in~\eqref{relldef} which behaves as $1/\xi$ in the limit $X \to -1/2$. In units of $\ell$, i.e., the scale factor in the domain wall metric the temperature is therefore regular in this limit, $T \sim 1/\ell \hat r_h$ as long as $\hat r_h$ is kept constant. From (\ref{fxidef}) we indeed see that $\hat r_h \to 1$.
In conclusion, we keep $\ell$ and $\hat r_h$ fixed when taking $\xi \to \infty$, so that $T$ remains fixed but $\ell' \to \infty$.

We will now sketch how the quasi normal modes at large but finite $\xi$ arise. The precise derivation is carried out below in Sec.~\ref{sec:preciselimit}. The modes will be found for approximately real $\varpi$. Existence of a mode requires matching of the normalizable UV solution (for $w^{1/\xi}\ll 1$) to the oscillating solution in the IR. Such matching is made possible by tuning the length of the intermediate interval where the solutions with the Schrodinger potential~\eqref{Vinterm} are oscillatory (i.e., $\varpi^2-q^2>1$). In this interval the solution behaves as
\be
 h \propto \sin \le(\frac 12 \sqrt{\varpi^2-q^2-1}\log w \ri) \ .
\ee
We expect that the $n$th quasi normal mode oscillates (roughly) $n$ times. The oscillations take place for $1 \ll |\log w | \ll \xi$, so we obtain
\be
 n \sim  \sqrt{\varpi_n^2-q^2-1}\, \xi \ . 
\ee
A sequence of quasi normal modes is therefore expected for $\varpi^2-q^2>1$ and for large $n$ we expect  $\varpi_n \sim n/\xi$. These results will be verified below.

\subsection{Precise analysis of the limit $X \to -1/2$} \label{sec:preciselimit}

We  solve the fluctuation equations in two domains, as suggested by the analysis of the Schrodinger potential in Sec.~\ref{sec:flucts}, in order to compute the two-point correlator of the spin-two components of the energy momentum tensor analytically at large (but finite) $\xi$.  First we analyze the fluctuations at large $\xi$ with fixed arbitrary $w$. Then we analyze them close to the boundary, i.e., for small $w$ with fixed arbitrary $\xi$. At large $\xi$ these results maybe combined to complete solutions for the fluctuations from the boundary to the horizon, and therefore to compute the correlator analytically, up to corrections suppressed by $1/\xi$.

It might look tempting to carry out the fluctuation analysis in the domain wall coordinates instead of the conformal coordinate $r$, because the background is regular at $X=-1/2$ in the former coordinates. This would not, however, lead to essential changes in the analysis below, and the fluctuation equations take a simpler form in the conformal coordinates. Also, while the background is regular in the domain wall coordinates, $X=-1/2$ is still a special point for the fluctuations (as we shall see below). 

\subsubsection{Fluctuations at $X=-1/2$} \label{sec:xiinfty}

Let us start by taking the limit $X \to -1/2$ keeping the rescaled frequency and momentum in~\eqref{rescaledpom} as well as the coordinate $w$ of~\eqref{wdef} fixed. For the latter one zooms in the region where $\hat r$ is close to $\hat r_h$ by defining $\hat r = \hat r_h - {\hat r_h} (1-w)/\xi$ and taking $\xi\to\infty$ which sends $\hat r_h\to 1$ (from equation (\ref{fxidef})) and keeps $w$ finite\footnote{It is important to note that this limit does not imply any limit in the domain wall coordinates. The limit $\hat r\to \hat r_h$ follows from $\xi\to\infty$ not from $u\to u_h$ in (\ref{relldef}) and (\ref{fxidef}). In particular the function $w$ can be expressed in terms of the domain-wall coordinates in the limit $\xi\to\infty$ as $w = 1- 3\log\le(\frac{u_0-u}{u_0-u_h}\ri)$ where $u$ hence $w$ is arbitrary.}.  
In this limit the equation~\eqref{nonhydroeq} becomes
\be
 \frac{1}{4w} \left(-q^2- 2 i  \varpi \right)\flf (w)+ \left(-w+i\varpi \, \right) \flf '(w) + w(1-w) \flf ''(w) =0 \ ,
\ee
where we dropped terms $\sim 1/\xi$.
The solution is given by
\bea \label{x12sol}
 \flf(w) &=& C_- w^{\frac{1}{2} \left(-\Sq-i \varpi +1\right)} \ _2F_1\left(\frac{1}{2}\left(- \Sq- i \varpi + 1\right),\frac{1}{2}\left(- \Sq- i \varpi + 1\right);1-\Sq;w\right) \nn\\
 &+&C_+ w^{\frac{1}{2} \left(\Sq-i \varpi +1\right)} \ _2F_1\left(\frac{1}{2}\left(\Sq- i \varpi + 1\right),\frac{1}{2}\left(\Sq- i \varpi + 1\right);1+\Sq;w\right)
\eea
where
\be \label{Sdef}
 \Sq = \sqrt{q^2 - \varpi^2+1} \ .
\ee
As we pointed out in Sec.~\ref{Xhalf}, the same solution has been found 
in the minisuperspace studies of the 2D black hole of the linear dilaton model \cite{Dijkgraaf:1991ba,Nakayama:2007sb}, for the reason that the background becomes linear dilaton in this limit as explained in section (\ref{Xhalf}).
We choose the branches of the square root factors such that the solution for negative $q^2 - \varpi^2+1$ is given by replacing $\Sq \mapsto - i\Sqt$ where
\be \label{Stdef}
 \Sqt = \sqrt{\varpi^2-q^2-1}\ ,
\ee
which corresponds to analytic continuation through the upper half of the complex $\varpi$-plane. Moreover for simplicity we restrict to frequencies with $\mathrm{Re}\,\varpi \geq 0$ below. The expressions for $\mathrm{Re}\,\varpi <0 $ can be obtained by applying reflection symmetry with respect to the imaginary $\varpi$-axis.

The expansion of~\eqref{x12sol} at small $w$ gives
\be \label{x12smallw}
 \flf(w) = C_- w^{\frac{1}{2} \left(-\Sq-i \varpi +1\right)} \left[1+\mathcal{O}\left({w}\right)\right] + C_+ w^{\frac{1}{2} \left(\Sq-i \varpi +1\right)} \left[1+\mathcal{O}\left({w}\right)\right]
\ee
This UV expansion obviously differs from the standard expansion $\sim C_1 + C_2 w$. This is not surprising because taking $\xi \to \infty$ takes us to the case of~\eqref{Vinterm} of the Schr\"odinger potential near the boundary, indicating a nonstandard boundary behavior. Notice also that the blackening factor $f(\hat r) = 1 - (\hat r/\hat r_h)^\xi$ equals one up to tiny corrections except for very close to the horizon $\hat r_h-\hat r \sim 1/\xi$ when $\xi$ is large and we took the limit of $\xi \to \infty$ such that $w =(\hat r/\hat r_h)^\xi$ is fixed. Therefore we were keeping the blackening factor nontrivial but losing the connection to the UV boundary. In other words,~\eqref{x12sol} is correct up to terms $\propto |\log(w)/\xi|$, the solution is valid for $e^{-\xi}\ll w$ for large but finite $\xi$, and the solution for $w \ll  e^{-\xi}$ would have a different UV behavior.

We then use the regularity condition at the horizon. Expanding~\eqref{x12sol} at $w=1$ we see the the outgoing wave is absent if
\be
\frac{C_- \, \Gamma \left(1-\Sq\right)}{\Gamma \left(\frac{1}{2} \left(1-i \varpi -\Sq\right)\right)^2}+\frac{C_+\, \Gamma \left(1+\Sq\right)}{\Gamma \left(\frac{1}{2} \left(1-i \varpi +\Sq\right)\right)^2}=0 \ .
\ee
Notice that for large $\varpi$ the UV expansion of the solution becomes
\be
 \frac{1}{\sqrt{w}}\flf(w) = C_-   \left[1+\mathcal{O}\left({w}\right)\right] + C_+ w^{-i \varpi} \left[1+\mathcal{O}\left({w}\right)\right] \ .
\ee
Therefore the first (second) term can be interpreted as an incoming (outgoing) wave at the boundary. The ratio of the coefficients defines the reflection amplitude
\be \label{Rampl}
 \mathcal{R}(\varpi,q) = \frac{C_+(\varpi,q)}{C_-(\varpi,q)} = - \frac{\Gamma \left(1+i\Sqt\right)\, \Gamma \left(\frac{1}{2} \left(1-i \varpi -i\Sqt\right)\right)^2}{ \Gamma \left(1-i\Sqt\right)\, \Gamma \left(\frac{1}{2} \left(1-i \varpi +i\Sqt\right)\right)^2} \ .
\ee
We expressed the amplitude in terms of  $\Sqt$ rather than $\Sq$ because this is more natural for large $\varpi$ where the interpretation as a scattering matrix element is clear. From this reflection amplitude one can derive the density of states as a derivative of the scattering phase 
\be \label{Rdensity}
\mathcal{R}(\varpi,q) = e^{i \Phi(\varpi,q)}, \quad \rho(\varpi,q) = \frac{d \Phi(\varpi,q)}{d\varpi} \, ,
\ee
nevertheless this is a limiting form for the density of states that holds in the exact $\xi \rightarrow \infty$ limit and for large frequencies $\varpi$. The exact result for finite $\xi$ shows a more intricate behaviour that will be analysed in the next subsection. A first indication comes from the regime of low $\varpi$, i.e., when $\varpi^2-q^2-1$ is negative, since then there is no wave propagation near the boundary hence $\mathcal{R}(\varpi,q)$ appears to be a retarded correlator rather than a reflection amplitude\footnote{This possibility is also related to the fact that we have the extra transverse directions and hence one can consider excitations with large transverse momentum $q$.}. These observations agree with the solutions in the Schr\"odinger form above.

The boundary behavior of the fluctuations and the matching with the solutions in the linear dilaton geometry for arbitrary $\xi$ are discussed in Appendix \ref{sec:bdry}. 
We notice a non-trivial point in the otherwise standard procedure: the naive separation of the general  solution into a source and a VEV term leads to a correlator that 
has singularities when $\xi$ is an even positive integer. However these singularities have a simple form and are unrelated to the quasinormal poles; we find it more convenient to 
split the correlator into a part $G_s$ which contains the singularities in $\xi$ but is regular as a function of $\varpi$, and a $G_\mathrm{reg}$ which is regular in $\xi$ and contains the information 
about the QNMs.

\begin{figure}[!tb]
\begin{center}
\includegraphics[width=0.506\textwidth,trim=0 10mm 0 0]{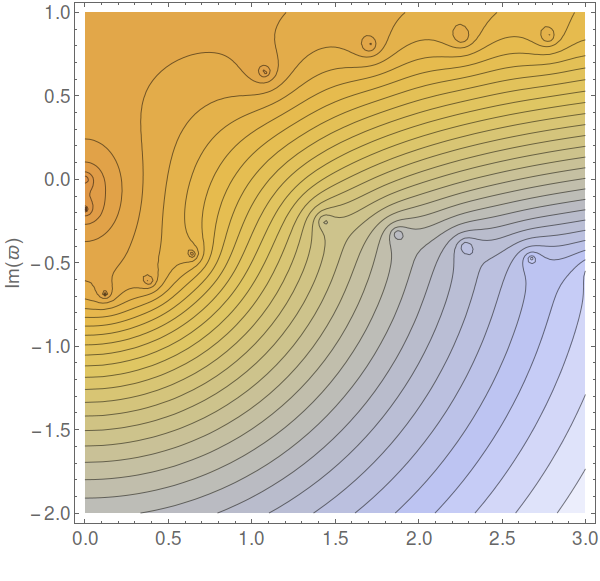}%
\hspace{2mm}\includegraphics[width=0.48\textwidth]{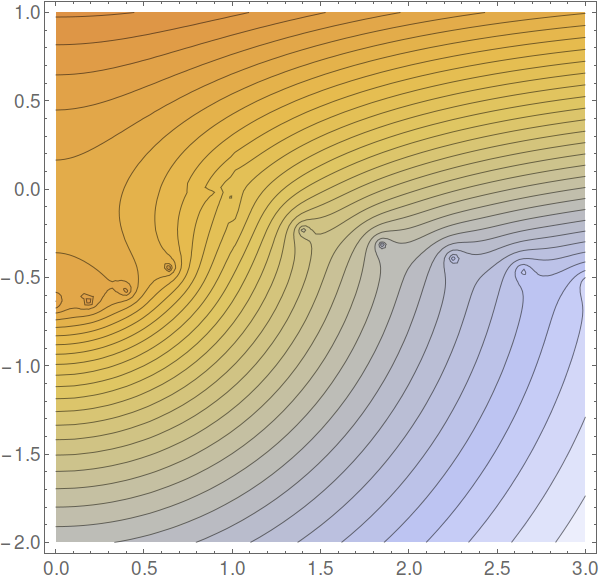}
\includegraphics[width=0.506\textwidth]{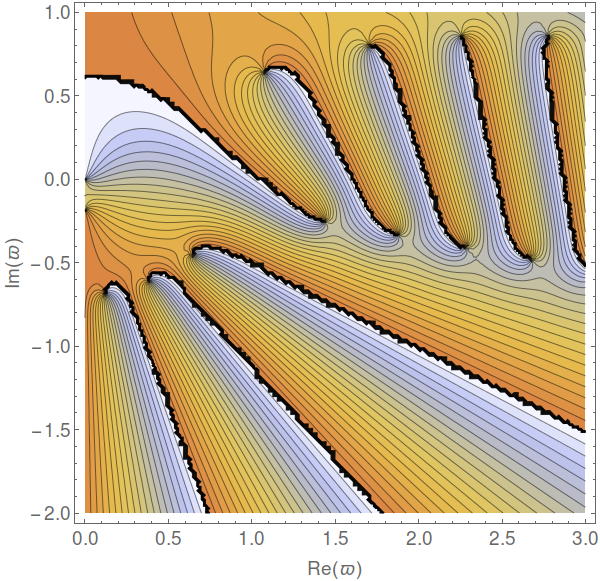}%
\hspace{2mm}\includegraphics[width=0.48\textwidth]{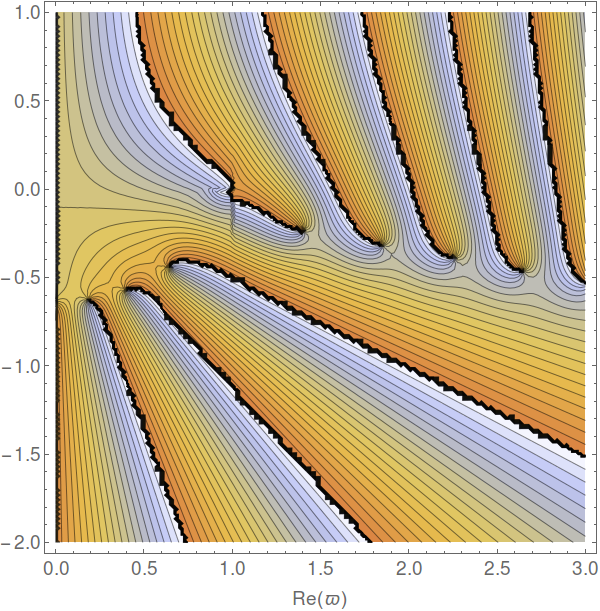}
\end{center}
\caption{The analytic correlator and comparison to direct numerical result, $X=-0.45$ and $q=0$. Left column: numerically extracted correlator on the complex $\varpi$ plane. Right column: the same plots for the analytic approximation at large $\xi$. Top row: logarithm of the absolute value of the correlator. Bottom row: phase of the correlator. }
\label{fig:corrcomparisonx045}
\end{figure}

\subsubsection{Analytic correlator and its properties}

The matching of the boundary and horizon expansion, performed in Appendix \ref{sec:bdry} gives us the 
regular part of the correlator
\bea \label{corrsmallomega}
  G_\mathrm{reg} &=& \frac{2\pi\, \xi^\xi \hat r_h^{-\xi}}{\Gamma\left(\frac{\xi}{2}\right)\Gamma\left(1+\frac{\xi}{2}\right)} \left(\frac{\left(\varpi ^2-q^2\right)}{16}\right)^\frac{\xi}{2}
  \left[\left(\frac{1+\Sq}{1-\Sq}\right)^\frac{\xi}{2}\, e^{-\xi \Sq}\, \mathcal{R}- i\theta(-\mathrm{Im}\,\varpi) \right]\nn\\
  &=&- \frac{2\pi\, \xi^\xi \hat r_h^{-\xi}\, e^{-\xi \Sq}}{\Gamma\left(\frac{\xi}{2}\right)\Gamma\left(1+\frac{\xi}{2}\right)}\left(\frac{\left(\varpi ^2-q^2\right)}{16}\right)^\frac{\xi}{2} \left(\frac{1+\Sq}{1-\Sq}\right)^\frac{\xi}{2} \frac{\Gamma \left(1-\Sq\right)}{\Gamma \left(1+\Sq\right)} \frac{\Gamma \left(\frac{1}{2} \left(1-i \varpi +\Sq\right)\right)^2}{\Gamma \left(\frac{1}{2} \left(1-i \varpi -\Sq\right)\right)^2} \nonumber\\
  &&- \theta(-\mathrm{Im}\,\varpi)\ \frac{2\pi i\, \xi^\xi \hat r_h^{-\xi}}{\Gamma\left(\frac{\xi}{2}\right)\Gamma\left(1+\frac{\xi}{2}\right)} \left(\frac{\left(\varpi ^2-q^2\right)}{16}\right)^\frac{\xi}{2}
\eea
for $0\leq \re \varpi \lesssim \sqrt{1+q^2}$, and
\bea \label{corrfinal}
 G_\mathrm{reg} &=& \frac{2\pi\, \xi^\xi \hat r_h^{-\xi} }{\Gamma\left(\frac{\xi}{2}\right)\Gamma\left(1+\frac{\xi}{2}\right)}\left(\frac{\left(\varpi ^2-q^2\right)}{16}\right)^\frac{\xi}{2} \left[i+\left(\frac{1+i \Sqt}{1-i \Sqt}\right)^\frac{\xi}{2} \frac{e^{-i \xi \Sqt}}{\mathcal{R}}\right]^{-1}\nn \\
 &=&  \frac{2\pi\, \xi^\xi \hat r_h^{-\xi} }{\Gamma\left(\frac{\xi}{2}\right)\Gamma\left(1+\frac{\xi}{2}\right)}\left(\frac{\left(\varpi ^2-q^2\right)}{16}\right)^\frac{\xi}{2} \nn \\
&\times& \left[i-\left(\frac{1+i \Sqt}{1-i \Sqt}\right)^\frac{\xi}{2} e^{-i \xi \Sqt}\  \frac{\Gamma \left(1-i\Sqt\right)}{\Gamma \left(1+i\Sqt\right)} \frac{\Gamma \left(\frac{1}{2} \left(1-i \varpi +i\Sqt\right)\right)^2}{\Gamma \left(\frac{1}{2} \left(1-i \varpi -i\Sqt\right)\right)^2}\right]^{-1}
\eea
for $\re \varpi \gtrsim \sqrt{1+q^2}$. The precise regime of validity of the two expressions is determined by the saddle point approximations discussed in Appendix~\ref{app:spbessel}. As in Sec.~\ref{sec:xiinfty} we defined the reflection amplitude $\mathcal{R}$ such that it is analytic in the upper half $\varpi$-plane (so that $\Sq$ is mapped to $-i\Sqt$ on the real line as one passes the branch point)\footnote{Notice that the $\theta$ term that is nonvanishing on the lower-half plane includes a nontrivial phase factor on the imaginary axis. This term however cancels with a similar factor in $G_s$ so the full correlator is real.}.

The result for the correlator is shown at $X=-0.45$ and $q=0$ in Fig.~\ref{fig:corrcomparisonx045} where we also compare it to the result obtained through a direct numerical solution to the fluctuation equation~\eqref{nonhydroeq} (see Appendix~\ref{app:comparison} for more details). Top row plots show the absolute value and bottom row plots the phase of the correlation. Several zeroes and poles are seen in the plots for the absolute values, where orange (blue) hues indicate the smallest (largest) values for the correlator. Only those singularities in the lower half plane for $\re \varpi > 1$ are poles, and the other singularities are zeroes. The phase of the correlator jumps by $2\pi$ at the black curves in the bottom row plots, so they are not physical branch cuts.

The analytic correlator and the numerical result agree well for most values of $\varpi$, but there are also some differences which appear to be larger than the expected $\morder{1/\xi}$ corrections. First, the disagreement for $\mathrm{Im}\,\varpi \gtrsim 0.5$ is due to the numerical computation failing in this region, which happens because the correlator is small and its extraction from the numerical solution to the fluctuation equation is challenging. 

Second, there are also differences close to the imaginary axis, in particular near $\varpi=0$. These differences signal the failure of our analytic approximation near some special points: The difference between the powers in the terms of the expansion~\eqref{x12smallw} equals $S$, and when it takes integer values, subleading terms in the source term are singular. This leads to the failure in matching of the IR fluctuations to the boundary behavior. The problematic points are given by 
\be \label{spuriouspoles}
 \omega = \pm i\sqrt{n^2-1-q^2}
 \ee 
with $n=1,2,\ldots$. For $n=1$ and $q=0$, the matching fails at $\varpi=0$ therefore explaining the differences between the plots of Fig.~\ref{fig:corrcomparisonx045} near the origin. For other values of $n$, the failure of matching produces spurious poles in the analytic approximation (not visible in the plots) which are absent in the full numerical result. These poles already appear in the reflection amplitude~\eqref{Rampl} and have been discussed in the literature: see, e.g.,~\cite{Parnachev:2005hh,Kovtun:2005ev}.

In addition, there is minor disagreement near $\varpi=\sqrt{1+q^2}$ which is special point of the saddle point approximation of Appendix~\ref{app:spbessel}. It is possible to derive analytic approximations which cover this point as well as the points~\eqref{spuriouspoles}, but this is not necessary for the purposes of this article.

The location of the quasi normal modes in the limit $X \to -1/2$ is given by the poles of the expressions~\eqref{corrsmallomega} and~\eqref{corrfinal}. The only poles of the former expression are, however, the spurious poles at the points $\omega = \pm i\sqrt{n^2-1-q^2}$ where our matching procedure fails. The physical poles are those of the latter expression, and they are determined by the equation
\be \label{qnmeq}
 \left( \frac{1+i \Sqt}{1-i \Sqt}\, e^{-2i \Sqt}\right)^\frac{\xi}{2}  = i\, \frac{\Gamma \left(1+i\Sqt\right)}{\Gamma \left(1-i\Sqt\right)} \frac{\Gamma \left(\frac{1}{2} \left(1-i \varpi -i\Sqt\right)\right)^2}{\Gamma \left(\frac{1}{2} \left(1-i \varpi +i\Sqt\right)\right)^2} = -i\mathcal{R} \,.
\ee
It is controlled, among other things, by the function
\be
 g(\Sqt) = \frac{1+i \Sqt}{1-i \Sqt}\, e^{-2i \Sqt} \ .
\ee
It is straightforward to check that the (absolute value of the) reflection amplitude $\mathcal{R}$ is analytic and does not contain poles or zeroes in the region of the validity of the equation, i.e., $\re \varpi \gtrsim \sqrt{1+q^2}$.
Therefore when taking $\xi \to \infty$ with other parameters fixed, solutions to~\eqref{qnmeq} can only be found when $|g(\Sqt)|$ deviates from one by at most corrections suppressed by $1/\xi$. Together with the regime of validity of~\eqref{corrfinal} this implies that $\Sqt$ is real and positive, up to $1/\xi$ corrections. 

In order to see in more detail where the quasi normal modes lie for large but fixed $\xi$, we can expand $g(\Sqt)$ at a point $\Sqt = \Sqt_0$ on the real axis:
\be
 g(\Sqt_0+\delta \Sqt) = g(\Sqt_0) \exp\left[-\frac{2i\Sqt_0^2 }{1+\Sqt_0^2}\delta \Sqt+ \mathcal{O}\left(\left(\delta \Sqt\right)^2\right) \right] \,.
\ee
Since $g(\Sqt_0)$ is a pure phase we may write\footnote{One should be careful here because there are several branch choices for the expression on the left hand side. Looking at, e.g.,~\eqref{corrfinal}, the relevant branch choice is seen to be $g(\Sqt_0)^{\xi/2} = \exp[\xi \log((1+i \Sqt_0)/(1-i \Sqt_0))/2-i \Sqt_0 \xi]$ where the standard branch is used for the logarithm.}
\be
 g(\Sqt_0)^{\xi/2} = e^{i\phi_0}
\ee
where $-\pi < \phi_0 < \pi$. It is natural keep $q$ fixed, denote $\Sqt_0 = \sqrt{\varpi_0^2-q^2-1}$, and assume that $\delta \Sqt$ is due to a (possibly complex) variation of $\varpi$ which we denote by $\delta \varpi$.  When $\delta \Sqt \sim 1/\xi$,~\eqref{qnmeq} boils down to
\bea\label{quant}
 \frac{\Sqt_0^2 }{1+\Sqt_0^2}\, \xi \delta \Sqt &=& \frac{\Sqt_0 \tilde \varpi_0 }{1+\Sqt_0^2}\, \xi \delta \varpi\\\nn
&=& \phi_0 - \frac{\pi}{2} - 2 \arg \Gamma \left(1+i\Sqt_0\right) + 2 i \log  \frac{\Gamma \left(\frac{1}{2} \left(1\!-\!i \varpi_0 \!-\! i\Sqt_0\right)\right)}{\Gamma \left(\frac{1}{2} \left(1\!-\!i \varpi_0 \!+\!i\Sqt_0\right)\right)} + 2 \pi n 
\eea
where $n$ takes integer values. We carry out a numerical analysis and check of this formula in Appendix~\ref{app:comparison}.

From the result~\eqref{quant} we see that the spacing between the nodes on the complex $\varpi$ plane is given by 
\be
 \frac{2\pi(1+\Sqt_0^2)}{\Sqt_0 \varpi_0\, \xi}=  \frac{2\pi(\varpi_0^2-q^2)}{\Sqt_0 \varpi_0\, \xi} \ .
\ee
The density of states\footnote{Notice that the density of states at finite $\xi$ arises from the phase factor $g(\Sqt)^{\xi/2}$ and therefore differs drastically from the expression defined in~\eqref{Rdensity} in terms of the reflection amplitude which holds when $\xi = \infty$. } is given as its inverse:
\be \label{rhores}
 \rho(\varpi_0) = \left.\frac{d n}{d \varpi}\right|_{\varpi=\varpi_0} = \frac{\Sqt_0 \varpi_0\, \xi}{2\pi(\varpi_0^2-q^2)} \ .
\ee
The imaginary parts of the nodes obey
\be \label{imom}
 \im\, \delta \varpi = \im\, \varpi = \frac{2(1+\Sqt_0^2)} {\Sqt_0 \varpi_0\, \xi}\log \left| \frac{\Gamma \left(\frac{1}{2} \left(1\!-\!i \varpi_0 \!-\! i\Sqt_0\right)\right)}{\Gamma \left(\frac{1}{2} \left(1\!-\!i \varpi_0 \!+\!i\Sqt_0\right)\right)}\right| \ .
\ee
This expression is negative. For large $\varpi_0$ we obtain $\im\, \varpi \simeq - \pi \Sqt_0/\xi $. Thus we have an analytic confirmation of the numerical results of 
section \ref{sec:flucts}: the QNM approach the real axis and become dense in the limit $\xi \to \infty$.

Finally let us compute the residues of the correlator at the nodes, i.e., at the roots $\varpi=\varpi_n$ of~\eqref{quant}. 
Inserting the roots back in the correlator~\eqref{corrfinal} we find the leading order result
\be \label{resres}
 \mathrm{Res}_{\varpi_n} = -\frac{2\pi\, \xi^\xi \hat r_h^{-\xi}}{\Gamma\left(\frac{\xi}{2}\right)\Gamma\left(1+\frac{\xi}{2}\right)} \left(\frac{\left(\varpi^2_0-q^2\right)}{16}\right)^\frac{\xi}{2} \frac{\varpi^2_0-q^2}{\Sqt_0 \varpi_0\, \xi} \ .
\ee
Notice that the rapidly oscillating phases are absent in this result. Therefore (and also because we assumed that the nodes are within an $\mathcal{O}\left(1/\xi\right)$ distance  of $\varpi=\varpi_0$) there is no dependence on the mode number $n$.  
Notice also that when weighted with the density of states~\eqref{rhores}, the last factor in~\eqref{resres} would cancel, leaving only the term coming from the multiplicative factor in~\eqref{corrfinal}.

\section{Completing the CR geometry in the UV}\label{sec:UVcompletion}

We have derived the behavior of the QNMs of the energy-momentum tensor in the transverse sector in the critical limit $\xi \to \infty$. We may wonder if these results have 
any physical significance, given that the CR geometry does not describe a UV-complete theory and can only be a good description of the IR physics. 
In order to answer this question we carry out a detailed analysis of the UV completion of the CR geometry in this section, i.e., geometries which approach asymptotically AdS$_5$ in the UV and the CR form in the IR. 
We show that the results of the previous section also apply to this more general class of geometries at low temperatures and for the lowest modes of the spectrum;
the corrections coming from the UV part of the geometry are suppressed. Moreover, we obtain a fully analytic approximation for the transverse correlator of the energy-momentum tensor by gluing a slice of the AdS$_5$ geometry in the UV directly to the CR geometry. This improved analytic approximation works for the whole spectrum for temperatures below a certain critical temperature, and is expected to be a good model for the spectrum of more general geometries (i.e., geometries with a smooth flow from AdS$_5$ to CR instead of a joint).

The fact that the UV corrections are suppressed at small temperatures, so that the lowest quasi normal modes are determined by the CR geometry in the IR, can be understood by considering how the dilaton potential behaves at large $\xi$. At the horizon the dilaton and the potential take the value
\be \label{horvalues}
 \phi_h = \hat r_h^{-\frac{3X}{1-4X^2}} \ , \qquad V(\phi_h) =   \frac{9}{\ell^2\le(1-1/\xi\ri)}\, \hat r_h^\frac{2(\xi-4)}{3} \ .
\ee
We see that $V(\phi_h)$ diverges or goes to zero depending on whether the location of the horizon is larger or smaller than one (in terms of the dimensionless conformal coordinate). 
Therefore we expect that when $\hat r_h>1$ the physics determined by the near-horizon geometry will be insensitive to finite deformations of the potential in the UV.

We will show this in detail for geometries which are asymptotically $AdS_5$ in the UV and smoothly deform to the CR form in the IR. This is important because for most realistic models of the YM theory~\cite{ihqcd1,ihqcd2}, the potentials are expected to be this type, with IR asymptotics corresponding to the critical value $X=-1/2$. Since the flow from AdS$_5$ to CR is not known analytically, as we already mentioned above,
we will also discuss a scenario where an exact AdS$_5$ UV geometry is glued directly to the CR solution, and argue that this is a good approximation to smooth geometries with the same UV and IR  asymptotics. 
Remarkably, in the glued geometry the limit $X \to -1/2$ is well-defined, and we can give analytic results for the QNM at $X =1/2$ and not just close to it\footnote{Actually most of the discussion on the gluing does not rely on $X$ being close to the critical value. However, it is only in this limit that  we have analytic control over the IR geometry and the fluctuations. Therefore we will assume the limit $X \to -1/2$ in the following.}.

\subsection{Background geometry for a generic dilaton potential} \label{sec:gluedbg}

We will start with a generic discussion, first at zero temperature.  It is useful to introduce the superpotential formalism and consider the domain wall coordinates \ref{dw}, which are regular in the critical limit $X \to -1/2$. With the normalization conventions of~\eqref{action}, we choose $W(\phi)$ such that
\be \label{VWrel}
 V(\phi) = \frac{64}{27}\, W(\phi)^2 - \frac{4}{3}\, W'(\phi)^2 \ .
\ee
In order to obtain a flow from the CR solution in the IR to an AdS$_5$ at the boundary, we require that\footnote{We can assume that $0 \leq \Delta \leq 2$ in the UV expansion of $W(\phi)$. Our calculations are also valid for $\Delta>2$, but generic dilaton potentials $V(\phi)$ do not lead to IR regular superpotentials with $\Delta$ in this range; instead $V(\phi)$ needs to be fine tuned. }
\be
 W(\phi) = \frac{9}{16X^2\ell} e^{-\frac{4 X}{3}\phi}\left[1 + \morder{\frac{1}{\phi}}\right] \ , \qquad W(\phi) = \frac{9}{4\ell_\mathrm{AdS}} + \frac{\Delta}{2\ell_\mathrm{AdS}} \phi^2 + \morder{\phi^3} \ .
\ee
The exact form of the corrections in the UV and in the IR is not important, we assume power law corrections for simplicity\footnote{When $X=-1/2$ the corrections are however important in that they determine the order of the continuous deconfinement transition in the dual plasma~\cite{Gursoy:2010kw}.  
Power law corresponds to BKT like transitions generically.}.
Notice also that we chose the minimum of the potential to lie at $\phi = 0$. We will also choose a potential such that $\ell_\mathrm{AdS}/\ell ,  \Delta \sim \morder{1}$ and that the asymptotic regions are smoothly connected through a simple, monotonic function.

In terms of the superpotential, the background equations of motion read
\be
 \phi'(u) = W'(\phi(u))\ , \qquad A'(u) = -\frac{4}{9} W(\phi(u)) \ .
\ee
Notice that we have readily fixed one integration constant of the full solution to the action~\eqref{action} by fixing the superpotential (which corresponds to restricting to the IR regular geometries). The two remaining constants of integration are ``trivial'': they appear through the invariance of the equations of motion under shifts of $u$ and $A$. 

The UV ($u \to -\infty$) asymptotics of the solutions are
\be 
 \phi(u) = \phi_0 e^{u \Delta /\ell_\mathrm{AdS}} +\morder{e^{2 u \Delta/\ell_\mathrm{AdS}}} \ , \quad A(u) = \tilde  A_0 -\frac{u}{\ell_\mathrm{AdS}} -\frac{1}{9}  \phi_0^2 e^{2 u \Delta /\ell_\mathrm{AdS}}+\morder{e^{3 u \Delta/\ell_\mathrm{AdS}}} \, ,
\ee
while the IR asymptotics read
\begin{align}
 \phi(u) &= \frac{3}{4 X} \log\left(-\frac{u}{\ell}\right) + \morder{\frac{1}{ \log\left(-\frac{u}{\ell}\right)}}& \\
 A(u) &= A_0 +\frac{1}{4 X^2} \log\left(-\frac{u}{\ell}\right) + \morder{\frac{1}{ \log\left(-\frac{u}{\ell}\right)}}& 
\end{align}
where we fixed one of the constants of integration ($C_1$ in~\eqref{BHmetX}) such that $u \to 0$ in the IR. The full solution would give two relations between $\tilde A_0$, $A_0$, and $\phi_0$. Actually we can use the freedom of shifting $A$ to set $\tilde A_0=0$. Then $A_0$ and $\phi_0$ are $\morder{1}$ for dilaton potentials $V$ which meet the requirements specified above. 

We then consider the transformation to conformal coordinates, which lead to singular behavior for $X=-1/2$. Therefore we will assume that $0>X>-1/2$ 
and that $X+1/2 \ll 1$ or equivalently that $\xi \gg 1$. We choose that the UV boundary is at $r=0$ so that
\be \label{rurelation}
 r = \int_{-\infty}^u d\tilde u\, e^{-A(\tilde u)} \ .
\ee
Near the boundary we therefore have 
\be
 r = \ell_\mathrm{AdS} e^{-\tilde A_0 + u/\ell_\mathrm{AdS}}\left[1  +  \morder{e^{2 u \Delta/\ell_\mathrm{AdS}}}\right] \ ,
\ee
but in the IR the expansion is more interesting, namely
\be
 r = \ell' \le[\le(-\frac{u}{\ell}\ri)^{-\frac{1-4X^2}{4X^2}}\le(1+\morder{\frac{1}{ \log\left(-\frac{u}{\ell}\right)}}\ri) - 1 \ri] + \morder{\xi^0} \ ,
\ee
where $\ell' = 4X^2e^{-A_0} \ell/(1-4X^2)$. Comparing to the CR solution in Sec.~\ref{sec:bg} we see that there is a shift of the $r$ coordinate by $\ell'$, given by the last term in the square brackets, which is $\morder{\xi}$. Such a shift is a global property of the definition of the conformal coordinate in~\eqref{rurelation}, and would not appear in a naive direct IR expansion of the relation. 

The shift ensures that $r$ becomes $\morder{\ell \xi^0}$ for $-u \sim \ell$, i.e., when the asymptotic IR expansion starts to fail, and matches smoothly with the UV expansion. We see that the UV region ($-u \gg \ell$) maps to $r \ll \ell$, and the IR region ($-u \ll \ell$) maps to $r \gg \ell$. 
The shift also ensures that the limit $X \to -1/2$ is smooth, as can be seen from the analysis in Sec.~\ref{sec:bg}: for the critical limit of~\eqref{relldef} to match with~\eqref{relldef12} an analogous shift of $r$ is needed in either of the definitions.

\begin{figure}[!tb]
\begin{center}
\includegraphics[width=0.7\textwidth]{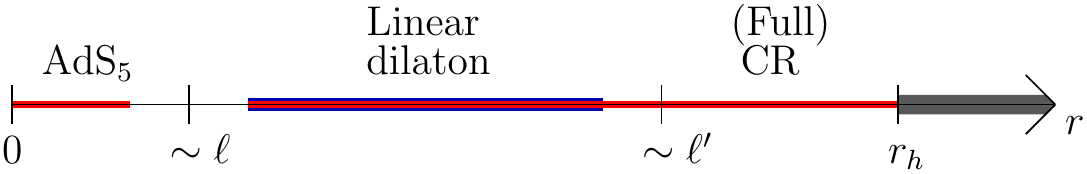}%
\end{center}
\caption{The structure of the (zero temperature) geometry for potentials asymptoting to CR behavior in the IR with large $\xi$. For $0<r\ll \ell$, the geometry is asymptotically AdS$_5$, and for $r \gg \ell$ it is asymptotically CR (regimes marked with red color). Further, when $\ell \ll r \ll \ell'$ (where $\ell' \sim \ell\xi$) the geometry is to a good approximation the linear dilaton geometry (the interval marked with blue color), whereas for $r \gg \ell'$ this approximation fails and only the (full) CR geometry can be used to describe the solution.}
\label{fig:geometry}
\end{figure}

The result of the shift is that in terms of the conformal coordinates, a large section of the length $\sim\ell'$ of the CR geometry is cut away from the UV, and then replaced by a short (length $\sim \ell$) section of the AdS geometry. 
Dropping the subleading terms, the warp factor in the CR part of the geometry becomes
\be
 e^{A(r)} \simeq e^{A_0} \le(1+\frac{r}{\ell'}\ri)^{-\frac{1}{1-4X^2}} \ , \qquad \le( r \gg \ell\ri) \ .
\ee
This ensures that taking $\xi \to \infty$ at fixed $r$ and $\ell$ such that $\ell' \to \infty$, the warp factor smoothly approaches the solution at $X=-1/2$, i.e., an exponential function of $r$ (i.e., the linear dilaton geometry). Collecting these observations, the structure of the geometry is that of Fig.~\ref{fig:geometry}. The AdS and CR geometries are glued together at $r \sim \ell$, but in addition for large $\xi$ there is a regime with $\ell \ll r \ll \ell'$ where the geometry is close to the linear dilaton geometry.

For finite temperature configurations at small enough temperatures, the above discussion is only modified by adding a nontrivial blackening factor in the CR geometry in the IR. We check now what is the condition for this to work.
In the conformal coordinates the blackening factor reads 
\be \label{fconf}
 f(r) = 1 - \le[\frac{\pi T\ell}{e^{A_0}(1-X^2)}\le(1+\frac{r}{\ell'}\ri)\ri]^\xi \ ,\qquad \frac{r_h}{\ell'} = \frac{e^{A_0}(1-X^2)}{\pi T\ell} -1 \ .
\ee
At large $\xi$, the horizon in the asymptotic IR region as $\xi$ and consequently $\ell'$ grow if the last factor in~\eqref{fconf} is positive, i.e., $ T < 3 e^A_0/4\pi\ell \equiv T_c$ (where we inserted $X \to -1/2$).
In order to write down a slightly more precise condition we should check when the $r$-dependence of the blackening factor becomes negligible at the regime where the geometry is nontrivial ($r \sim \ell$). From~\eqref{fconf} we see that this is the case when $r_h/\ell \gg 1$. Therefore the horizon may lie in any location of the CR-part of the geometry in Fig.~\ref{fig:geometry}, including the regime where the geometry resembles that of the linear dilaton background.
The temperature deviates significantly from $T_c$ when $r_h \sim \xi$, but the approximation holds under the weaker condition $\xi ( T_c - T) /T_c \gg 1$. 

In summary, for temperatures lower than 
\be
 T_c = \frac{3 e^{A_0}}{4\pi \ell} \ ,
\ee
the blackening factor only modifies the CR part of the geometry, and we expect that the analytic results of Sec.~\ref{sec:Xonehalf} describe accurately the correlators, up to some critical frequency. This critical frequency is analyzed in more detail in Appendix~\ref{app:UVmodflucts} and found to be $\morder{1/\ell}$. Notice that this temperature agrees with the temperature of the black hole~\eqref{tempsp} when $X=-1/2$ exactly.

Fluctuations for the case of a dilaton potential which interpolates smoothly between the CR asymptotics in the IR and AdS behavior in the UV are discussed in detail in Appendix~\ref{app:UVmodflucts}. Here we shall proceed directly to the case of directly gluing the AdS and CR geometries together, in which case there is much better analytic control. We expect, even though this cannot be proven, that the analytic results for the correlators and quasi normal modes obtained though the gluing procedure are qualitatively similar to those for simple geometries interpolating smoothly between AdS and CR.

\subsection{Gluing together the UV and IR geometries} \label{sec:gluing}

Here we develop a fully analytic approximation for the correlator of the transverse spin two modes (which also applies to other fluctuations of the metric at zero momentum) by considering a background where the UV AdS geometry is glued directly to the IR CR geometry. In order to do this it is convenient to start from the superpotential, which we take to be continuous:\footnote{The solution which we write down does not probe the superpotential for $\phi<\phi_c$. It is anyhow natural to take a the superpotential to be constant in this region in which case the background in~\eqref{bgsolglue} is the single consistent solution.}
\be
 W(\phi) = \frac{9}{4\ell_\mathrm{AdS}} \theta(\phi_c-\phi) + \frac{9}{4\ell_\mathrm{AdS}} e^{-\frac{4 X}{3}\le(\phi-\phi_c\ri)}  \theta(\phi-\phi_c) \ .
\ee
Notice that the potential $V(\phi)$ in~\eqref{VWrel} will be discontinuous at $\phi=\phi_c$.
For this superpotential the background solution is
\begin{align} \label{bgsolglue}
 A(r) &= - \log r +\log\ell_\mathrm{AdS} \ , \qquad \phi(r) = \phi_c \ ,&  \quad &(r\leq r_c) \ ,&\\
 A(r) &= - \frac{1}{1-4X^2}\log\le[(1-4X^2)\le(\frac{r}{r_c}-1\ri)+1\ri] -\log\frac{r_c}{\ell_\mathrm{AdS}}\ ,&  & \nn\\
 \phi(r) &= -\frac{3 X}{1-4X^2}\log\le[(1-4X^2)\le(\frac{r}{r_c}-1\ri)+1\ri] + \phi_c\ ,&  \quad &(r\geq r_c) \ . &
\end{align}
The general solution has two integration constants: one related to shifts of $r$ which we have already fixed by requiring that the boundary is located at $r=0$, and $r_c$ which can be changed by rescaling the $r$ coordinate and therefore plays the role of $A_0$ in Sec.~\ref{sec:bg}. The precise connection to the solution there is 
\be
 A_0 = -\log\frac{r_c}{\ell_\mathrm{AdS}} - \frac{1}{3X} \phi_c \ ,\qquad \ell' = \frac{\ell}{e^{A_0}(1-4X^2)} = \frac{r_c\,e^{\frac{1-4X^2}{3X}\phi_c}}{1-4X^2} \ .
\ee
In addition there is, in agreement with the analysis in 
Sec.~\ref{sec:gluedbg}, a shift of the coordinate $r$ in the CR part of the metric ($r \ge r_c$) given by
\be
 \Delta r = r_c\le(\frac{1}{1-4X^2}-1\ri) =\frac{4 X^2 r_c}{1-4X^2} \ .
\ee
Notice that as $X \to -1/2$ we find $\Delta r \to \ell'$ as argued above and that $A$, $A'$, and $\phi$ are continuous at $r=r_c$, whereas $\phi'$ is discontinuous.  The temperature, the scaled frequency, and the scaled momentum become
\be
\qquad T = \frac{\xi}{4\pi(\Delta r +r_h)}\ ,\qquad  \varpi = \frac{2\omega(\Delta r +r_h)}{\xi}\ , \qquad \mathrm{and} \qquad q = \frac{2k(\Delta r +r_h)}{\xi} \ ,
\ee
respectively.

The fluctuations (around a generic dilaton potential) are considered in Appendix~\ref{app:UVmodflucts}. It is convenient to use ordinary time instead of the Eddington-Finkelstein coordinate $v$ which causes some changes in the fluctuation equations with respect to Sec.~\ref{sec:flucts}. 
For the current setup the solutions for the fluctuation equations are given in~\eqref{UVexps} (setting $\Delta=0$) and~\eqref{IRexps} 
for $r<r_c$ and $r>r_c$, respectively:
\begin{align} \label{gluedexps}
\Xi (r) &= C_\mathrm{UV}^{(1)} \frac{i \pi m^2}{4} r^{2}H^{(1)}_{2}( m\, r )  + C_\mathrm{UV}^{(2)} \frac{4}{ m^{2}} r^{2} J_{2}(m\,r)\ , & (r<r_c) \, \\
\Xi (r) &=C_\mathrm{IR}^{(1)} \frac{i \pi  \tilde r^{\xi /2}  H_{\xi/2}^{(1)}\le(m\tilde r\ri)}{2^{\xi /2} m^{-\xi/2} \Gamma \left(\frac{\xi }{2}\right)} + C_\mathrm{IR}^{(2)} \frac{2^{\xi /2}\Gamma \left(\frac{\xi }{2}+1\right) \tilde r^{\xi /2} J_{\xi/2}\le(m\tilde r\ri)}{ m^{\xi /2}{\Delta r}^\xi}\ , & (r>r_c) \, \label{gluedexps2}
\end{align}
where $\tilde r =r +\Delta r$.
The UV and IR coefficients are related through a transition matrix
\be
 C_\mathrm{UV} = M C_\mathrm{IR} 
\ee
where $C_\mathrm{UV/IR} = (C_\mathrm{UV/IR}^{(1)},C_\mathrm{UV/IR}^{(2)})$ (see also Appendix~\ref{app:UVmodflucts}). By requiring the continuity of the solutions and their derivatives at $r=r_c$, we can compute the transition matrix $M$. 
It is tempting to use the limit $X \to -1/2$ to simplify the results but as it turns out it is better to use the exact expressions to avoid precision issues due to cancellations of large factors. The result reads
\begin{align} \label{Mglued}
 M_{11} &= \frac{  2 i  \pi \left(\frac{\mu}{2}\right)^{\xi/2} }{\tilde m  \Gamma \left(\frac{\xi }{2}\right)} \left[J_1(\tilde m ) H_{\xi /2}^{(1)}(\mu)-J_2(\tilde m ) H_{\xi/2-1}^{(1)}(\mu)\right]& \nn\\
 M_{12} &= \frac{2  \Gamma \left(\frac{\xi }{2}+1\right)}{ \left(\frac{\mu}{2}\right)^{\xi/2}  \tilde m}\left(\frac{\xi-1}{\xi-4}\right)^\xi \left[J_1(\tilde m) J_{\xi/2}(\mu )-J_2(\tilde m) J_{\xi/2-1}(\mu )\right] &\nn\\ 
 M_{21} &= \frac{ \pi ^2  \left(\frac{\mu}{2}\right)^{\xi/2}  \tilde m^{3} }{16 r_c^4 \Gamma \left(\frac{\xi }{2}\right)}\left[H_1^{(1)}(\tilde m) H_{\xi/2}^{(1)}(\mu )-H_2^{(1)}(\tilde m) H_{\xi/2 -1}^{(1)}(\mu )\right] &\nn\\
 M_{22} &= \frac{i    \pi  \tilde m ^3  \Gamma \left(\frac{\xi }{2}+1\right)}{16 r_c^4\left(\frac{\mu}{2}\right)^{\xi/2} } \left(\frac{\xi-1}{\xi-4}\right)^\xi 
  \left[H_2^{(1)}(\tilde m ) J_{\xi/2-1}(\mu)-H_1^{(1)}(\tilde m ) J_{\xi /2}(\mu)\right] &
\end{align}
where $\tilde m = m r_c$ and $\mu = m r_c (\xi -1)/3$.

There is a small subtlety when comparing to the generic expressions above in Sec.~\ref{sec:Xonehalf} and in Appendix~\ref{app:UVmodflucts}. Namely, we included a factor $\Delta r^\xi$ in the definition the normalizable IR wave function in~\eqref{gluedexps2}, which is most convenient for the gluing procedure, whereas $\ell'^\xi$ was used in~\eqref{IRexps}. As a result the $m \to 0$ limit of $M_{22}$ differs from~\eqref{M22spintwo} by a factor of $(\ell'/\Delta r)^\xi$, which is finite and in general different from one in the limit $\xi \to \infty$. Also, the proper normalization of $G_\mathrm{reg}$ for the normalization of~\eqref{gluedexps2}, and when using the matrix in~\eqref{Mglued}, is to replace $\hat r_h^{-\xi}$ in the results of Sec.~\ref{sec:preciselimit} by $(1+r_h/\Delta r)^{-\xi}$ rather than  $(1+r_h/\ell')^{-\xi}$.

The analytic result for the transverse spin-two correlator is given in terms of the transition matrix by
\be \label{Gtildetext}
 \widetilde G_\mathrm{reg}  = \frac{M_{21}+M_{22}  G_\mathrm{reg}}{M_{11}+M_{12}  G_\mathrm{reg}} \ ,
\ee
which is regulated in the same way as $G_\mathrm{reg}$ in Sec.~\ref{sec:Xonehalf}, see Appendix~\ref{app:UVmodflucts} for details.

\begin{figure}[!tb]
\begin{center}
\includegraphics[width=0.49\textwidth]{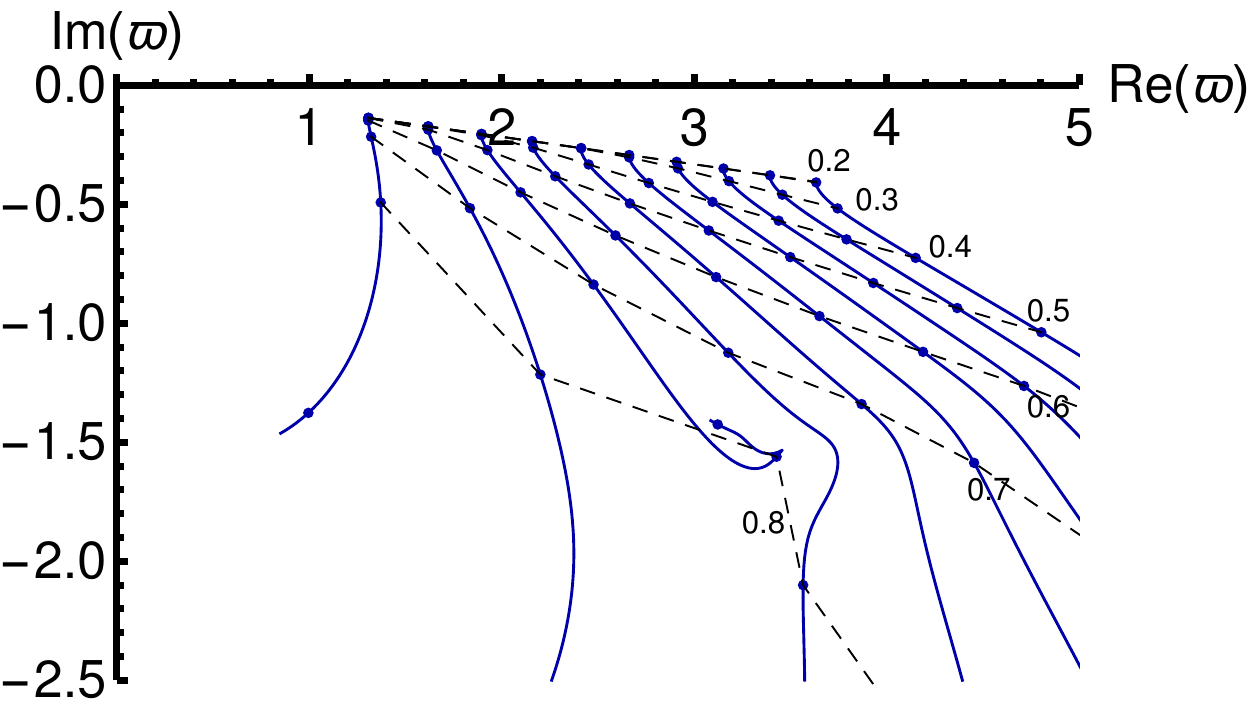}%
\includegraphics[width=0.49\textwidth]{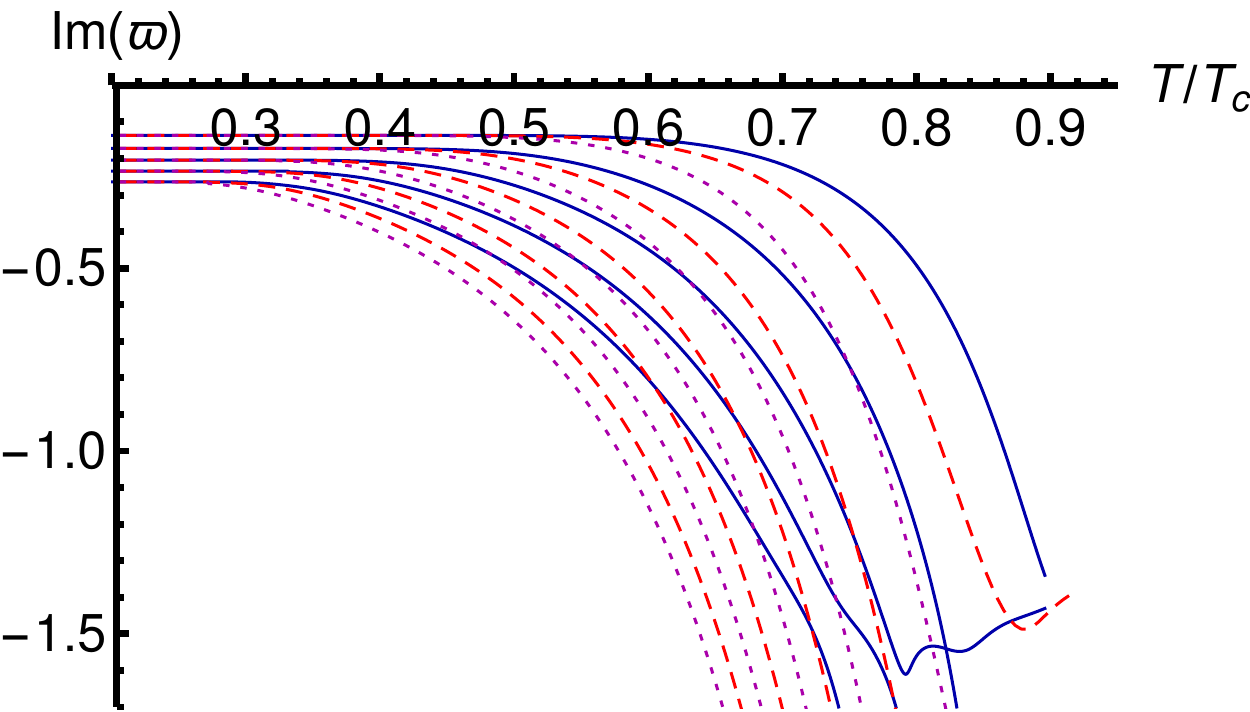}
\end{center}
\caption{The dependence of the location of quasi normal modes on temperature at $X=-0.47$ ($\xi\simeq 26.77$) and at $q=0$ in the setup where AdS and CR geometries were glued together. Left: The trajectories of the ten lowest QNMs on the complex $\varpi$-plane as $T$ grows from $T=0.2 T_c$ to $T=0.91 T_c$. The dashed curves are at constant $T/T_c$ with values of the ratio indicated by the labels. The markers are at $T/T_c=0.2$, $0.3$, \ldots $0.9$ for all curves. Right: Comparison of the result to simple boundary conditions at $r=r_c$. Blue, dashed red, and dotted magenta curves give the imaginary part of $\varpi$ as a function of the temperature for the QNMs with glued, Dirichlet, and Neumann boundary conditions, respectively.}
\label{fig:QNMevolution}
\end{figure}

\subsection{Temperature dependence of the QNMs} \label{sec:QNMTdep}

Let us then study the dependence of the location of the quasi normal modes on temperature by using the exact solutions for the fluctuations. Their location is given by the equation $M_{11}+M_{12}  G_\mathrm{reg} = 0$ with $M_{ij}$ given in~\eqref{Mglued} and the correlator in~\eqref{corrfinal}. We take $q=0$ so the results apply for all fluctuations of the metric (not just the transverse spin-two modes), and choose $X=-0.47$ close to the critical value so that $\xi \simeq 26.77$ is relatively high and corrections in $1/\xi$ are suppressed. We show how the trajectories of the QNMs on the complex $\varpi$-plane in Fig.~\ref{fig:QNMevolution} (left). At the lowest temperature $T=0.2T_c$, the locations of the QNMs are indistinguishable from their zero temperature limit, which is governed by the CR geometry with the explicit expressions given in Sec.~\ref{sec:preciselimit}. As the temperature grows, the higher QNMs start to move first. As $T$ approaches $T_c$, all QNMs move toward larger negative values of $\mathrm{Im}\,\varpi$. The slope of the locations of QNMs and their separation on the complex $\varpi$-plane also grow. 
There is additional structure related to the third mode, the movement of which changes direction at $T/T_c \simeq 0.8$ and becomes much slower. This reflects the existence of an additional set of modes with weaker temperature dependence, which we will study in more detail in Sec.~\ref{sec:modesxiinfty}.

Recall that the essence of our gluing procedure in Secs.~\ref{sec:gluedbg} and~\ref{sec:gluing} was that a long section (length $\sim \xi$ in $r$-coordinates) of the CR geometry was replaced by a short section (length $\sim\xi^0$) of an AdS geometry in the UV. This can be viewed as a smoothed out cutoff of the CR geometry at $r=r_c$. Therefore we compare the QNMs to those obtained in a simple setup where a hard wall is placed at $r=r_c$ instead of a glued geometry in Fig.~\ref{fig:QNMevolution} (right). The blue curves are the trajectories in the glued setup, whereas the red dashed and magenta dotted curves are the trajectories for the hard wall with Dirichlet and Neumann boundary conditions for the fluctuations at the wall, respectively.  We see that the direction of the movement is roughly the same in all cases, but the QNMs deviate from their zero temperature limit clearly faster in the hard wall setup. This suggests that the gluing procedure is indeed necessary to describe realistic trajectories.

\subsection{The critical case $X=-1/2$} \label{sec:modesxiinfty}

Interestingly, many of the expressions derived above in this section 
remain well-defined in the limit $X \to -1/2$ (or $\xi \to \infty$). That is, making the geometry asymptotically AdS in the UV regulates the $\xi \to \infty$ limit of the results for the CR geometry in Sec.~\ref{sec:Xonehalf} which could not be directly generalized to $\xi = \infty$. In order to highlight the behavior at $\xi = \infty$, we discuss here the results in the case of gluing, i.e, the results of Sec.~\ref{sec:gluing} and Sec.~\ref{sec:QNMTdep} in the limit $\xi \to \infty$.
 
For $X=-1/2$ and $\xi=\infty$ the background solution of~\eqref{bgsolglue} becomes
\begin{align} \label{bgsolxiinfty}
 A(r) &= - \log r +\log\ell_\mathrm{AdS} \ ,& \qquad \phi(r) &= \phi_c \ ,&  \quad &(r\leq r_c) \ ,&\\
 A(r) &= 1-\frac{r}{r_c} -\log\frac{r_c}{\ell_\mathrm{AdS}}\ ,&\quad \phi(r) &= -\frac{3}{2} +\frac{3r}{2r_c}+ \phi_c\ ,&  \quad &(r\geq r_c) \ . &
\end{align}
Adding a horizon far from the gluing point with a blackening factor $f(r) = 1 - \exp\le(3(r-r_h)/r_c\ri)$ and $r_h \gg r_c$, we notice that the temperature $T = 3/(4\pi r_c)$ is independent of $r_h$ as expected for this geometry~\cite{ihqcd4,Gursoy:2015nza}. We remark that $r_h$ should be understood as a proxy for the temperature in spite of this: 
the temperature is only independent of $r_h$ up to highly suppressed corrections $\sim \exp(-3 r_h/r_c)$ which we will ignore below, and the dependence would also be present for more generic dilaton potentials interpolating between the AdS and linear dilaton behaviors.

The UV solution to the (zero temperature) fluctuation equations is unchanged, while the IR solution simplifies to
\be \label{IRx12}
\Xi(r) =  C_\mathrm{IR}^{(1)}\,  e^{\frac{3 r}{2r_c} \left(1 - \sqrt{1-\hat \mu^2}\right)}+C_\mathrm{IR}^{(2)}\,  e^{\frac{3 r}{2r_c} \left(1 + \sqrt{1-\hat \mu^2}\right)}
\ee
where $\hat \mu = 2 m r_c/3$ which becomes $\hat \mu = m/2\pi T =\sqrt{\varpi^2 -q^2}$ after adding a horizon deep in the IR. Notice that there is a branch cut for $\hat \mu>1$.
The transition matrix takes the relatively simple form
\begin{align} \label{MgluedX12}
 M_{11} &= \frac{4 e^{\frac{3}{2} \left(1-\sqrt{1-\hat \mu ^2}\right)} \left[\hat \mu  J_1\left(\frac{3 \hat \mu }{2}\right)-\left(1-\sqrt{1-\hat \mu ^2}\right) J_2\left(\frac{3 \hat \mu }{2}\right)\right]}{3 \hat \mu ^2} & \nn\\
 M_{12} &= \frac{4 e^{\frac{3}{2} \left(1+\sqrt{1-\hat \mu ^2}\right)} \left[\hat \mu  J_1\left(\frac{3 \hat \mu }{2}\right)-\left(1+\sqrt{1-\hat \mu ^2}\right) J_2\left(\frac{3 \hat \mu }{2}\right)\right]}{3 \hat \mu ^2}
 & \nn\\
 M_{21} &= -\frac{27 i \pi  e^{\frac{3}{2} \left(1-\sqrt{1-\hat \mu ^2}\right)} \hat \mu ^2 \left[\hat \mu  H_1^{(1)}\left(\frac{3 \hat \mu }{2}\right)-\left(1-\sqrt{1-\hat \mu ^2}\right) H_2^{(1)}\left(\frac{3 \hat \mu }{2}\right)\right]}{128 r_c^4} & \nn \\
 M_{22} &= -\frac{27 i \pi  e^{\frac{3}{2} \left(1+\sqrt{1-\hat \mu ^2}\right)} \hat \mu ^2 \left[\hat \mu  H_1^{(1)}\left(\frac{3 \hat \mu }{2}\right)- \left(1+\sqrt{1-\hat \mu ^2}\right) H_2^{(1)}\left(\frac{3 \hat \mu }{2}\right)\right]}{128 r_c^4} &
\end{align}

The correlator is then given as in~\eqref{Gtildetext}, but as $G_\mathrm{reg}$ of Sec.~\ref{sec:preciselimit} does not have smooth $\xi \to \infty$ limit, we need to study the finite temperature solutions to the fluctuation equations for the IR geometry in order find the correct expression for $G_\mathrm{reg}$ when $X=-1/2$. The solutions are readily given in~\eqref{x12sol} in terms of the coordinate $w$. From the blackening factor $f(r)=1-\exp\le(3(r-r_h)/r_c\ri)$ we read that $w=\exp\le(3(r-r_h)/r_c\ri)$, and factors of $\exp\le[i\omega \int_0^r f^{-1}(\tilde r) d\tilde r\ri]$ should be added due to change from the Eddington-Finkelstein time coordinate to ordinary time so that
\begin{align}
 \Xi(r) &= C_- e^{\frac{3(r-r_h)}{2r_c} \left(1-\Sq\right)} f(r)^{-i\varpi/2} & \nn\\
 &\phantom{=}\times\ _2F_1\left(\frac{1}{2}\left(- \Sq- i \varpi + 1\right),\frac{1}{2}\left(- \Sq- i \varpi + 1\right);1-\Sq;e^\frac{3(r-r_h)}{r_c}\right) & \nn\\
 &\phantom{=}+C_+ e^{\frac{3(r-r_h)}{2r_c} \left(1+\Sq\right)} f(r)^{-i\varpi/2} & \nn\\
 &\phantom{=}\times\ _2F_1\left(\frac{1}{2}\left(\Sq- i \varpi + 1\right),\frac{1}{2}\left(\Sq- i \varpi + 1\right);1+\Sq;e^\frac{3(r-r_h)}{r_c}\right) &
\end{align}
where $\Sq = \sqrt{1-\hat \mu^2} = \sqrt{1 - \varpi^2+q^2}$. Taking the UV limit we pin down the connection to~\eqref{IRx12}
\be
 C_\mathrm{IR}^{(1)} = C_- e^{-\frac{3r_h}{2r_c} \left(1-\Sq\right)} \ , \qquad  C_\mathrm{IR}^{(2)} = C_+ e^{-\frac{3r_h}{2r_c} \left(1+\Sq\right)} \ ,
\ee
so that 
\be
 G_\mathrm{reg} = e^{-\frac{3r_h \Sq}{r_c}} \mathcal{R} = - e^{-\frac{3r_h \Sq}{r_c}} \, \frac{\Gamma \left(1- \Sq\right)\, \Gamma \left(\frac{1}{2} \left(1-i \varpi + \Sq\right)\right)^2}{ \Gamma \left(1+ \Sq\right)\, \Gamma \left(\frac{1}{2} \left(1-i \varpi - \Sq\right)\right)^2}
\ee
where we inserted the reflection amplitude from~\eqref{Rampl}. The regular part of the full correlator simplifies to
\begin{align} \label{xiinftyG}
&\widetilde G_\mathrm{reg} = -\frac{81 i \pi  \hat \mu^4}{512 r_c^4 } &\\\nn
&\times \frac{\hat \mu  \left[1+e^{3 \Sq\le(1-\frac{r_h}{r_c}\ri)} \mathcal{R}\right]H_1^{(1)}\left(\frac{3 \hat \mu}{2}\right)+ \left[(\Sq-1) -e^{3 \Sq\le(1-\frac{r_h}{r_c}\ri)} (\Sq+1) \mathcal{R}\right]H_2^{(1)}\left(\frac{3 \hat \mu}{2}\right)}{\hat \mu  \left[1+e^{3 \Sq\le(1-\frac{r_h}{r_c}\ri)} \mathcal{R}\right]J_1\left(\frac{3 \hat \mu}{2}\right)+ \left[(\Sq-1)-e^{3 \Sq\le(1-\frac{r_h}{r_c}\ri)} (\Sq+1) \mathcal{R}\right]J_2\left(\frac{3 \hat \mu}{2}\right)} \ . &
\end{align}
Notice that the branch cut\footnote{There is also another branch cut $\sim \hat \mu^4\log \hat \mu$ due to the nonanalyticity of the Hankel functions, but this cancels against a similar branch cut in $\widetilde G_0$ of~\eqref{G0expr} (where one should take the limit $\Delta \to 0$).} arising from the square root in the definition of $\Sq$ cancels in this expression: it is invariant under $\Sq \mapsto -\Sq$ (which implies $\mathcal{R} \mapsto \mathcal{R}^{-1}$). 

\begin{figure}[!tb]
\begin{center}
\includegraphics[width=0.505\textwidth,trim=0 11mm 0 0]{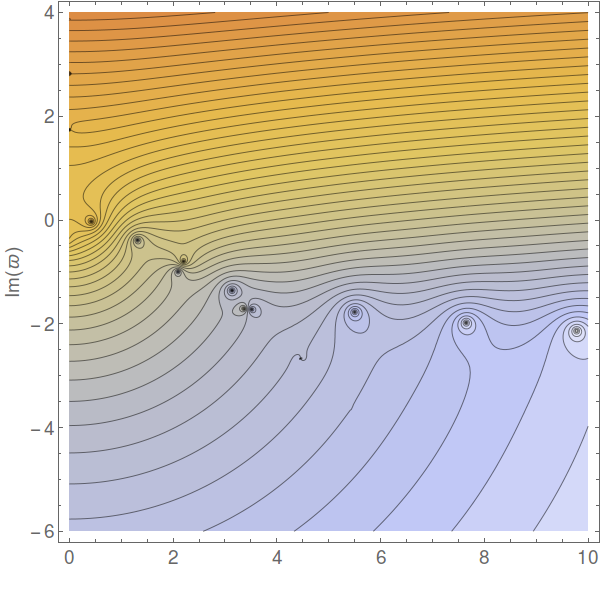}%
\hspace{2mm}\includegraphics[width=0.48\textwidth]{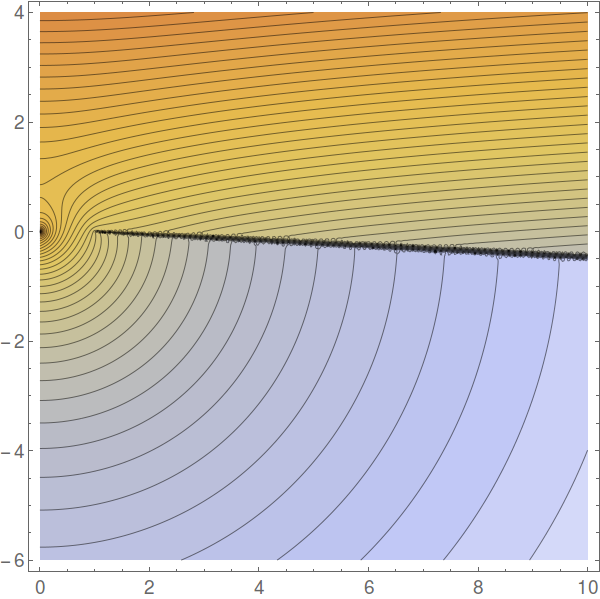}
\includegraphics[width=0.505\textwidth]{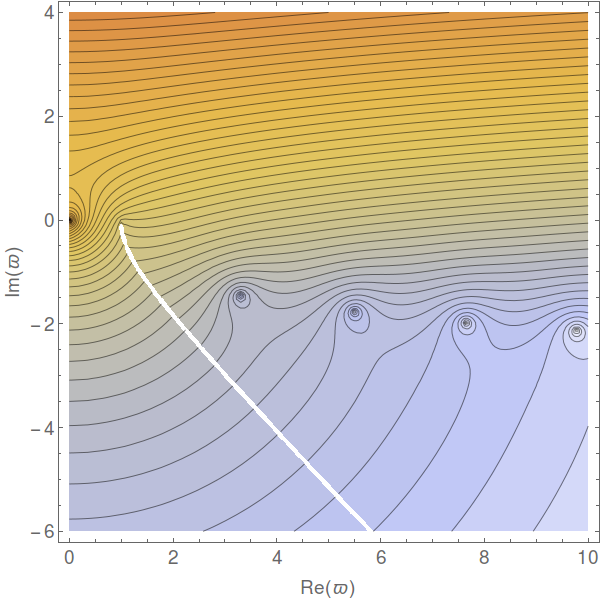}%
\hspace{2mm}\includegraphics[width=0.48\textwidth]{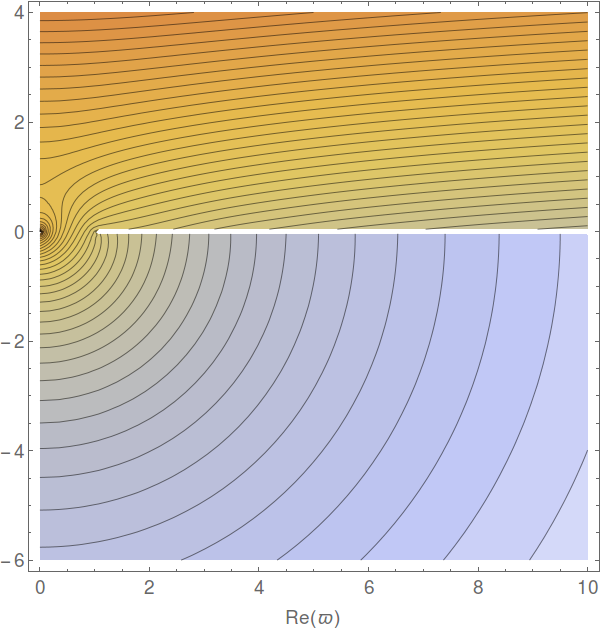}
\end{center}
\caption{The (logarithm of the) absolute value of the correlators of the energy-momentum tensor on the complex $\varpi$-plane in various analytic approximations. The plots are for $X=-1/2$ ($\xi=\infty$) and $q=0$. The contours are at constant values of $|\widetilde G_\mathrm{reg}|$, with orange/yellow colors (mostly top parts of the plots) indicating small values and blue/white colors  (mostly bottom parts of the plots) indicating large values. Top left: the ``glued'' correlator of~\protect\eqref{xiinftyG} at $r_h/r_c = 2$. Top right: the correlator of~\protect\eqref{xiinftyG} at $r_h/r_c = 20$. Bottom left: the large $\varpi$ approximation of the correlator~\protect\eqref{Glargepi} with $r_h/r_c = 2$. Bottom right: the limit of large black hole~\protect\eqref{rhinftylimit} of the glued correlator. }
\label{fig:gluedcorrelator}
\end{figure}

We plot the correlator for $q=0$ in Fig.~\ref{fig:gluedcorrelator} (top row) for $r_h/r_c = 2$ (top left plot) and for $r_h/r_c = 20$ (top right plot). The plotted quantity is the logarithm of the absolute value so both poles and zeros of the correlator appear as singularities in the left hand plot (those singularities which have more whitish or bluish colors than the surroundings are poles). There are two kind of modes: 
\begin{enumerate}
 \item Modes appearing at small $\mathrm{Re}\varpi \lesssim 4$, which have similar structure as the modes of the CR geometry. We will call them ``CR modes''.
  \item Modes for $\mathrm{Re}\varpi \gtrsim 4$, which have larger residues, and form a line with almost negligible slope. We will call them ``AdS modes''.
\end{enumerate}
In the right hand plot the black line is actually a dense set of poles and zeroes which accumulate near the real axis as $r_h$ increases. This suggests that there is a branch cut in the limit $r_h \to\infty$. We will elaborate on these observations below.

\begin{figure}[!tb]
\begin{center}
\includegraphics[width=0.7\textwidth]{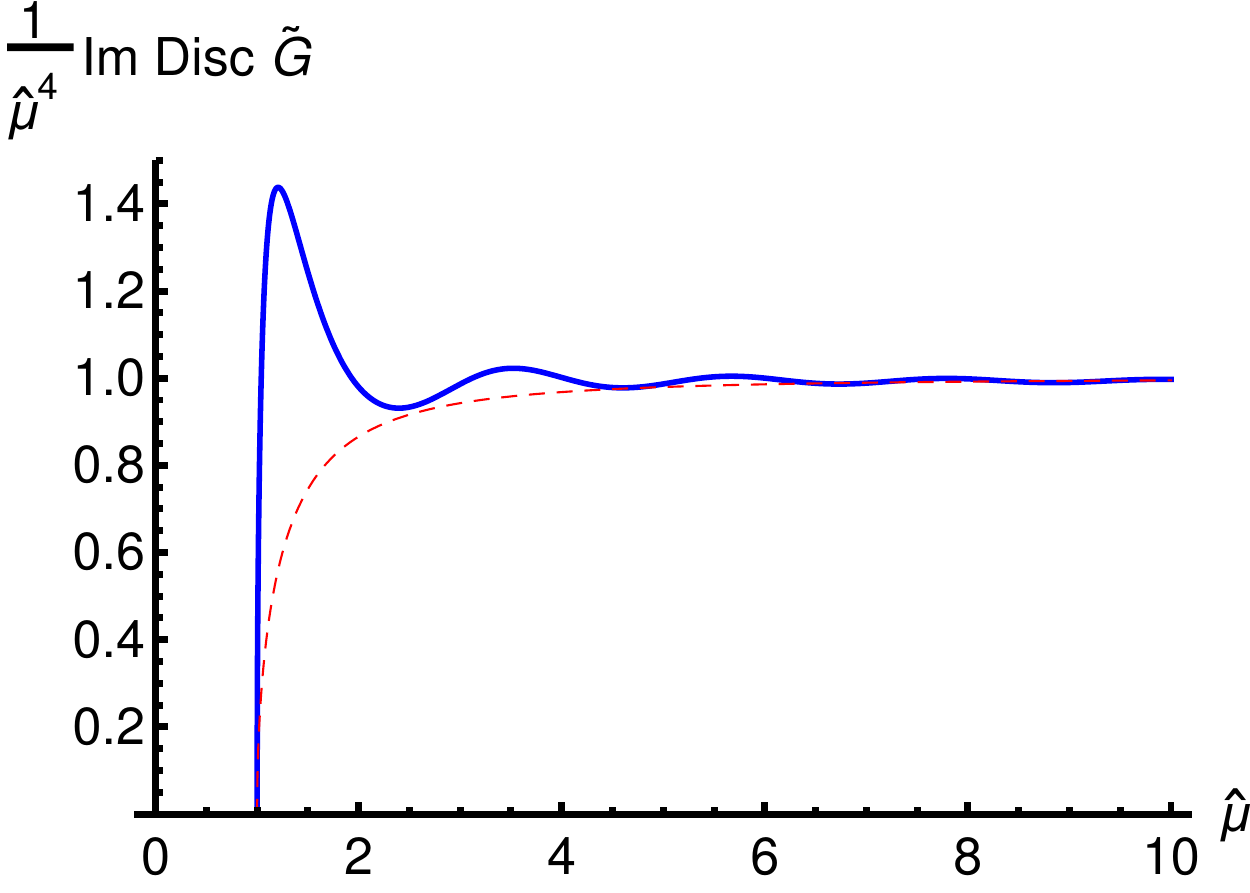}%
\end{center}
\caption{The discontinuity of the correlator, normalized by the factor $\hat \mu^{-4}$, as a function of $\hat \mu$. The result (solid blue curve) is compared to the function $\sqrt{1-\hat \mu^{-2}}$ (thin dashed red curve).}
\label{fig:disc}
\end{figure}

The limit of large black hole $r_h \to \infty$ is indeed interesting. In this limit we find
\be \label{rhinftylimit}
 \widetilde G_\mathrm{reg} \to -\frac{81 i \pi  \hat \mu^4}{512 r_c^4 }  \frac{\hat \mu H_1^{(1)}\left(\frac{3 \hat \mu}{2}\right)+ (\Sq-1) H_2^{(1)}\left(\frac{3 \hat \mu}{2}\right)}{\hat \mu J_1\left(\frac{3 \hat \mu}{2}\right)+ (\Sq-1) J_2\left(\frac{3 \hat \mu}{2}\right)} \ .
\ee
In the limiting expression, the branch cut has become physical, as expected: it results from the accumulation of the poles from the QNMs of finite size black holes. It is placed on the real $\varpi$-axis. The discontinuity is given by
\be
 \mathrm{Disc}\,\widetilde G_\mathrm{reg}= \frac{27 i \hat \mu ^3 \sqrt{\hat \mu ^2-1}}{64 \left(\hat \mu  J_1\left(\frac{3 \hat \mu }{2}\right){}^2-2 J_2\left(\frac{3 \hat \mu }{2}\right) J_1\left(\frac{3 \hat \mu }{2}\right)+\hat \mu  J_2\left(\frac{3 \hat \mu }{2}\right){}^2\right)} \ .
\ee
We plot the discontinuity (divided by the dominant factor $\hat \mu^4$) as a function of $\hat \mu$ in Fig.~\ref{fig:disc}. Interestingly, it is well approximated by $\mathrm{Disc}\,\widetilde G_\mathrm{reg}= i \hat \mu^4$ for all $\hat \mu>1$.

\begin{figure}[!tb]
\begin{center}
\includegraphics[width=0.7\textwidth]{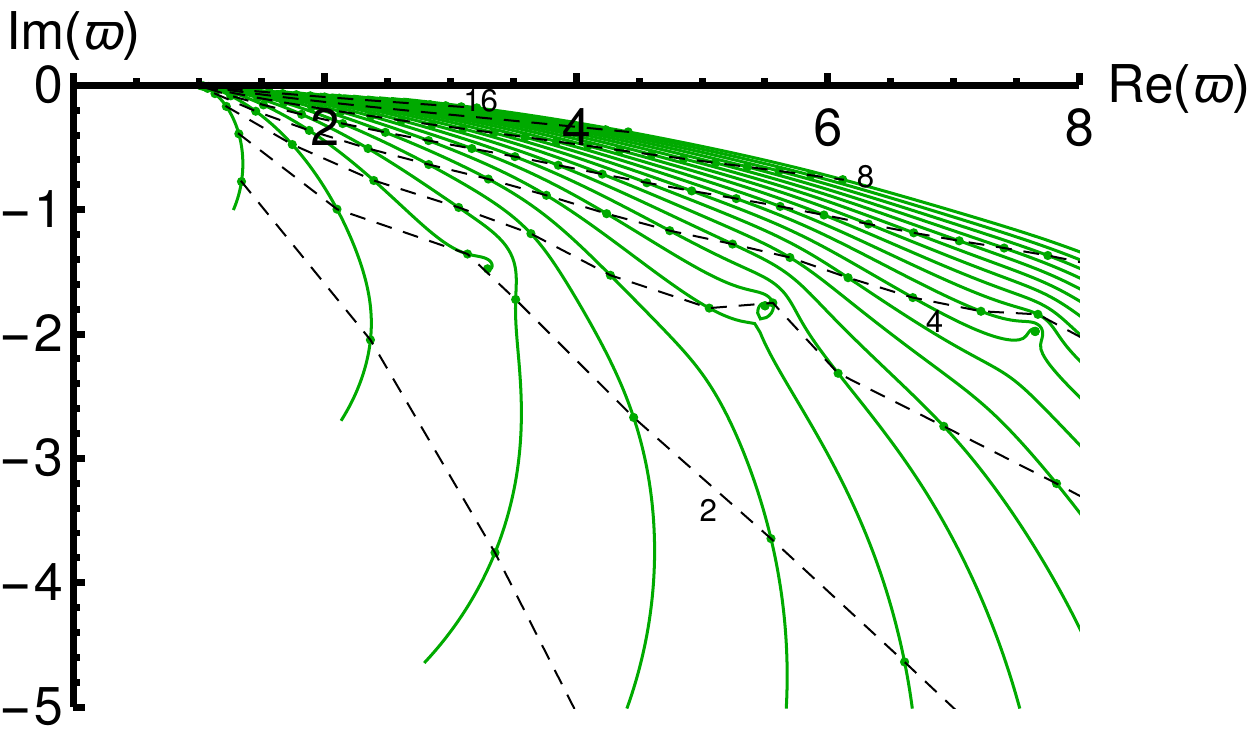}%
\end{center}
\caption{The trajectories of the QNMs on the complex $\varpi$-plane as $r_h$ is varied from $r_h = 1.2 r_c$ (lower end points of the curves) to $r_h=20 r_c$ (upper end points). The dots are the locations of the QNMs for $r_h/r_c = \sqrt{2}$, $2$, $2\sqrt{2}$, \ldots $16$. The dashed lines were added to guide the eye, they connect the locations at the same $r_h$ as indicated by the labels.}
\label{fig:qnmmovxiinfty}
\end{figure}

The behavior of the correlator~\eqref{xiinftyG} changes depending on whether the combination $e^{3 \Sq\le(1-\frac{r_h}{r_c}\ri)} \mathcal{R}$ is smaller or larger than one. It is instructive to study this in the limit of large $\varpi$. Choosing the branch such that $\Sq \simeq i \varpi$, we see that $\mathcal{R} \sim \exp(\pi \varpi)$, and consequently
\be
 e^{3 \Sq\le(1-\frac{r_h}{r_c}\ri)} \mathcal{R} \sim \exp\le[ -3 i \varpi \le(\frac{r_h}{r_c}-1\ri) +\pi\varpi \ri] \ .
\ee
This is either exponentially enhanced or suppressed at large $|\varpi|$ depending on the phase of $\varpi$. The critical line is given by the equation
\be \label{criticalphi}
 \frac{\mathrm{Im}\varpi}{\mathrm{Re}\varpi} = - \frac{\pi}{3\le(\frac{r_h}{r_c}-1\ri)} \equiv \tan \phi_\varpi
\ee
with $-\pi/2<\phi_\varpi<0$. Up to exponentially suppressed corrections, the large-$\varpi$ approximation for the correlator therefore amounts to setting the reflection amplitude either $0$ or $\infty$ depending on the phase of $\varpi$. We find that~\eqref{xiinftyG} approaches Eq.~\eqref{rhinftylimit} but with a modified branch choice for $\Sq$, so that the change in the branch is determined by~\eqref{criticalphi}. Using the standard branch for the square root, the result may be written as
\be \label{Glargepi}
 \widetilde G_\mathrm{reg} \simeq -\frac{81 i \pi  \hat \mu^4}{512 r_c^4 }  \frac{\hat \mu H_1^{(1)}\left(\frac{3 \hat \mu}{2}\right)+ \le[e^{i\phi_\varpi}\sqrt{e^{-2i\phi_\varpi}\le(\hat \mu ^2-1\ri)}-1\ri] H_2^{(1)}\left(\frac{3 \hat \mu}{2}\right)}{\hat \mu J_1\left(\frac{3 \hat \mu}{2}\right)+ \le[e^{i\phi_\varpi}\sqrt{e^{-2i\phi_\varpi}\le(\hat \mu ^2-1\ri)}-1\ri] J_2\left(\frac{3 \hat \mu}{2}\right)} 
\ee
as $|\varpi| \to \infty$.
Notice that as $r_h \to \infty$ it follows that $\phi_\varpi \to 0$ and the large-$\varpi$ result approaches~\eqref{rhinftylimit}.

We compare the limiting expressions in the limits of large $\varpi$ and large $r_h$ numerically to the full glued correlator~\eqref{xiinftyG} in Fig.~\ref{fig:gluedcorrelator}. The left and right plots on the bottom row are given by~\eqref{Glargepi} and~\eqref{rhinftylimit}, respectively. For the left hand plots (which both have $r_h =2 r_c$) we see that the line of the CR modes 
in the top left plot has been replaced by a branch cut in the bottom left plot. The AdS modes are precisely reproduced by the limiting expression. The accumulation of the modes to the branch cut in the limit of large $r_h$ is clear from the right hand plots (top right plot has $r_h=20 r_c$, and bottom right plots $r_h =\infty$).   

Interpreting the branch cut in~\eqref{Glargepi} as the line of the CR modes, and noting that the AdS modes appear at roughly constant $\mathrm{Im} \,\varpi \sim 1$, we can estimate where the lines of the two sets of modes meet. Using~\eqref{criticalphi} we find that this happens roughly at $\mathrm{Re}\,\varpi \sim r_h/r_c$. 

Finally we study the dependence of the locations of the QNMs on the size of the horizon $r_h$. The trajectories of the modes as $r_h$ is varied are shown in Fig.~\ref{fig:qnmmovxiinfty}. The results should be compared to those at finite $\xi \simeq 26.77$ in Fig.~\ref{fig:QNMevolution} (left) where the variation of $r_h$ also changed the temperature. Notice that Fig.~\ref{fig:QNMevolution} shows a smaller region of the complex $\varpi$-plane. 
The trajectories at small $r_h$ (higher temperatures) are strikingly similar. The main difference between the two plots is seen at high $r_h$: at finite $\xi$ the evolution of the QNMs stops at fixed location (given by the QNMs of the CR geometry) whereas at infinite $\xi$ each mode approaches $\varpi = 1$ in the limit $r_h \to \infty$. The evolution also has similarities with that observed in the case of global AdS~\cite{Jokela:2015sza} where the QNMs approach the real line as $T \to 0$ but the spectrum remains discrete.

An interesting feature in Fig.~\ref{fig:qnmmovxiinfty} is that some of the modes stop at finite $\mathrm{Im}\,\varpi$ as $r_h$ decreases while other evolve smoothly towards smaller values of  $\mathrm{Im}\,\varpi$. By comparing to Fig.~\ref{fig:gluedcorrelator} we see that (after they have stopped) these former modes can be identified as the AdS modes, whereas the evolving modes are the CR modes. 

\section{Discussion} \label{sec::conclusion}

We studied the quasi normal modes of a strongly interacting non-conformal plasma, based on prototype holographic theories given by Chamblin-Reall blackhole solutions. The solutions are parametrized by a non-conformality parameter $\xi$ that ranges from $\xi=4$ (conformal plasma) to $\xi=\infty$ (critical plasma). 

One of our main goals was to determine the approach to criticality by studying fluctuations in a $1/\xi$ expansion. We showed that the fluctuations in this limit are controlled by a linear dilaton blackhole. Furthermore they can be obtained analytically directly by solving the fluctuation equations in the limit $\xi\to\infty$. In particular we showed that the ratio of the incoming and outgoing waves at the horizon in this critical limit matches precisely the reflection amplitude on the linear dilaton blackhole which can be obtained by Liouville CFT - WZW model techniques. This provides a non-trivial check on our calculations. 

More importantly this connection suggests that the plasma in the critical limit is governed by such an exact CFT. This can be taken further by conjecturing that the entire critical phenomena around the strongly interacting continuous transition is governed by the Liouville CFT - WZW model, suggesting for example that the associated critical exponents in the plasma can be fixed by exact CFT techniques. We plan to investigate this connection, and its implications for critical phenomena in strongly interacting condensed matter theories, e.g. spin models \cite{Gursoy:2010kw} in future work.  

Here, we observe an interesting connection of the critical limit to the large D limit recently considered by Emparan and collaborators \cite{Emparan:2013moa,Emparan:2015rva}. In fact the 
exponential potential for the dilaton can be obtained as a generalized dimensional reduction from a theory of Einstein gravity with cosmological constant compactified on 
a torus \cite{Kanitscheider:2009as}. The parameter appearing in the dilaton potential is related to the number of extra dimensions, which is allowed to be continuous. In this description, the limit of large $\xi$ corresponds to the number of extra dimensions going to infinity. That this limit is analytically tractable appears to be related to the large-D limit studied in \cite{Emparan:2013moa,Emparan:2015rva}. A fundamental insight in the work of \cite{Emparan:2013moa,Emparan:2015rva} is that the physics of QNM is related to the emergence of different scales: the horizon radius $r_h$ and $r_h/D$, 
that become parametrically separated in the large D limit. 
In particular  \cite{Emparan:2015rva}  noticed that in the large D limit there is a set of quasi-normal modes that are localized in a near-horizon region of size $r_h/D$. They called them the decoupled sector, and computed the frequencies up to fourth order in $1/D$. Furthermore, in \cite{Emparan:2013xia} they showed that the quasi normal modes localized near the horizon correspond to the same fluctuations in the linear dilaton blackhole, because as shown in \cite{Soda:1993xc,Grumiller:2002nm}, the space-time action of a D dimensional blackhole geometry dimensionally reduced on a D-2 dimensional sphere becomes the 2D string action with the linear dilaton blackhole as the corresponding solution. 

In section \ref{sec:Xonehalf} we observed very similar phenomena. The limit $\xi\to\infty$ focuses in a strip of the geometry of size $1/\xi$ near the horizon where one can think of the modes near the horizon as analogs of the decoupled modes of  \cite{Emparan:2015rva}. We also showed that the physics of these decoupled modes near the horizon is determined by the linear dilaton blackhole which whose properties are in turn determined by the exact CFT data. These decoupled modes indeed exist in the 
shear and sound channels, but not in the tensor channel that we consider in our paper.  Indeed we saw explicitly that our QNM  are not 
decoupled but depend on the full geometry, and in fact many of the important features result from the matching between the near-horizon and the asymptotic region.  Therefore our results are complementary to those in \cite{Emparan:2015rva}. One difference between the large D and large $\xi$ limits is that, whereas in the uplifted description there is no sense in going beyond $D = \infty$, in the aforementioned dimensional reduction the corresponding critical value $\xi=\infty$  is not a limiting value. One can easily see this is by noting that the critical limit in parameter $X$ corresponds to the value $X=-1/2^+$ and one is free to consider blackhole solutions with $X < -1/2$. Whereas the range $0>X> -1/2$ corresponds to $4<\xi<\infty$, the range $-1<X<-1/2$ corresponds to $0>\xi>-\infty$. 
The corresponding black hole solutions with $X<-1/2$ do not have good UV asymptotics however, hence the UV completion becomes indispensable in this case. We also  leave treatment of this case to future work. 

We have observed coalescence of infinitely many non-hydro modes on the real axis in the critical limit $\xi\to\infty$ dominating over the hydro mode whose imaginary part always stays finite at fixed momentum. This strongly invalidates the applicability of hydrodynamics below the momentum range $\bar{q} \lesssim 1/\sqrt{\xi}$. This is indeed what one would expect from a system near a strongly interacting fixed point. As the correlation length diverges near a continuous phase transition, all modes except fluctuations of the order parameter become irrelevant whereby  dynamics reorganize itself in a non-trivial fashion. However, we know in the end that systems at criticality should be amenable to a hydrodynamic description precisely given by the gapless effective  theory comprised of the fluctuations of the order parameter. It is tempting to conjecture that breakdown of hydro in  our system is due to omission of these fluctuations. Indeed, as discussed in detail in \cite{Gursoy:2010kw} the order parameter in these type of holographic theories is the Polyakov loop which acquires a non-trivial expectation value above $T_c$. Polyakov loop is dual to a string that winds around the Euclidean time circle and its excitations are related to string excitations some of which, e.g. the winding tachyon, become massless at criticality~\cite{Atick:1988si}. These string degrees of freedom should presumably be included to reestablish a hydrodynamic description. Let us also mention in passing, that, manifestation of this continuous phase transition in Lorentzian time is again appearance of massless poles in the two-point function of Wilson loops, holographically dual to the fluctuations of a string stretching between two space-like separated end points on the boundary with its tip close to horizon \cite{Gursoy:2010kw}. 

It might also be of interest to try to connect the findings of the present paper, with the known hydrodynamic description of the linear dilaton fluctuations in terms of density perturbations of an incompressible fermi sea composed out of the elementary fermions that describe the physics of 2d string theory\footnote{This description holds for two dimensions, but one can imagine that it is still valid in the case of no momentum dependence in the extra transverse directions, or that one can extend it appropriately in a BMN-like limit akin to what happens in the $1/2$ BPS geometries of~\cite{Lin:2004nb}.}. 

This also motivates a more thorough study of the role of symmetries near criticality, since the present model might offer a concrete universal example for strongly coupled systems developing an infinite number of conserved currents.

Finally we should also mention that the regulating procedure we adopt gluing an AdS spacetime, will provide an explicit breaking of such symmetries that now hold only in the extreme IR. Ours and similar~\cite{Anninos:2017hhn} centaur-like geometries might then have holographic duals with similar properties to those of the recently studied SYK model~\cite{Maldacena:2016hyu}, where extra symmetries arise in the IR limit and are broken both explicitly and spontaneously by UV effects.

\acknowledgments

We would like to thank D. Anninos, R. Janik, N. Jokela, E. Kiritsis, D. Mateos, L. Silva Pimenta, and D.T. Son for discussions and helpful suggestions. This work is partially supported by the Netherlands Organisation for Scientific Research (NWO) under the VIDI grant 680-47-518, the Delta-Institute for Theoretical Physics (D-ITP) that is funded by the Dutch Ministry of Education, Culture and Science (OCW) and the Scientific and Technological Research Council of Turkey (TUBITAK). PB is supported by the Advanced ERC grant SM-grav, No 669288. UG is grateful for the hospitality of the Bo\~gazi\c ci University and the Mimar Sinan University in Istanbul.

\appendix

\section{WKB analysis of the transverse fluctuations} \label{WKB}

We study the fluctuation of the transverse spin-2 modes in the WKB approximation, following the analysis of \cite{Fuini:2016qsc}. 
We start from the equation in the Schr\"odinger form \eqref{eqSchr}. 
The fluctuation has been redefined and one can check that the boundary conditions are the following: 
\begin{eqnarray} 
h(w \sim 0) & \sim& C_1 \left(1+ {\cal O}(w^{1/\xi})\right) + C_2 w \left(1+ {\cal O}(w^{1/\xi})\right) \,, \nonumber\\
h(w \sim 1) &\sim& (1-w)^{{1\over 2} - i {\varpi \over 2}} \,.
\end{eqnarray}  
We introduce $s = \varpi/q$, and we anticipate that in the large $q$ limit the leading behavior is $s = 1 + s_\alpha q^{- \alpha}$, for some value of $\alpha$. 
We can write the potential as 
\bea
V & = & q^2 Q_0 + q^{2-\alpha} Q_\alpha + Q_2 + \ldots  \\
Q_0 & = &  - {w^{{2\over \xi}-1} \over 4 (1-w)^2} \,,  \nonumber\\
Q_\alpha & = &  - {s_\alpha \over 2}  {w^{{2\over \xi}-2} \over  (1-w)^2} \,,  \nonumber\\
Q_2 &= & - {1 \over 4 (1-w)^2} \,,  \nonumber
\eea
and we assume for the moment that $0<\alpha<2$. 
We see that the term $Q_\alpha$ is subleading in $q$ but it dominates close to the boundary. We distinguish then a boundary region and an intermediate region: 
\bea 
\textrm{boundary: } \quad w \ll q^{-\alpha}\,, \quad h = w^{1\over 2} J_{\pm \xi/2} \left(\xi \sqrt{s_\alpha \over 2} q^{1-{\alpha \over 2}} w^{1\over \xi} \right) \,, \\
\textrm{intermediate: } \quad  q^{-\alpha} \ll w \ll 1 \,,\quad h = w^{1\over 2} J_{\pm {\xi \over \xi+2}} \left({\xi \over \xi+2} q w^{\xi+2 \over 2 \xi} \right) \,.
\eea
The WKB approximation starts with an Ansatz of the form 
\be
h = A \, e^{(q T_0 + q^{1-\alpha} T_\alpha + \ldots)} \,,
\ee
Plugging in the equation and expanding in $q$ we find that $A = Q_0^{-1/4}$, and $T_0, T_\alpha$ can be found explicitly in terms of hypergeometric functions: 
\bea 
T_0 &= i {\xi \over \xi+2} w^{\xi+2 \over 2 \xi} {}_2 F_1 (1, {2+\xi \over 2 \xi}, {2+3\xi \over 2\xi}, w) \,, \\
T_\alpha & = -i s_\alpha {\xi \over \xi-2} w^{-\half + {1 \over \xi}} {}_2 F_1 (1, {\xi-2 \over 2 \xi}, {\xi+2 \over 2\xi}, w) \,.
\eea
For small $w$, the WKB expansion reduces to 
\be 
h \sim w^{\xi-2 \over 4 \xi} \, \textrm{exp} \left( {i \xi \over \xi+2} q  w^{\xi+2 \over 2 \xi} + {i s \over \xi-2}  q^{1-\alpha} w^{2-\xi \over 2 \xi} \right) \,.
\ee
The change of variable $y = w q^{2 \xi \over \xi+2}$ applied to the near-boundary equation and the WKB solution shows that the solution is consistent if 
\be 
\alpha = {2 \xi \over \xi +2 } \,. 
\ee
We recover $\alpha = 4/3$ in the conformal case, and $\alpha<2$ as we assumed in the beginning. 
As in the conformal plasma, we see that the large-$q$ modes are long lived, and even more so than in the conformal case. 
Rotating $y$ in the complex plane, we can also find that the phase of $s_\alpha$ is
\be
s_\alpha = |s_\alpha| e^{-i {2\pi \over \xi+ 2}}
\ee  
These results have also been confirmed by numerical calculations, as shown in Fig. \ref{fig:largeq}. In this plot, the values of $|s_\alpha|$ were determined through the normalizable modes to the near-boundary Schrodinger equation, which after the above change of variables reads
\be
 h''(y)= \frac{1}{4}y^{2/\xi-2}\left(y-2 |s_\alpha|\right)h(y) \ .
\ee

\section{Near-boundary behavior and matching with the linear dilaton}\label{sec:bdry}

We  analyze here the boundary behavior of the fluctuations for generic $\xi$.
If terms including the factor $\hat r^\xi$ are dropped, all fluctuation equations take the (zero temperature) form
\be
- \left(\ell'^2 k^2 \hat r + (\xi-1) \ell' i \omega \right) \flf(\hat r)+\left(2 i \ell' \hat r  \omega +1 -\xi\right)\flf'(\hat r) +\hat r \flf''(\hat r) \simeq 0\ ,
\ee
which is related to the Bessel equation. Therefore the fluctuations have the following behavior near the boundary:
\begin{align} \label{besselfs}
 \flf(\hat r) &= C_1 \,  \Gamma \left(1-\frac{\xi}{2}\right) \left(\frac{\ell'^2 \left(\omega ^2-k^2\right)}{4}\right)^\frac{\xi}{4}e^{-i \ell' \omega \hat r } \hat r^{\xi/2} J_{-\xi/2}\left(\ell'\sqrt{\omega ^2-k^2} \hat r\right)\left[1+\mathcal{O}\left({\hat r^{\xi+1}}\right)\right] &\nn\\
  & \phantom{=}+ C_2\,  \Gamma \left(1+\frac{\xi}{2}\right) \left(\frac{\ell'^2 \left(\omega ^2-k^2\right)}{4}\right)^{-\frac{\xi}{4}}e^{-i \ell' \omega \hat r } \hat r^{\xi/2} J_{\xi/2}\left(\ell'\sqrt{\omega ^2-k^2} \hat r\right)\left[1+\mathcal{O}\left({\hat r^{\xi}}\right)\right] &\nn\\
  &\equiv C_1 \flf^{(1)}(\hat r) +C_2 \flf^{(2)}(\hat r)
\end{align}
Omitted terms are $\mathcal{O}((\hat r/\hat r_h)^\xi)$ or equivalently $\mathcal{O}(w)$. We choose the branch of non-integer powers such that they are real for $\omega>|k|$.
The normalization was chosen such that the boundary expansion has the standard form:
\be \label{UVexp}
 \flf(\hat r) = C_1\left[1+\mathcal{O}\left({\hat r}\right)\right] + C_2 \hat r^\xi \left[1+\mathcal{O}\left({\hat r}\right)\right] \ .
\ee
The expansion of $\flf^{(1)}$ does not contain a term proportional to $\hat r^\xi$ for generic $\xi$. Dimensional reduction~\cite{Gouteraux:2011qh} suggests that the VEV should therefore be identified with the coefficient $C_2$, as was demonstrated in~\cite{Gursoy:2015nza}. 

There is, however, an apparent problem with the definitions~\eqref{besselfs}. Namely, whenever $\xi$ is an even integer, the Bessel functions $J_{\pm\xi/2}(z)$ are proportional, and do not form a proper basis for the solutions near the boundary. Both functions vanish as $J_{\pm\xi/2}(z) \sim z^{\xi/2}$ as $z \to 0$. Forcing the source function $\flf^{(1)}$ in~\eqref{besselfs} to approach a constant in the UV results therefore in its normalization factor being divergent. In the Taylor series around $r = 0$, the leading divergent term of $\flf^{(1)}$ is $\propto \hat r^\xi$.

Notably the singularities cancel in a well chosen linear combination of the two functions:
\bea \label{sourceredef}
 \flf_\mathrm{reg}(\hat r) &\equiv& \flf^{(1)}(\hat r) - e^{-i \pi \xi/2}\frac{\Gamma \left(1-\frac{\xi}{2}\right)}{\Gamma \left(1+\frac{\xi}{2}\right)} \left(\frac{\ell'^2 \left(\omega ^2-k^2\right)}{4}\right)^\frac{\xi}{2} \flf^{(2)}(\hat r) 
\eea
where we could have also chosen the opposite sign in the phase factor $e^{-i \pi \xi/2}$ (or a suitable linear combination of the phase factors with opposite signs). Therefore $\flf^{(2)}$ and $\flf_\mathrm{reg}$ form a proper basis of the solutions for any value of $\xi$.

In order to understand what this cancellation means, it is useful to compare these functions to the vacuum solutions of a massless scalar in AdS$_{\xi+1}$. First, we denote the Euclidean (rescaled) momentum as $p^2=\ell'^2(q^2-\varpi^2)$. Then the basis functions take a simple form in terms of the modified Bessel functions $I$ and $K$:
\begin{align} \label{Iexpr}
 \flf^{(2)}(\hat r) &= \left(\frac{2\hat r}{p}\right)^{\xi/2} \Gamma\left(1+\frac{\xi}{2}\right)\,e^{-i \ell' \omega \hat r }\,I_{\xi/2}\left(p \hat r\right) & \\
 \flf_\mathrm{reg}(\hat r) &= \frac{ 2 \left(\frac{p\hat r}{2}\right)^{\xi/2}}{ \Gamma\left(\frac{\xi}{2}\right)}\,e^{-i \ell' \omega \hat r }\,K_{\xi/2}\left(p \hat r\right) & 
 \label{Kexpr}
\end{align}
up to corrections suppressed by $\hat r^\xi$. These functions are recognized as the standard expressions for vacuum fluctuations of a massless scalar in AdS$_{\xi+1}$ up to the factor $e^{-i \ell' \omega \hat r }$ which arises due to working in the Eddington-Finkelstein coordinates. In particular we notice that the Bessel function $K$, with all phase factors canceling in~\eqref{Kexpr}, is only obtained with the current choice of the sign in the phase factor of~\eqref{sourceredef}. As correlators are expected to be analytic in the upper complex $\omega$-plane, this choice is natural as (with real $p$ mapping to positive imaginary $\omega$) it leads to the basis function $\flf_\mathrm{reg}$ admitting natural analytic extension to the whole  upper $\omega$-plane.

Notice that the coefficient multiplying $\flf^{(2)}$ in~\eqref{sourceredef} is the coefficient of the $r^\xi$-term for generic $\xi$ in the regulated function $\flf_\mathrm{reg}$, and therefore gives the vacuum correlator for a massless scalar in AdS$_{\xi+1}$. This is also seen from~\eqref{Iexpr} and~\eqref{Kexpr} as $K$ is the IR-regular solution. To be precise, this is only true when $\xi/2$ is noninteger. If $\xi/2 = m$ with $m$ an integer, the terms $\propto \hat r^\xi$ and $\propto \hat r^{2m}$ in the expansion of $K$ both contribute and partially cancel, so that the divergences of the Gamma function $\Gamma \left(1-\frac{\xi}{2}\right)$ are regulated.

An IR-regular solution is in general written as
\begin{align} \label{Hdecomp}
 \flf(\hat r) &= C_1 \flf^{(1)}(\hat r) + C_2 \flf^{(2)}(\hat r) & \nn\\
 &= C_1 \flf_\mathrm{reg}(\hat r) + \left(C_2 - G_s C_1\right) \flf^{(2)}(\hat r) \equiv C_1 \flf_\mathrm{reg}(\hat r) + C_\mathrm{reg} \flf^{(2)}(\hat r)
\end{align}
where
\be
 G_s(\omega,k) = -\frac{\Gamma \left(1-\frac{\xi}{2}\right)}{\Gamma \left(1+\frac{\xi}{2}\right)} \left(\frac{\ell'^2 \left(k^2-\omega ^2\right)}{4}\right)^\frac{\xi}{2}
\ee
is the vacuum scalar correlator for generic $\xi$. The full correlator is then
\be \label{Gdef}
 G(\omega,k) = \frac{C_2(\omega,k)}{C_1(\omega,k)} = \frac{C_\mathrm{reg}(\omega,k)}{C_1(\omega,k)} + G_s(\omega,k) \equiv  G_\mathrm{reg}(\omega,k) + G_s(\omega,k)\ .
\ee
Notice that here $G_\mathrm{reg}$ is defined in terms of the well-behaved basis functions and is therefore regular for any value of $\xi$. That is, we have isolated the divergences of the correlator in the trivial piece $G_s$, whereas all nontrivial temperature dependent effects (in particular the poles due to the quasi normal modes) remain in $G_\mathrm{reg}$. When $\xi/2$ is an integer only the latter decomposition in~\eqref{Hdecomp} is well defined: in this case the full correlator is automatically regulated in exactly the same way as the vacuum scalar correlator, i.e, thanks to the cancellation of two terms in the boundary expansion of $\flf_\mathrm{reg}$.

As a final remark, we notice that the origin of the divergences in $G_s$ can be seen to be the Fourier transform (see, e.g.,~\cite{Bzowski:2013sza}): 
\be
 \int \frac{d\omega d^{\xi-1} k}{(2\pi)^\xi}\, G_s(\omega,k)\, e^{-i\omega x_0 +i\vec k\cdot \vec x} = \frac{1}{\pi^{\frac{\xi}{2}}\left(\vec x^2-x_0^2\right)^{\xi}} \frac{ \Gamma (\xi)}{\Gamma \left(\frac{\xi}{2}\right)} \ .
\ee
In the coordinate space correlator the singularities are absent for all positive $\xi$.

The near-boundary solutions~\eqref{besselfs} then need to be matched with the large-$\xi$ solutions in the bulk~\eqref{x12sol} in order to obtain the complete solution in the limit of large $\xi$. 

We notice that terms $\mathcal{O}(1/\xi)$ we omitted and we approximated $w^{1/\xi}\simeq 1$ in Sec.~\ref{sec:xiinfty}. That is, the expression~\eqref{x12sol} is valid for $\xi \gg 1$ and $e^{-\xi}\ll  w$. In 
the above boundary analysis
we dropped terms $\mathcal{O}(w)$, so the expressions in~\eqref{besselfs} are valid for $w\ll 1$ for any value of $\xi$. Therefore both~\eqref{besselfs} and~\eqref{x12sol} are valid when $\xi \gg 1$ and $e^{-\xi}\ll  w\ll 1$ (at fixed $q$ and $\varpi$). Consequently, we can match the two solutions in this intermediate region, and as their combination obtain a solution which holds for any (potentially extremely small) value of $w$, and has corrections suppressed at large $\xi$.

The solutions~\eqref{x12sol} for $\xi \gg 1$ and $e^{-\xi}\ll  w\ll 1$ have already been computed in~\eqref{x12smallw}. What remains to be done is the analysis of the expressions~\eqref{besselfs} and~\eqref{sourceredef} in this limit, i.e., compute their behavior at large $\xi$ with fixed $w$, $\varpi$, and $q$. This is done by applying saddle point approximation to an integral representation for the Bessel functions in Appendix~\ref{app:spbessel}.
Inserting these in~\eqref{besselfs} and~\eqref{sourceredef} we find
\bea
 \flf_\mathrm{reg}(w) &\simeq&  
 \frac{2 \sqrt{\pi}}{\sqrt{\xi \Sq}\, \Gamma\left(\frac{\xi}{2}\right)}  \left(\frac{\xi^2 \left(\varpi ^2-q^2\right)}{16}\right)^\frac{\xi}{4} \Bigg[  e^{\frac{1}{2} \xi  \left(-\Sq-i \varpi\right)} \left(\frac{1+\Sq}{1-\Sq}\right)^\frac{\xi}{4} w^{\frac{1}{2} \left(1-i \varpi-\Sq\right)}\nn\\
 &&\ + i\, \theta\left(-\mathrm{Im}\varpi\right)\ e^{\frac{1}{2} \xi  \left(\Sq-i \varpi\right)} \left(\frac{1-\Sq}{1+\Sq}\right)^\frac{\xi}{4} w^{\frac{1}{2} \left(1-i \varpi+\Sq\right)} \Bigg]\\ 
 \flf^{(2)}(w) &\simeq& \frac{\Gamma\left(1+\frac{\xi}{2}\right)\hat r_h^\xi}{\sqrt{\pi  \xi \Sq} }\left(\frac{\xi^2 \left(\varpi ^2-q^2\right)}{16}\right)^{-\frac{\xi}{4}} e^{\frac{1}{2} \xi  \left(\Sq-i \varpi\right)} \left(\frac{1-\Sq}{1+\Sq}\right)^\frac{\xi}{4} w^{\frac{1}{2} \left(1-i \varpi+\Sq\right)} \phantom{aaaaa}
\eea
for $\re \varpi \lesssim \sqrt{1+q^2}$ up to $\mathcal{O}(1/\xi)$ corrections. 
For $\re \varpi \gtrsim \sqrt{1+q^2}$ we find similarly
\bea
  \flf_\mathrm{reg}(w) &\simeq&  \frac{2 e^{\frac{i \pi }{4}} \sqrt{\pi}}{\sqrt{\xi \Sqt} \, \Gamma\left(\frac{\xi}{2}\right)}\left(\frac{\xi^2 \left(\varpi ^2-q^2\right)}{16}\right)^\frac{\xi}{4} e^{\frac{1}{2} \xi  \left(i \Sqt-i \varpi\right)} \left(\frac{1-i \Sqt}{1+i \Sqt}\right)^\frac{\xi}{4} w^{\frac{1}{2} \left(1-i \varpi+i \Sqt\right)}\nn\\
 \flf^{(2)}(w) &\simeq&
 \frac{\Gamma\left(1+\frac{\xi}{2}\right)\hat r_h^\xi}{\sqrt{\pi  \xi \Sqt}}\left(\frac{\xi^2 \left(\varpi ^2-q^2\right)}{16}\right)^{-\frac{\xi}{4}}\Bigg[e^{-\frac{i \pi}{4}  } e^{\frac{1}{2} \xi  \left(i \Sqt-i \varpi\right)} \left(\frac{1-i \Sqt}{1+i \Sqt}\right)^\frac{\xi}{4} w^{\frac{1}{2} \left(1-i \varpi+i \Sqt\right)} \nn\\
 &&\qquad \qquad+ e^{\frac{i \pi }{4}} e^{\frac{1}{2} \xi  \left(-i \Sqt-i \varpi\right)} \left(\frac{1+i \Sqt}{1-i \Sqt}\right)^\frac{\xi}{4} w^{\frac{1}{2} \left(1-i \varpi-i \Sqt\right)}\Bigg] \ .
\eea

Matching the above expressions with~\eqref{x12smallw} and using the definitions of the coefficients from~\eqref{Hdecomp} leads to
\bea
 C_- &=& \frac{2 \sqrt{\pi}}{\sqrt{\xi \Sq}\, \Gamma\left(\frac{\xi}{2}\right)}  \left(\frac{\xi^2 \left(\varpi ^2-q^2\right)}{16}\right)^\frac{\xi}{4}  e^{\frac{1}{2} \xi  \left(-\Sq-i \varpi\right)} \left(\frac{1+\Sq}{1-\Sq}\right)^\frac{\xi}{4} C_1 \\
C_+ &=& \frac{\Gamma\left(1+\frac{\xi}{2}\right)\hat r_h^\xi}{\sqrt{\pi  \xi \Sq} }\left(\frac{\xi^2 \left(\varpi ^2-q^2\right)}{16}\right)^{-\frac{\xi}{4}} e^{\frac{1}{2} \xi  \left(\Sq-i \varpi\right)} \left(\frac{1-\Sq}{1+\Sq}\right)^\frac{\xi}{4} C_\mathrm{reg} \nn\\
&& + \frac{2 i\, \theta\left(-\mathrm{Im}\varpi\right) \sqrt{\pi}}{\sqrt{\xi \Sq}\, \Gamma\left(\frac{\xi}{2}\right)}  \left(\frac{\xi^2 \left(\varpi ^2-q^2\right)}{16}\right)^\frac{\xi}{4}  e^{\frac{1}{2} \xi  \left(\Sq-i \varpi\right)} \left(\frac{1-\Sq}{1+\Sq}\right)^\frac{\xi}{4} C_1
 \eea 
at small $\re \varpi$. 

At large $\re\varpi$ (with $\Sq \mapsto -i\Sqt$ in~\eqref{x12smallw}) we obtain
\bea
 C_- &=& \frac{1}{\sqrt{\pi \xi \Sqt}}e^{\frac{1}{2} \xi  \left(i \Sqt-i \varpi\right)} \left(\frac{1-i \Sqt}{1+i \Sqt}\right)^\frac{\xi}{4} \Bigg[\frac{2\pi e^{\frac{i \pi }{4}}}{\Gamma\left(\frac{\xi}{2}\right)}\left(\frac{\xi^2 \left(\varpi ^2-q^2\right)}{16}\right)^\frac{\xi}{4} C_1\nn \\
 &&+ e^{-\frac{i \pi }{4}}\Gamma\left(1+\frac{\xi}{2}\right) \hat r_h^\xi\left(\frac{\xi^2 \left(\varpi ^2-q^2\right)}{16}\right)^{-\frac{\xi}{4}}C_\mathrm{reg}\Bigg]\\
 C_+ &=&\frac{\Gamma\left(1+\frac{\xi}{2}\right)\hat r_h^\xi}{\sqrt{\pi  \xi \Sqt}}\left(\frac{\xi^2 \left(\varpi ^2-q^2\right)}{16}\right)^{-\frac{\xi}{4}}e^{\frac{i \pi}{4}  } e^{\frac{1}{2} \xi  \left(-i \Sqt-i \varpi\right)} \left(\frac{1+i \Sqt}{1-i \Sqt}\right)^\frac{\xi}{4} C_\mathrm{reg} \ .
\eea

\section{Saddle point approximation of the Bessel functions} \label{app:spbessel}

The Bessel functions for generic arguments can be defined by the integral 
\be
 J_\nu(z) = \frac{1}{2\pi i} \left(\frac{z}{2}\right)^\nu\int_\mathcal{L} dt\, e^{t-\frac{z^2}{4t}} t^{-\nu-1}
\ee
where the integration contour $\mathcal{L}$ starts from $t=-\infty$, circles the origin in counterclockwise direction, and returns to $t = -\infty$. We wish to evaluate this for 
\be
 \nu =\xi/2 \ , \qquad z = \frac{\xi}{2}\sqrt{\varpi^2-q^2} w^{1/\xi}
\ee
taking $\xi \to \infty$ with $\varpi$, $q$, and $w$ fixed. In this limit the integral representation will be dominated near its saddle points. The result can be found in integral tables, but since the analysis is relatively simple, we derive the result here. 

To identify the saddle points, we study the argument of the exponential
\be
 f(t) = t - \frac{z^2}{4t} - (\nu+1) \log t \ .
\ee
The saddle points are defined by $f'(t) = 0$. This gives 
\be
 t = \frac{\nu+1}{2} \pm \frac{1}{2}\sqrt{\left(\nu+1\right)^2 - z^2 } \equiv t_\pm  \ .
\ee
Second derivatives are given by
\be
 f''(t_\pm) = \frac{2}{t_\pm^2}\left(t_\pm-\frac{\nu+1}{2}\right) = \pm\frac{\sqrt{\left(\nu+1\right)^2 - z^2 }}{t_\pm^2} \ .
\ee
Since $t_\pm \sim \xi$ and $f(t_\pm)\sim 1/\xi$, the contributions to the integrals near the saddle points are limited to $t-t_\pm \sim \sqrt{\xi}$, so that the saddle point approximation works (corrections from higher order terms in the expansion are suppressed by $1/\xi$). The integral can therefore be computed as
\be
 J_\nu(z) = \sum_{s.p.}\frac{1}{\sqrt{2\pi f''(t_\pm)}}  \left(\frac{z}{2}\right)^\nu e^{f(t_\pm)}\left[1+\mathcal{O}\left(\xi^{-1}\right)\right]
\ee
where the sum is only over those saddle points which lie on the suitably deformed (steepest descent) integration contour. 

At large $\xi$ the saddle points are found at
\be
 t_\pm = \frac{\xi}{4}\left({1 \pm \sqrt{1-\varpi^2+q^2}}\right) +\mathcal{O}\left(\xi^0\right) \ .
\ee
When $0\leq \varpi<\sqrt{q^2+1}$, the points are on the real axis, and the integration contour only goes through the point $t=t_+$. In this case we find that
\be \label{smallomsp}
 e^{-\frac{i\varpi \xi w^{1/\xi}}{2}}J_{\xi/2}\left(\frac{\xi}{2}\sqrt{\varpi ^2-q^2} w^{1/\xi}\right) = \frac{e^{\frac{1}{2} \xi  \left(\Sq-i \varpi\right)} \left(\frac{1-\Sq}{1+\Sq}\right)^\frac{\xi}{4} w^{\frac{1}{2} \left(-i \varpi+\Sq\right)}}{\sqrt{\pi  \xi \Sq } }\left[1+\mathcal{O}\left(\frac{1}{\xi}\right)\right] \ ,
\ee
where we also included the phase factor which appears in each solution in the text, and $S$ was defined in~\eqref{Sdef}. 
When $\varpi>\sqrt{q^2+1}$, we have $t_+ = (t_-)^*$ and the contour passes both saddle points. In this case
\bea \label{largeomsp}
&& e^{-\frac{i\varpi \xi w^{1/\xi}}{2}}J_{\xi/2}\left(\frac{\xi}{2}\sqrt{\varpi ^2-q^2} w^{1/\xi}\right) = \frac{1}{\sqrt{\pi  \xi \Sqt}}\Bigg[e^{-\frac{i \pi}{4}  } e^{\frac{1}{2} \xi  \left(i \Sqt-i \varpi\right)} \left(\frac{1-i \Sqt}{1+i \Sqt}\right)^\frac{\xi}{4} w^{\frac{1}{2} \left(i \Sqt-i \varpi\right)} \nn\\
 &&\qquad \qquad+ e^{\frac{i \pi }{4}} e^{\frac{1}{2} \xi  \left(-i \Sqt-i \varpi\right)} \left(\frac{1+i \Sqt}{1-i \Sqt}\right)^\frac{\xi}{4} w^{\frac{1}{2} \left(-i \Sqt-i \varpi\right)}\Bigg]\left[1+\mathcal{O}\left(\frac{1}{\xi}\right)\right] \ .
\eea
where we used the quantity $\Sqt$ of~\eqref{Stdef} the principal branch of which is analytic for $\varpi>\sqrt{q^2+1}$.

When $\varpi$ is not real, it is nontrivial to figure out which of the saddle points should be included. The ambiguous contribution is, however, from the saddle point which is exponentially subleading, and therefore it is not important to compute exactly when it should or should not be included in the sum. Because of reflection symmetry over the imaginary axis we may restrict to $\mathrm{Re}\,\varpi\geq 0$. Numerically we can verify that the former (latter) result is a good approximation when $0\leq \re \varpi \lesssim \sqrt{1+q^2}$ ($\re \varpi \gtrsim \sqrt{1+q^2}$). Actually, while the above conditions are enough for our purposes the region of validity is larger for each expression: Eq,~\eqref{smallomsp} holds everywhere (for $\mathrm{Re}\,\varpi\geq 0$) except for the immediate vicinity of the line $[\sqrt{1+q^2},\infty[$, where as Eq.~\eqref{largeomsp} holds everywhere except in the vicinity of $[0,\sqrt{1+q^2}]$. Both expressions fail in within $1/\xi$ distance from the point $\varpi = \sqrt{1+q^2}$ where the two saddle points merge. This issue could be fixed by considering a different analytic approximation, but this is not necessary for the scope of this article.

For the source term, we find similarly
\bea \label{sourcesmallom}
&& e^{-\frac{i\varpi \xi w^{1/\xi}}{2}}J_{-\xi/2}\left(\frac{\xi}{2}\sqrt{\varpi ^2-q^2} w^{1/\xi}\right) =  \frac{1}{\sqrt{\pi  \xi \Sq} }\Bigg[ 2 \sin\left(\frac{\pi \xi}{2}\right) e^{\frac{1}{2} \xi  \left(-\Sq-i \varpi\right)} \left(\frac{1+\Sq}{1-\Sq}\right)^\frac{\xi}{4}  \nn\\
&&\qquad \times w^{\frac{1}{2} \left(-\Sq-i \varpi\right)} + e^{-\frac{i \pi  \xi }{2}} e^{\frac{1}{2} \xi  \left(\Sq-i \varpi\right)} \left(\frac{1-\Sq}{1+\Sq}\right)^\frac{\xi}{4} w^{\frac{1}{2} \left(\Sq-i \varpi\right)}\Bigg]\left[1+\mathcal{O}\left(\frac{1}{\xi}\right)\right] 
\eea
when $0 \leq \re \varpi \lesssim \sqrt{1+q^2}$ and
\bea
&& e^{-\frac{i\varpi \xi w^{1/\xi}}{2}}J_{-\xi/2}\left(\frac{\xi}{2}\sqrt{\varpi ^2-q^2} w^{1/\xi}\right) =  \frac{1}{\sqrt{\pi  \xi \Sqt}}\Bigg[e^{\frac{i \pi  \xi }{2}-\frac{i \pi}{4}  } e^{\frac{1}{2} \xi  \left(i \Sqt-i \varpi\right)} \left(\frac{1-i \Sqt}{1+i \Sqt}\right)^\frac{\xi}{4} \nn \\\nn
 &&\qquad \times w^{\frac{1}{2} \left(i \Sqt-i \varpi\right)} + e^{-\frac{i \pi  \xi }{2}+\frac{i \pi }{4}} e^{\frac{1}{2} \xi  \left(-i \Sqt-i \varpi\right)} \left(\frac{1+i \Sqt}{1-i \Sqt}\right)^\frac{\xi}{4} w^{\frac{1}{2} \left(-i \Sqt-i \varpi\right)}\Bigg]\left[1+\mathcal{O}\left(\frac{1}{\xi}\right)\right] 
\eea
for  $\re \varpi \gtrsim \sqrt{1+q^2}$. In the second term of~\eqref{sourcesmallom}, the phase factor $e^{-\frac{i \pi  \xi }{2}}$ is correct for $\im \varpi>0$. For $\im \varpi<0$, this factor should be\footnote{Following precisely the contours of steepest descent leads to a more complicated structure on when the various saddle point contributions should be included, which is not analytically tractable. All results presented here agrees with the exact result (the Bessel functions) up small corrections for the values of $\varpi$ specified in the text.  The fact that we do not follow the precise contours of steepest descent in principle leads to ambiguities in the results, but within the specified regimes these involve only saddle point contributions which are strongly suppressed with respect to the leading result. } $e^{\frac{i \pi  \xi }{2}}$.

\section{Numerical check of the transverse correlator of the CR plasma} \label{app:comparison}

In this Appendix we check numerically the analytic results in Sec.~\ref{sec:Xonehalf}.
In order to compare the analytic approximation  
to the full correlator, we evaluated it numerically (see Sec.~\ref{sec:numerics}).  
The results for the comparison were already presented in Fig.~\ref{fig:corrcomparisonx045} in the main text. As we argued in Sec.~\ref{sec:flucts}, the result only depends on $\hat r_h$ trivially, and for this plot we have set $\hat r_h = 1$. In the rest of this Appendix we do several additional numerical checks.

\begin{figure}[!tb]
\begin{center}
\includegraphics[width=0.49\textwidth]{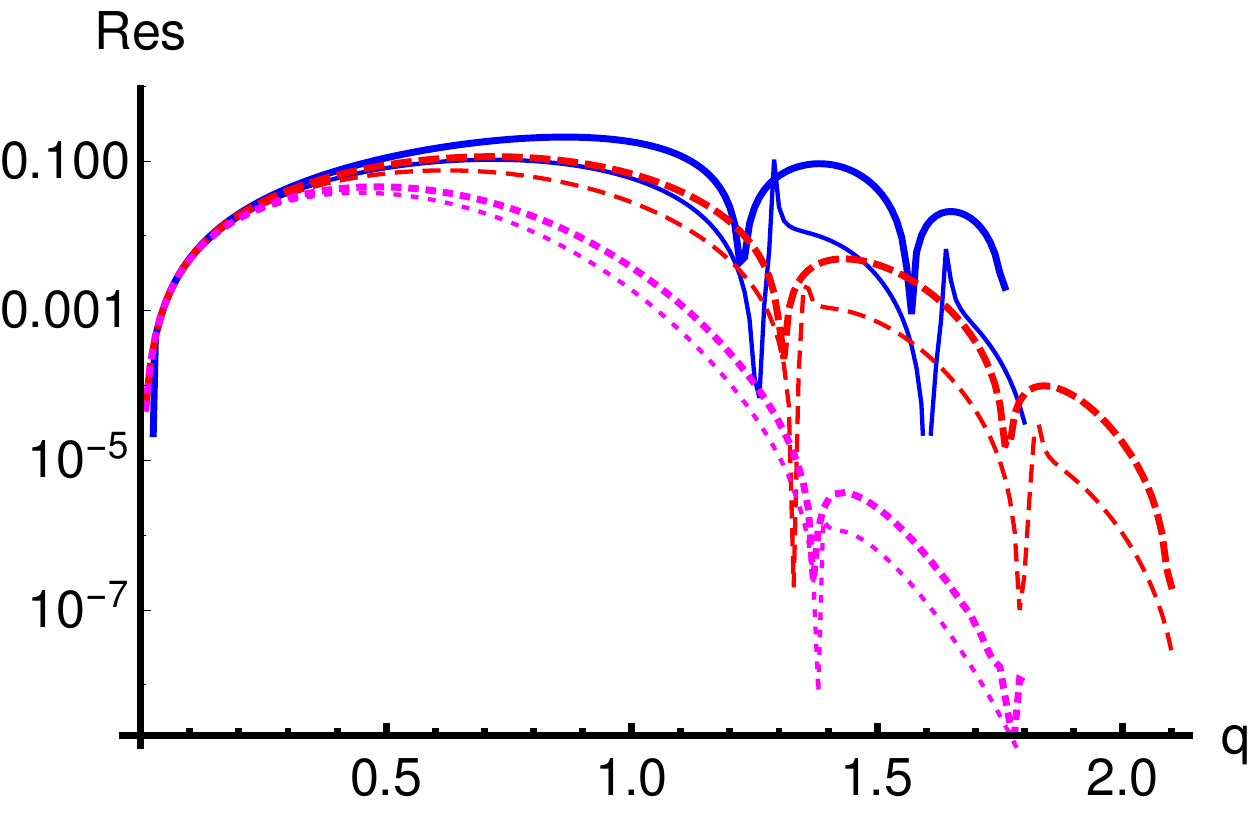}%
\hspace{2mm}\includegraphics[width=0.49\textwidth]{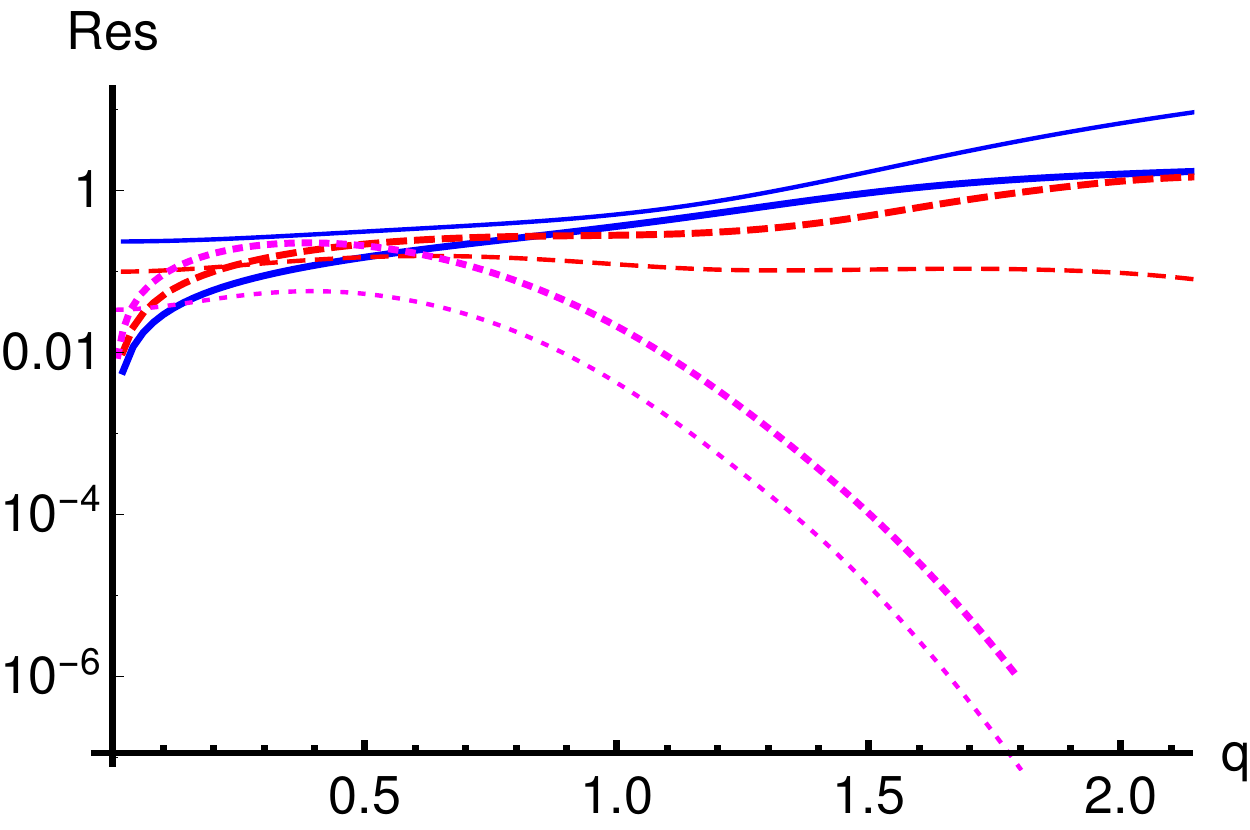}
\end{center}
\caption{The (absolute values of the) residues of the hydrodynamic modes compared to the analytic approximation for the (transverse tensor) correlator at the location of the mode as a function of $q$. Left: shear mode. Right: sound mode. The thick curves are the values of the residues and the thin curves show the correlator at the location of the mode. The values of $X$ are (roughly from top to bottom) for the solid blue, dashed red, and dotted magenta curves are $X=0$, $X=-0.35$, and $X=-0.45$, respectively. }
\label{fig:residuecomp}
\end{figure}

We have carried out a further rough check in the region of large negative $\mathrm{Im}\, \varpi$, by comparing the residues of the hydrodynamic modes in the shear and sound channels to the analytic approximation of the transverse tensor correlator. This makes sense because, as it turns out, the residues of the hydrodynamic modes are numerically easier to compute than full numerical result for the correlator, and therefore can be used to probe the regime of large negative $\mathrm{Im}\, \varpi$. A natural  expectation is that at large $q$ the residues are of the same order as the average values of the correlators near the modes. If the correlators in the shear and sound channels are further comparable to the transverse tensor correlator, the analytic approximation for the latter, evaluated at the location of the hydro mode, should be close to the residue for $X$ close to $-1/2$. We have tested this numerically and show the results in Fig.~\ref{fig:residuecomp}. For the shear mode (left plot) the comparison works remarkably well even at $X=0$ (blue curves). In particular, the zeroes of the residues appear very close to the zeroes of the analytic result, and the agreement improves as $X \to -1/2$: for $X=-0.45$ (magenta curves). For the sound mode (right plot) a good agreement is only seen for $X=-0.45$.

We have also carried out the comparison of the above formulae~\eqref{quant} and~\eqref{imom} to the nodes of the analytic approximation for the correlator 
and to the modes extracted numerically directly from the fluctuation equation. 

\begin{figure}[!tb]
\begin{center}
\includegraphics[width=0.49\textwidth]{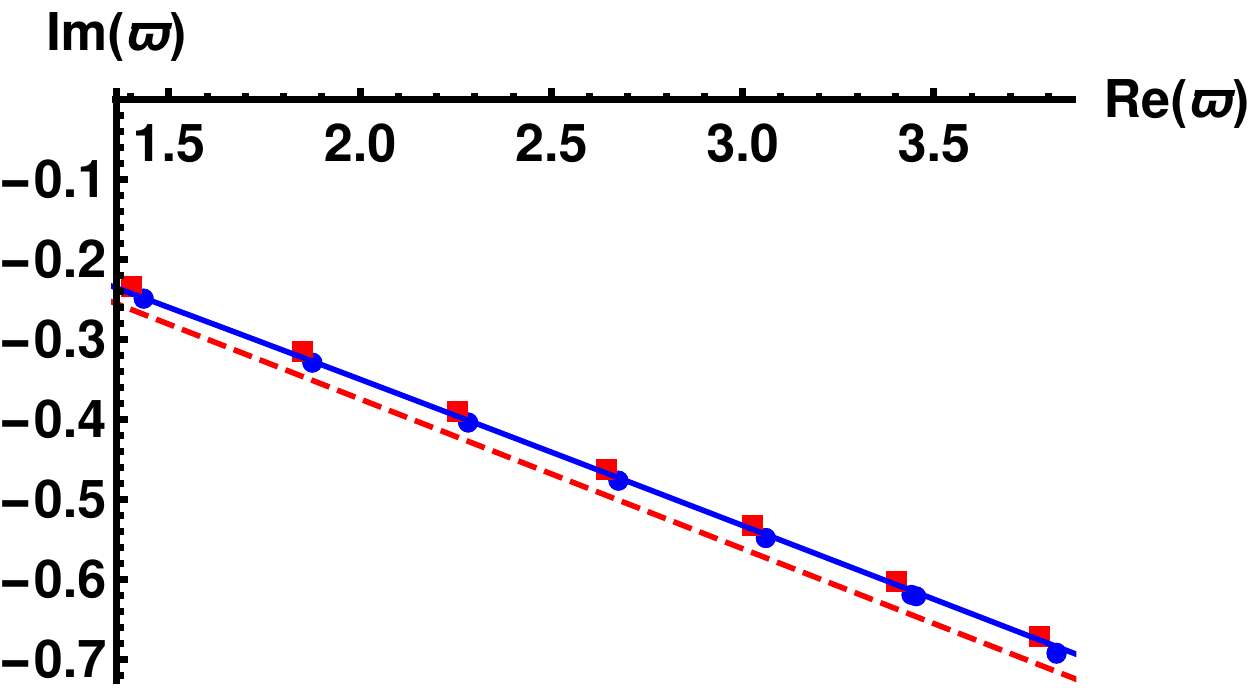}%
\includegraphics[width=0.49\textwidth]{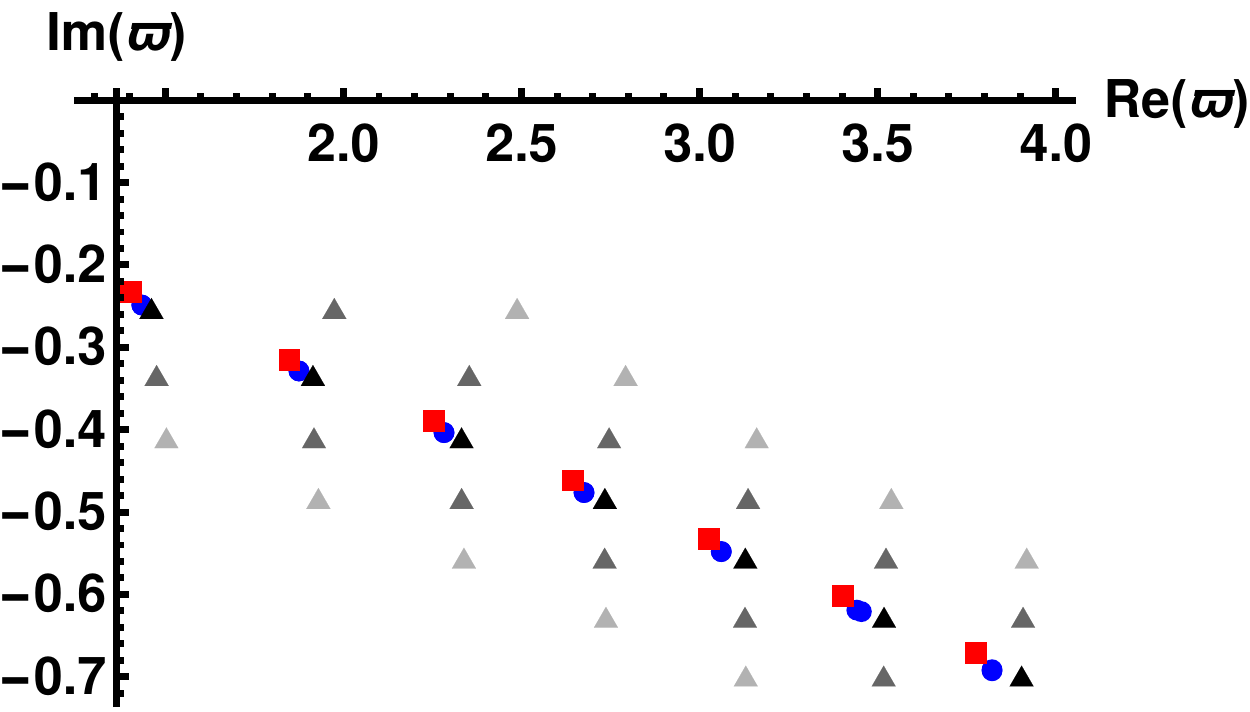}

\vspace{2mm}

\includegraphics[width=0.49\textwidth]{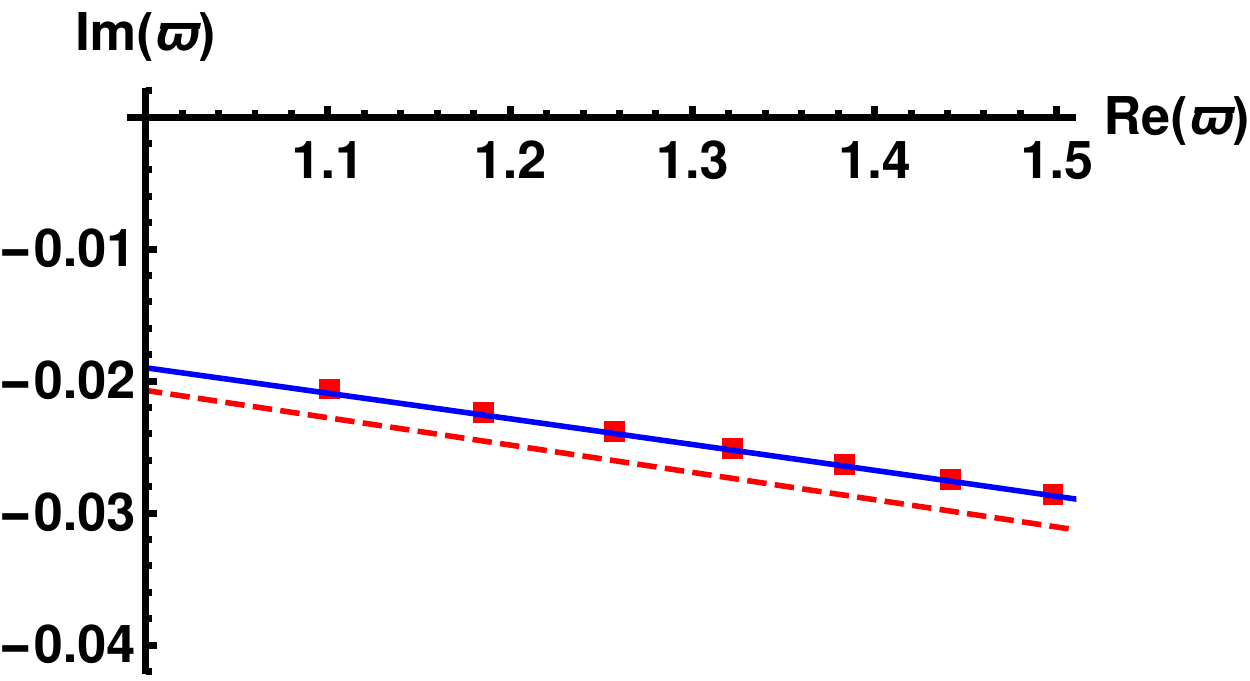}%
\includegraphics[width=0.49\textwidth]{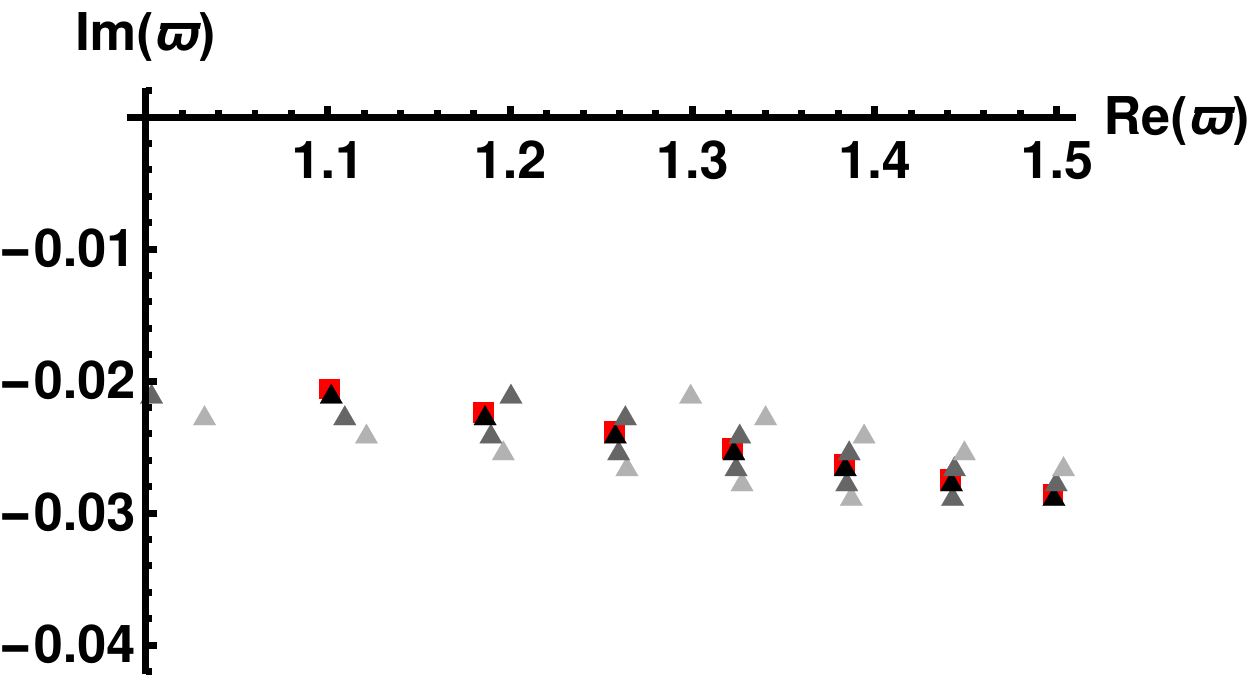}
\end{center}
\caption{Comparison of the numerical results for the location of quasi normal modes to analytic approximations. Top row: $X=-0.45$. Bottom row: $X=-0.495$. See text for details.}
\label{fig:comparisonx045}
\end{figure}

We plot the locations of some modes (relatively low in the spectrum) on the complex $\varpi$-plane for $X=-0.45$ ($\xi = 16.7895$) and for $q=0$ in Fig.~\ref{fig:comparisonx045} (top row). The blue disks are the numerical results from solving the full equation~\eqref{nonhydroeq} in both plots. Similarly, the red boxes are the nodes of the analytic approximation~\eqref{corrfinal} (which have been extracted numerically). The blue curve in the left-hand plot is given by~\eqref{imom}, or more precisely by varying the point of expansion as $\varpi_0 = \re \varpi$ in this formula. The red dashed curve is the approximation at large $\varpi$: $\im \varpi = - \pi \re \varpi$. In the left-hand plot, we show the locations of the nodes obtained from~\eqref{quant} as triangles. The triangles with different shades of gray correspond to different expansion points $\Sqt = \Sqt_0$. The black triangles are for ``optimal'' choices of the expansion point, defined such that $\re\, \delta \varpi = 0 $ for $n=0$ in~\eqref{quant}. That is, the expansion point lies directly above the predicted location of the node. The dark gray triangles are then the nodes for $n=\pm 1$ and light gray triangles for $n = \pm 2$. The agreement is good, given that the approximation~\eqref{quant} is based on linear expansion around points on the real $\varpi$-axis.

We show similar results for $X=-0.495$ in Fig.~\ref{fig:comparisonx045} (bottom row) (so that $\xi \simeq 151.75$). This value lies so close to the critical point $X=-0.5$ that the direct numerical solution of the fluctuation equations was not possible. Therefore we show only the predictions from the various analytic approximations. As $\xi$ grows, the analytic approximation becomes better as expected. In particular the locations of the QNMs in the bottom right plot now shows convergence.

\section{Fluctuations analysis for a  generic dilaton potential} \label{app:UVmodflucts}

In this Appendix we discuss the fluctuations for  
the case of a smooth, generic dilaton potential $V(\phi)$ which produces asymptotically AdS (CR) geometry in the UV (IR). We use the conformal coordinates in terms of which the fluctuation equation takes a simple form. In order to describe the UV behavior, we need the fluctuation equations for a generic potential $V(\phi)$. As we constrain ourselves to the case of small temperatures, so that the horizon lies in the IR asymptotic region of the metric and the blackening factor equals one up to $r \sim \ell$, it is enough to study the fluctuations for generic $V(\phi)$ at zero temperature.

At zero temperature, all fluctuations of the metric satisfy the same equation, given by\footnote{In Sec.~\ref{sec:flucts} we wrote the fluctuation equations in the Eddington-Finkelstein coordinates. However, here it is convenient to use the true time coordinate $t$ instead of the tortoise coordinate $v$ in order to restore Lorentz covariance explicitly.}
\be
 \flf''(r) +3 A'(r) \flf'(r) + m^2 \flf(r) = 0 \ ,
\ee
where $m^2 = \omega^2-k^2$.
The dilaton fluctuation mixes with the metric, and its equation can be written as 
\be
\zeta''  (r) + \le(3A'(r) + 2 \frac{z'(r)}{z(r)} \ri) \zeta'(r) + m^2 \zeta(r) = 0 
\ee
where $z(r) = \phi'(r)/A'(r)$. 
Here $\zeta$ is related to the definitions of Sec.~\ref{sec:flucts} by
\be
 e^{i \omega r}\zeta(r) = \frac{1}{2} \le(\tilde H_{22}(r) +\tilde H_{33}(r)\ri) - \frac{2}{z(r)} \tilde \psi(r) 
\ee
where the exponential factor arises due to the change of the time coordinate from $t$ to $v$.

Near the boundary the fluctuation equations take the AdS$_5$ form:
\begin{align}
 \flf''(r)  -\frac{3}{r} \flf'(r) + m^2 \flf(r) = 0 \ , \\
 \zeta '' + \frac{2\Delta-3}{r} \zeta'(r) + m^2 \zeta(r) = 0 \ , 
 \label{AdSflucts}
 \end{align}
where we dropped $\morder{(r/\ell_\mathrm{AdS})^\Delta}$ corrections. 

Because the metric behaves smoothly around the transition region ($r \sim \ell$), the fluctuation wave function will be smooth as well. This implies that we can require approximate continuity between the wave functions in the asymptotic AdS ($r \ll \ell$) and CR ($r \gg \ell$) regimes, up to factors $\morder{1}$ which arise from the nontrivial evolution of the functions over the transition region. This can be used to estimate the locations of the QNMs for any value of $m$ so long as $T<T_c$. We will make the matching procedure more precise in the main text.

When, in addition, $m$ is small we can make a generic precise statement about the QNMs. Namely, in Sec.~\ref{sec:Xonehalf} we demonstrated that the locations of the QNMs are determined by the geometry in the vicinity of the horizon. Therefore we expect that at small temperatures and for small enough $m$ the QNMs of the UV modified geometry match with those of the CR geometry. In particular, when $\xi$ is large, the analytic results of Sec.~\ref{sec:Xonehalf} for their locations apply. We can read from the above equations when this is the case as follows.

In the UV, the wave function of the quasi normal mode is normalizable. However, it is not necessary to compute the wave function down to $r=0$ exactly in order to determine the QNMs at finite precision. Depending on $r$, $\omega$, and $k$, the basis of solutions to the fluctuation equations can be chosen to be either a pair where one of the functions rapidly increases and the other decreases with $r$, or a pair corresponding to an ingoing and an outcoming wave. Close to the UV boundary we expect the former, and near the horizon the latter. 
We need the change in the behavior of the fluctuations to happen far in the IR with respect to the transition region $r \sim \ell$ for the QNMs to be determined by the IR part of the geometry only. Since, as we have demonstrated above, the transition of the geometry is smooth, it is enough to look at the AdS equation~\eqref{AdSflucts}. We see that the wavy behavior is absent for $m \ll 1/\ell$. In conclusion, the results for the QNMs in Sec.~\ref{sec:Xonehalf} are reliable if $T < T_c$ and $m \ll 1/\ell$.

The lowest quasi normal modes have $m \sim T$ as we have seen in Sec.~\ref{sec:Xonehalf}. Therefore for the conditions $T < T_c$ and $m \ll 1/\ell$ to hold simultaneously we actually need that $T \ll T_c$ (since $T_c \sim 1/\ell$). The condition for the CR results to hold for a certain mode can be written as $T \ll 1/((m/T)\ell)$ where the ratio $m/T$ is independent of $T$ to a good approximation.

We will now formulate the statement on continuity more precisely. We discuss the wave function $\zeta$; the definitions for the transverse spin two modes are obtained by setting $\Delta = 0$. 

A generic solution to the fluctuation equations (at zero temperature) can be expressed either in UV or IR bases, which are defined as follows. In the UV ($r \to 0$), the equation~\eqref{AdSflucts} is solved by
\begin{align} \label{UVexps}
\zeta (r) &= C_\mathrm{UV}^{(1)} \frac{i \pi }{2^{2-\Delta } \Gamma (2-\Delta ) m^{\Delta -2}} r^{2-\Delta}H^{(1)}_{2-\Delta}( m\, r )  + C_\mathrm{UV}^{(2)} \frac{\Gamma (3-\Delta )}{2^{\Delta -2} m^{2-\Delta }} r^{2-\Delta} J_{2-\Delta}(m\,r) & \nn\\
&\equiv  C_\mathrm{UV}^{(1)} \zeta_\mathrm{UV}^{(1)}(r) + C_\mathrm{UV}^{(2)} \zeta_\mathrm{UV}^{(2)}(r) &
\end{align}
up to corrections $\morder{(r/\ell_\mathrm{AdS})^\Delta}$. Here $H^{(1)}$ is the Hankel function of the first kind. In the IR we find that
\begin{align} \label{IRexps}
\zeta (r) &=C_\mathrm{IR}^{(1)} \frac{i \pi  (r+\ell')^{\xi /2}  H_{\xi/2}^{(1)}\le(m(r+\ell')\ri)}{2^{\xi /2} m^{-\xi/2} \Gamma \left(\frac{\xi }{2}\right)} + C_\mathrm{IR}^{(2)} \frac{2^{\xi /2}\Gamma \left(\frac{\xi }{2}+1\right) (r+\ell')^{\xi /2} J_{\xi/2}\le(m(r+\ell')\ri)}{ m^{\xi /2}{\ell'}^\xi} & \nn\\
&\equiv  C_\mathrm{IR}^{(1)} \zeta_\mathrm{IR}^{(1)}(r) + C_\mathrm{IR}^{(2)} \zeta_\mathrm{IR}^{(2)}(r) &
\end{align}
up to logarithmically suppressed corrections. Notice that we included the shift by $\ell'$ which also appears in the background above, and the normalization of the vev term is the same is in~\eqref{Iexpr} and~\eqref{Kexpr} (where $p^2 = - m^2$) -- the factor of ${\ell'}^\xi$ was absorbed into a rescaling of $r$ in Secs.~\ref{sec:bg} and~\ref{sec:Xonehalf}. The presence of this factor ensures that the UV and IR fluctuations can be matched in the transition regime $r\sim \ell$ with $\morder{1}$ coefficients at large $\xi$.

The UV and IR coefficients are related through a transition
\be
 C_\mathrm{UV} = M C_\mathrm{IR} 
\ee
where $C_\mathrm{UV/IR} = (C_\mathrm{UV/IR}^{(1)},C_\mathrm{UV/IR}^{(2)})$ and, thanks to linearity, the $2\times 2$ transition matrix $M$ is the same for all solutions $\zeta$. 
The form of the matrix $M$ will of course depend on the details of the evolution of the fluctuations over the transition region $r \sim \ell$. This can in principle be computed numerically but the computation turns out to be challenging due to precision issues. However we can find an approximate form simply using the asymptotic solutions for the fluctuations and requiring continuity (and continuity of the derivatives) to hold between the IR and UV expansions near $r \sim \ell$. 

We develop the continuity argument for an explicit analytic approximation in the case of the fluctuations of the metric in Sec.~\ref{sec:gluing}. In the rest of this Appendix we point out some general properties of $M$ in the limit of small $m$. Namely, for zero $m$, the fluctuation equations admit an exact solution:
\be \label{zetaex}
 \zeta(r) = C_1 + C_2 \int_0^r \frac{e^{-3A(\tilde r)}}{z(\tilde r)^2} d\tilde r \ .
\ee
This implies the following relations at $m=0$
\begin{align}
C_\mathrm{UV}^{(1)} &= C_\mathrm{IR}^{(1)} + \morder{1} C_\mathrm{IR}^{(2)}  &\\
e^{2 \Delta \tilde A_0} \phi_0^2 \Delta^2(4-2\Delta) \ell_\mathrm{AdS}^{3-2\Delta} C_\mathrm{UV}^{(2)} &= 4(1-X^2) X^2 e^{4A_0} \ell^{-1} C_\mathrm{IR}^{(2)} &
\end{align}
That is, three of the components of the matrix $M$ could be solved. At finite $m$ this implies that 
\begin{align}
 M_{11} &= 1 + \morder{m^2} \ ,  \qquad M_{21} = \morder{m^2} \ ,&\nonumber\\
  M_{22} &= \frac{4(1-X^2) X^2 e^{4A_0} }{e^{2 \Delta \tilde A_0} \phi_0^2 \Delta^2(4-2\Delta) \ell_\mathrm{AdS}^{3-2\Delta}\ell} + \morder{m^2} \ .&
\end{align}
For the case of the transverse spin-two correlator ($\Delta =0$) the expression for the element $M_{22}$ is singular. In this case the exact solution is the same as in~\eqref{zetaex} but without the $z$-dependent factor. Repeating the calculation for this solution yields 
\be \label{M22spintwo}
   M_{22} = \frac{e^{4A_0}(1-X^2)}{\ell\,\ell_\mathrm{AdS}^3}  + \morder{m^2} \ .
\ee

Finally we express the correlators in terms of the transition matrix and the analytic expressions derived in Sec.~\ref{sec:Xonehalf}. The correlator is given as the ratio of the coefficient of the terms $\propto r^0$ and $\propto r^{4-2\Delta}$ in the UV expression~\eqref{UVexps}. Taking into account the subleading terms of the Hankel function in~\eqref{UVexps} this gives
\be \label{G0expr}
 \widetilde G = \widetilde G_0 + \widetilde G_\mathrm{reg} \equiv -\frac{2^{2 \Delta -4} e^{i \pi  \Delta } m^{4-2 \Delta } \Gamma (\Delta -1) }{\Gamma (3-\Delta )} + \frac{C_\mathrm{UV}^{(2)}}{C_\mathrm{UV}^{(1)}}
\ee
where the regular term was separated as in Appendix~\ref{sec:bdry}. There is a singularity at $\Delta = 0$ in the first term which is regulated (due to cancellation of two terms in the UV expansions when $\Delta= 0$ exactly) as in Sec.~\ref{sec:bdry}. In terms of the IR quantities we find
\be \label{Gmatrix}
 \widetilde G = \widetilde G_0 + \frac{M_{21}+M_{22} \frac{C_\mathrm{IR}^{(2)}}{C_\mathrm{IR}^{(1)}}}{M_{11}+M_{12} \frac{C_\mathrm{IR}^{(2)}}{C_\mathrm{IR}^{(1)}}} =\widetilde G_0 + \frac{M_{21}+M_{22}  G_\mathrm{reg}}{M_{11}+M_{12}  G_\mathrm{reg}} \ .
\ee
Here $G_\mathrm{reg}$ is the analytic expression derived in Sec.~\ref{sec:Xonehalf}. One should recall, however, that the origin of the coordinate $r$ is shifted in the IR regime with respect to the definitions of Sec.~\ref{sec:Xonehalf}, and that the $\hat r$ coordinate of  Sec.~\ref{sec:Xonehalf} is moreover related to $r$ by rescaling $\hat r =r/\ell'$ (i.e., by a factor which diverges as $\xi \to \infty$). This implies that the factor $\hat r_h^{-\xi}$ in $G_\mathrm{reg}$ of Sec.~\ref{sec:preciselimit} should be replaced by $(1+r_h/\ell')^{-\xi}$.

\end{document}